\documentclass[12pt,oneside,a4paper]{article}

\usepackage{hyperref}
\hypersetup{colorlinks,bookmarksopen,bookmarksnumbered,citecolor=blue,
linkcolor=black,pdfstartview=FitH,urlcolor=blue}

\usepackage{graphicx}
\usepackage{amssymb}
\usepackage{cite}
\usepackage{bm}
\usepackage{indentfirst}
\usepackage{amsmath}
\usepackage{hhline}
\usepackage{multirow}
\usepackage{color}
\usepackage[hang,small,bf]{caption}
\usepackage[subrefformat=parens]{subcaption}
\usepackage{comment}
\usepackage{ulem}

\usepackage{braket}
\usepackage{textcomp}

\allowdisplaybreaks
\numberwithin{equation}{section}

\renewcommand{\bar}{\overline}

\definecolor{dred}{rgb}{0.7,0.15,0.09}
\definecolor{dblue}{rgb}{0,0.12,0.64}
\definecolor{dgreen}{rgb}{0.2,0.51,0.19}

\newcommand{\Slash}[1]{{\ooalign{\hfil/\hfil\crcr$#1$}}}

\begin{document}

\begin{titlepage}


\begin{center}

\vspace{1cm}
{\large\textbf{
Distance-dependent interaction between cosmic strings \\
inspired by higher-dimensional gauge theory
}
}

\vspace{1cm}

\renewcommand{\thefootnote}{\fnsymbol{footnote}}
Takuya Hirose$^{1}$\footnote[1]{t.hirose@ip.kyusan-u.ac.jp}
and
Yukihiro Kanda$^{2}$\footnote[2]{kanda.y@eken.phys.nagoya-u.ac.jp}
\vspace{5mm}

\textit{
 $^1${Faculty of Science and Engineering, Kyushu Sangyo University, \\
 Fukuoka 813-8503, Japan}\\
 $^2${Department of Physics, Nagoya University, Nagoya 464-8602, Japan}
}

\vspace{8mm}

\abstract{

We discuss Abrikosov-Nielsen-Olesen (ANO) strings with the one-loop effective potentials induced by higher-dimensional gauge theory. 
As our starting point, we consider a five-dimensional $SU(2)$ gauge theory with an extra-dimensional space $S^1/Z_2$.
We numerically show the properties of the strings in our model.
Especially, we investigate the interaction force between two parallel strings. 
We find that the interaction force switches from attraction to repulsion as two strings approach each other at a certain parameter region. 
This interaction is not observed for the ANO strings with the Mexican hat potential. 
Furthermore, we find that each different factor determines the interaction between large and small interstring distances. 
We interpret this difference as the origin of the distance-dependent interaction. Our interpretation can also be applied to other scalar potentials. 
The results in our study give us a new perspective to understand the interaction between the ANO strings with various scalar potentials. 

}

\end{center}
\end{titlepage}

\renewcommand{\thefootnote}{\arabic{footnote}}
\newcommand{\bhline}[1]{\noalign{\hrule height #1}}
\newcommand{\bvline}[1]{\vrule width #1}

\setcounter{footnote}{0}

\setcounter{page}{1}

\section{Introduction}

Cosmic strings are one-dimensional topological defects in quantum field theory, which is a line-like solution due to the non-triviality of the fundamental group of the vacuum manifold \cite{Kibble:1976sj}. 
(For a review, see Ref. \cite{Vilenkin:1984ea}.) 
They are realized in particle models that predict $U(1)$ symmetry breaking and serve as probes for these models \cite{Kibble:1982ae, Jeannerot:2003qv, Marsh:2015xka, Croon:2019kpe}.
Furthermore, the classical solutions corresponding to cosmic strings have been applied in many studies within the condensed matter physics \cite{Mermin:1979zz, Volovik, Svistunov, Pismen, Bunkov, Blatter:1994zz, Kawaguchi:2012ii}.

After the phase transition in the early universe,
cosmic strings are produced through the Kibble-Zurek mechanism \cite{Kibble:1976sj,Zurek:1985qw}, and form a network (called string network system). 
The string network system can be detected through the various cosmological observables such as the cosmic microwave background \cite{Albrecht:1997nt, Wyman:2005tu, Bevis:2007qz}, gravitational lens \cite{Vilenkin:1984ea} and the gravitational wave background \cite{Damour:2000wa, Damour:2001bk}. 
Therefore, the cosmic strings are good tools to probe the past $U(1)$ symmetry breaking predicted in a lot of models beyond the Standard Model (SM) \cite{Kibble:1982ae, Jeannerot:2003qv, Marsh:2015xka, Croon:2019kpe}.

A famous example of the cosmic strings is the Abrikosov-Nielsen-Olesen (ANO) string \cite{Abrikosov:1957wnz,Nielsen:1973cs} realized in the Abelian-Higgs model. 
The ANO string solution is obtained as a configuration of the scalar field $\phi$ and the gauge field $A_\mu$.
This configuration features a line-like region where $\phi$ and $A_\mu$ are excited. 
Conventionally, the ANO string is discussed under the scalar potential of the Mexican hat type: $V(\phi)=\lambda(|\phi|^2-v^2)^2$. 
In this case, the features of the ANO string are characterized by the ratio between the scalar and gauge masses, denoted as $\beta\equiv m_\phi^2/m_A^2$. 
Especially, the value of $\beta$ determines the interaction forces between two parallel ANO strings: 
when $\beta < 1$, the ANO strings attract each other, whereas when $\beta > 1$, they repel each other \cite{Bogomolny:1975de, Goodband:1995rt, Speight:1996px}. 
This property is used to explain the characteristics of superconducting materials \cite{Abrikosov:1957wnz, Blatter:1994zz}. 
By this analogy, the ANO strings with $\beta<1$ are referred to as type-I strings, while those with $\beta>1$ are referred to as type-II strings. 
When $\beta=1$, there are no forces between the ANO strings.
This situation is known as the Bogomol’nyi-Prasad-Sommerfield (BPS) state \cite{Bogomolny:1975de,Prasad:1975kr}. 

Besides the ANO string, there are many types of string solutions such as the global strings \cite{Vilenkin:1982ks}, superconducting strings \cite{Witten:1984eb, Jackiw:1981ee}, semi-local strings \cite{Vachaspati:1991dz, Achucarro:1999it}, and embedded strings \cite{Nambu:1977ag, Vachaspati:1992fi,Vachaspati:1992pi, Barriola:1993fy, Lepora:1995rg, Lepora:1995ri, Kanda:2023yyz}. 
It is important to investigate their properties because these strings may cause the different evolution of the string networks from those of the ANO string. 
For example, the network evolution for global strings and ``global-like'' strings---ANO strings with a very small gauge coupling---deviates from the scaling law \cite{Gorghetto:2018myk, Kawasaki:2018bzv, Buschmann:2019icd, Klaer:2019fxc, Kitajima:2022lre}, leading to a gravitational wave spectrum that differs from conventional ANO strings \cite{Kitajima:2022lre}. 
\footnote{On the other hand, Ref. \cite{Hindmarsh:2019csc, Hindmarsh:2021vih} claims that the network of the global strings obeys the conventional scaling law. } 
This deviation potentially explains the Pulsar Timing Array results in 2023 \cite{Kitajima:2023vre}, unlike the conventional stable ANO string \cite{NANOGrav:2023hvm}. 
Investigating the interaction between two parallel strings is a first step toward understanding the evolution of networks for various types of cosmic strings. 
Numerous studies have been conducted on this topic to better understand the dynamics of cosmic string networks \cite{Shellard:1987bv, Bettencourt:1994kf, Speight:1996px, Jacobs:1978ch, Eto:2022hyt, Fujikura:2023lil}.

In recent years, it has been pointed out that the features of interactions between strings change due to deviations of the scalar potential from the Mexican hat potential. 
In Ref. \cite{Eto:2022hyt}, the authors discussed the ANO string with the Coleman-Weinberg (CW) potential \cite{Coleman:1973jx} and call this string as the CW-ANO string.
In this model, the scalar field obtains a vacuum expectation value (VEV) due to the CW potential. 
The authors found that the CW-ANO string exhibits two interesting interaction properties that differ from those of the ANO string with the Mexican hat potential. 
First, a state corresponding to the BPS state does not exist for the CW-ANO string. 
Hence, the boundary between the type-I (attractive) and type-II (repulsive) regions of $\beta$ is not well-defined. 
Second, for the values of $\beta$ above the type-I region, the interaction between strings switches between attraction and repulsion depending on the distance between them. 
As mentioned in Ref. \cite{Eto:2022hyt}, it is challenging to understand why this distance dependence arises for the CW-ANO string. 
One of our motivations for this study is to provide a new perspective on this issue by examining the ANO string with different scalar potentials for Mexican hat type and CW type. 

To investigate a new cosmic string based on our motivations, higher-dimensional gauge theories \cite{Manton:1979kb, Hosotani:1983xw, Hosotani:1988bm, Hatanaka:1998yp} are the most interesting model.
The higher-dimensional gauge theory is a good candidate for a model beyond the SM.
It has known that the higher-dimensional gauge theories induce zero modes from extra components of higher-dimensional gauge field (so-called Wilson-line (WL) scalar field).
If the WL scalar field acquire the VEV in the higher-dimensional theories, the spontaneous symmetry breaking (SSB) of the gauge symmetry occurs, which is called the Hosotani mechanism \cite{Hosotani:1983xw, Hosotani:1988bm}.
\footnote{Of the higher-dimensional gauge theories, gauge-Higgs unification (GHU) \cite{Kubo:2001zc, Scrucca:2003ra, Haba:2004qf, Maru:2006wa, Hosotani:2007kn, Adachi:2018mby} is an attractive model since the WL scalar field is identified as a SM Higgs boson in the GHU.
The GHU gives us a solution of the gauge hierarchy problem and the finite Higgs mass \cite{Hatanaka:1998yp, Maru:2006wa, Hosotani:2007kn}. }
It has known that there are some studies on the application of the topological defect in the context of the higher-dimensional gauge theories \cite{Lim:2018lgg,Hasegawa:2019gpc,Adachi:2022kqs}.

In the higher-dimensional gauge theory, its one-loop effective potential is roughly cosine type potential\cite{Hosotani:1983xw}, and
are different from both the Mexican hat potential and the CW potential.
Typically, this potential tends to take larger values around the origin compared to the Mexican hat potential, while the CW potential is flatter in the same region. 
Hence, it is expected that the string produced with such a potential exhibits new features in the interactions between the strings, which are distinct from those observed for the conventional ANO string and the CW-ANO string.

In this paper, we consider the string solutions under the one-loop effective potentials induced by the higher-dimensional gauge theory, and investigate the interactions between the strings.
We start from a five-dimensional $SU(2)$ gauge theory with an extra-dimensional space $S^1/Z_2$ orbifold. 
Due to the orbifold, the Abelian-Higgs model without a potential at the classical level is realized as the four-dimensional effective theory. To break the $U(1)$ gauge symmetry via Hosotani mechanism, we introduce some fermions.
We consider three cases with different heights of the potential around the origin, and discuss the corresponding string solutions for each case.
To examine the interactions between the strings in the higher-dimensional gauge theory, we estimate the interaction energy of the two-string system as a function of the interstring distance $d$. 
As a result, We find that a novel property like the type-1.5 superconductor \cite{Babaev:2004hk, PhysRevLett.102.117001} is emerged.
By comparing the results in Ref. \cite{Eto:2022hyt} with our results, we discuss the relationship between string interactions and the scalar potential.

The outline of this paper is as follows: 
We briefly review the ANO string in Sec. \ref{Sec.ANOstring} and the five-dimensional gauge theory in Sec. \ref{5dgaugetheory}.
In Sec. \ref{Sec.1stringresult}, we examine single string solutions in the higher-dimensional $SU(2)$ gauge theory and their properties. 
We investigate the interaction between the two strings in Sec. \ref{2strings} and discuss the relationship between string interactions and the scalar potential. 
We present our conclusion in Sec. \ref{conclusion}.

\section{Abrikosov-Nielsen-Olesen string}
\label{Sec.ANOstring}
Before we investigate the cosmic string for several potentials, it is instructive to introduce Abrikosov-Nielsen-Olesen (ANO) string \cite{Abrikosov:1957wnz,Nielsen:1973cs}. 
In this section, we review the ANO string solution and the interaction between the ANO strings. 

\subsection{ANO string solution}

The ANO string is originally realized as a vortex string solution in the Abelian-Higgs model. 
The Lagrangian density is given by
\begin{align}
    \label{AHmodel}
    \mathcal{L} &= -\frac{1}{4} F_{\mu\nu} F^{\mu\nu} - |D_\mu \phi|^2 - V(\phi),
\end{align}
where the field strength is defined by $F_{\mu\nu} = \partial_\mu A_\nu - \partial_\nu A_\mu$ and the covariant derivative is given by $D_\mu = \partial_\mu - ig A_\mu$ with a gauge coupling $g$. 
In this paper, we use the metric $\eta=(-,+,\cdots,+)$.
Applying the variational principle from for Eq. \eqref{AHmodel}, the equations of motion in this model are derived as
\begin{align}
    &D_\mu D^\mu \phi - \frac{\partial V(\phi)}{\partial\phi^\dag} = 0, \label{EOM_phi} \\
    &\partial^\nu F_{\nu\mu} - ig \left(\phi^\dagger D_\mu \phi - (D_\mu \phi)^\dagger \phi\right) = 0. \label{EOM_gauge}
\end{align}

To find the ANO solution, we assume that a solution is static and $z$-independent. Hereafter, we use cylindrical coordinates $(r, \theta, z)$.
In this situation, such the solution is expressed as $\phi(x)=\phi(r,\theta)$ and $A_\mu(x)=A_\mu(r,\theta)$.
Moreover, we work in the temporal gauge $A_0(x) = 0$, and assume $A_z(x) = 0$ for simplicity.
The energy per unit length along the $z$ axis, which is denoted as $\mu$, is expressed as
\begin{align}
    \label{mu_abelianhiggs}
    \mu\equiv\frac{dE}{dz} = \int rdrd\theta \left[ \frac{1}{4} F_{ij}^2 + \left| D_i\phi \right|^2 + V(\phi) \right] \qquad (i,j=1,2) .  
\end{align}
The necessary conditions for $\mu$ to be finite are that the integrand in Eq. \eqref{mu_abelianhiggs} asymptotically approaches to zero for $r\rightarrow \infty$.
In other words, all fields satisfy the following conditions
\begin{align}
    F_{ij}(x) = 0, \quad D_i\phi(x)=0 , \quad |\phi(x)|=v,
\end{align}
for $r\rightarrow \infty$.
Here, the $v$ is a vacuum expectation value (VEV) of the scalar field $\phi$.
Taking these assumptions into account, the ansatz for a solution is as follows:
\begin{align}
    \label{NOstring}
    \phi(x) = f(r) v e^{in\theta} , \quad A_\theta(x) = \frac{n a(r)}{gr} , \quad A_r(x) = A_z(x) = 0,
\end{align}
where $n\in\mathbb{Z}$ is called the winding number.
$f(r)$ and $a(r)$ are monotonically increasing functions which satisfy the conditions
\begin{align}
\label{BC}
    f(0)=a(0)=0,\quad
    f(\infty)=a(\infty)=1.
\end{align}

Substituting Eq. \eqref{NOstring} into Eq. \eqref{mu_abelianhiggs}, the energy density is deformed as
\begin{align}
    \mu=v^2\int\rho d\rho d\theta\left[(f')^2+\frac{n^2(a')^2}{2\rho^2}+\frac{n^2(1-a)^2 f^2}{\rho^2}+\widetilde{V}\right],
    \label{mu_normalized_general_single}
\end{align}
where we normalize the radial coordinate as $\rho\equiv gvr$, and $\widetilde{V}\equiv V/(g^2v^4)$. 
Taking variations of $f(r)$ and $a(r)$, we obtain the equations that they must satisfy: 
\begin{align}
\label{eom_f_nmlzd}
    &f'' + \frac{f'}{\rho} - \frac{n^2}{\rho^2} (1-a)^2 f -\frac{1}{2}\frac{\partial \widetilde{V}}{\partial f} = 0, \\
\label{eom_a_nmlzd}
    &a'' - \frac{a'}{\rho} + 2f^2 (1-a) = 0.
\end{align}
The concrete shape of ANO solutions can be derived by solving Eqs. \eqref{eom_f_nmlzd} and \eqref{eom_a_nmlzd} with the boundary conditions (\ref{BC}). 
In general, these equations are solved numerically. 

In the Abelian-Higgs case, the potential is
\begin{align}
    \label{AHpotential}
    V(\phi)=V_{\text{AH}}(\phi)=\lambda \left(|\phi|^2 - v^2\right)^2,
\end{align}
where $\lambda$ is a quartic coupling.
Eqs. \eqref{eom_f_nmlzd} and \eqref{eom_a_nmlzd} are expressed as
\begin{align}
\label{eom_f_AH}
    &f'' + \frac{f'}{\rho} - \frac{n^2}{\rho^2} (1-a)^2 f + \beta (1-f^2) f = 0, \\
\label{eom_a_AH}
    &a'' - \frac{a'}{\rho} + 2f^2 (1-a) = 0.
\end{align}
Here, we introduced the parameter $\beta$ as
\begin{align}
    \label{beta}
    \beta=\frac{2\lambda}{g^2}=\frac{m^2_\phi}{m^2_A},
\end{align}
where $m^2_\phi=4\lambda v^2$ or $m^2_A=2g^2v^2$ are a mass square of scalar field or gauge field, respectively.
This is a significant parameter since it determines the shape and the character of the cosmic string.

\subsection{Evaluation of the interaction energy} \label{Sec.Rev_Int_ANO}

The interaction between the ANO strings can be understood in terms of the competition between magnetic pressure and the force to minimize the potential energy of the scalar field. 
A qualitative description is as follows. 
The width of each excited region for the scalar and gauge fields is approximated by the inverse of their respective masses. 
Based on this assumption, let us consider the situation where two ANO strings are approaching each other. 
For $\beta>1$, it means that $m_A^{-1}>m_\phi^{-1}$, so the magnetic fluxes influence each other first. 
Consequently, the interaction between the ANO strings is repulsive in this case. 
In contrast, for $\beta<1$, the scalar tubes interact first, leading to an attractive interaction between ANO strings. 
In addition, for $\beta=1$, there is no interaction between ANO strings. 
Although this description is somewhat abstract, many studies have verified this $\beta$ dependence of the string interaction \cite{Goodband:1995rt,Jacobs:1978ch, Speight:1996px,Eto:2022hyt,Fujikura:2023lil}. 
In these studies, an analytical method for calculating the interaction energy of two sufficiently separated ANO strings has been developed \cite{Speight:1996px,Fujikura:2023lil}. 

Before reviewing this analytical method, we explain the asymptotic behavior of $f(r)$ and $a(r)$ in the limit $r\rightarrow\infty$. 
In this limit, $f$ and $a$ are close to one since the solutions \eqref{NOstring} take the very close values to the VEVs.
As a deviation from $f=a=1$, we define the perturbation $\delta f$ and $\delta a$ as
\begin{align}
    \delta f\equiv1-f,\quad
    \delta a\equiv1-a.
\end{align}
Noting that $\delta f\ll1$ and $\delta a\ll1$ at $r\rightarrow\infty$, we expand the equations of motion \eqref{eom_f_nmlzd}, \eqref{eom_a_nmlzd} by $\delta f$ and $\delta a$ at the leading order.
The results have 
\begin{align}
    &\delta f''+\frac{1}{\rho}\delta f'-2\beta\delta f=0, \label{asymptoeq_f}\\
    &\delta a''-\frac{1}{\rho}\delta a'-2\delta a=0.
    \label{asymptoeq_a}
\end{align}
These differential equations can be attributed to the differential equation with the modified Bessel function $K_\alpha(x)$ as a solution.
Therefore, $\delta f$ and $\delta a$ asymptotically behave as
\begin{align}
    \label{fa_asympt}
    \delta f(\rho) \sim \frac{c_\phi}{\sqrt{2}} K_0(\sqrt{2\beta}\rho) 
    , \quad
    \delta a(\rho) \sim c_A \rho K_1(\sqrt{2}\rho) ,
\end{align}
where $c_\phi$ and $c_A$ are constants determined by numerical calculations.

Note that this asymptotic behavior applies to models with any scalar potential where $\beta\equiv\frac{m^2_\phi}{m^2_A}$. 
Eqs. (\ref{asymptoeq_f}) and (\ref{asymptoeq_a}) are derived as linear perturbations of Eqs. \eqref{eom_f_nmlzd} and \eqref{eom_a_nmlzd} with respect to $\delta f$ and $\delta a$. 
Since $\phi\propto f$ and $A_\theta\propto a$, these equations are determined solely by the kinetic and mass terms of $\phi$ and $A_\mu$ in the Lagrangian. 
Therefore, when we normalized the radial coordinate as $\rho=gvr$, Eqs. (\ref{asymptoeq_f}) and (\ref{asymptoeq_a}) hold for any scalar potential that induces spontaneous $U(1)$ symmetry breaking. 
We will later apply this result to higher-dimensional gauge theories. 

Using the approximation in Eq. (\ref{fa_asympt}), the interaction between two parallel straight strings is estimated. This method has been known as the point source formalism \cite{Speight:1996px}. 
In this formalism, vortex solutions are regarded as the point sources on the two-dimensional plane.
Through this approximation, we can calculate the energy of the system. 
We summarize the detail derivation in Appendix \ref{PSF}.
As a result, the interaction energy of two strings separated by $d$ is derived as
\begin{align}
    \label{2strings_energy}
    E_{int} = 2\pi v^2 \int dz \left[ n_1n_2 c_{A1}c_{A2}  K_0(m_Ad) -  c_{\phi1}c_{\phi2} K_0(m_\phi d) \right],
\end{align}
where $n_1(n_2)$, $c_{\phi 1}(c_{\phi2})$ and $c_{A 1}(c_{A2})$ are the winding number and the constants in Eq. (\ref{fa_asympt}) for the first (second) string, respectively. 
The first and second terms of the integrand in Eq. (\ref{2strings_energy}) correspond to the contribution from the gauge field and the scalar field, respectively. In light of this, the interaction of the gauge field can be understood as repulsive, while the interaction of the scalar field is attractive. For the case of $n_1=n_2=1$, the values of $c_{\phi 1}, c_{\phi2}, c_{A 1}$ and $c_{A2}$ have been investigated in \cite{Speight:1996px}, and found that the interaction becomes repulsive for $\beta>1$, and attractive for $\beta<1$. 
Note that the point source formalism cannot be applied at short distance because the approximation in Eq. (\ref{fa_asympt}) breaks down the behavior of two strings in that case. 


For $\beta=1$, the ANO string is not affected by any interaction forces from other strings. 
Thus, $\beta=1$ is a critical coupling for the ANO strings. 
This can also be confirmed analytically. 
We rewrite the energy per unit length, which we call tension, of the ANO string as follows. 
\begin{align}
    \mu&=2\pi v^2\int\rho d\rho\left[(f')^2+\frac{n^2(a')^2}{2\rho^2}+\frac{n^2(1-a)^2 f^2}{\rho^2}+\frac{1}{2}\beta(f^2-1)^2\right] \nonumber \\
    &=2\pi v^2|n| \nonumber \\
    &\quad+2\pi v^2\int\rho d\rho\left[\left(f'+|n|\frac{a-1}{\rho}f\right)^2+\frac{n^2}{2\rho^2}\left(a'+\frac{\rho}{|n|}(f^2-1)\right)^2+\frac{1}{2}(\beta-1)(f^2-1)^2\right].
    \label{Bogomolnyi completion}
\end{align}
Since the first and second terms in the integrand are positive, we obtain the inequality as
\begin{align}
    \mu\ge 2\pi v^2|n|+2\pi v^2\int\rho d\rho\left[\frac{1}{2}(\beta-1)(f^2-1)^2\right] \label{mu_inequality} .
\end{align}
If $\beta\geq1$, the second term in r.h.s of \eqref{mu_inequality} becomes non-negative, so that the tension is bounded by $2\pi v^2|n|$. This is known as the Bogomol'nyi bound\cite{Bogomolny:1975de}. 
Equality in Eq. \eqref{mu_inequality} holds for if and only if $f$ and $a$ satisfy the following equations. 
\begin{align}
\label{Bogomolnyi_eq}
    f'+|n|\frac{a-1}{\rho}f=0,\quad
    a'+\frac{\rho}{|n|}(f^2-1)=0.
\end{align}
Eqs. (\ref{Bogomolnyi_eq}) are called the Bogomol'nyi equations. 
For $\beta=1$, the Bogomoln'yi equations become equivalent to Eqs. (\ref{eom_f_AH}) and (\ref{eom_a_AH}), and the second term in r.h.s of \eqref{mu_inequality} vanishes. 
Thus, the tension is obtained as 
\begin{align}
\label{BPS_tension}
    \mu = 2\pi v^2|n| .
\end{align}
Eq. \eqref{BPS_tension} indicates that the tension of the stable ANO strings remains unchanged before and after the fusion or separation of the strings. 
This state is known as the BPS state \cite{Bogomolny:1975de,Prasad:1975kr}. 

As indicated by the above discussions, $\beta$ determines the interaction properties of the ANO strings. 
In our research for the interaction of strings in five-dimensional gauge theories, we also focus on the value of $\beta$. 
The detail discussions are provided in Sec. \ref{Sec.1stringresult}.

\section{Five-dimensional gauge theory}\label{5dgaugetheory}

To consider the cosmic string based on five-dimensional gauge theory, we review a five-dimensional $SU(2)$ gauge theory on $M^4\times S^1/Z_2$ \cite{Haba:2004qf} in this section.
$M^4$ means the four-dimensional Minkowski spacetime and the orbifold $S^1/Z_2$ is a circle with a radius $R$ imposed $Z_2$ parity. 
As we will discuss later, a spontaneous breaking of $U(1)$ gauge symmetry occurs in this model with the addition of some fermions.
Therefore, the five-dimensional $SU(2)$ gauge theory on $M^4\times S^1/Z_2$ is an excellent toy model for examining an initial example of a cosmic string in extra-dimensional models.

\subsection{$SU(2)$ gauge theory on $M^4\times S^1/Z_2$}

The five-dimensional Lagrangian is given by
\begin{align}
    \label{5Dlag}
    \mathcal{L}_{5D}&=-\frac{1}{4}F^a_{MN}F^{a MN}+\mathcal{L}_m, \\
    \label{matterLag}
    \mathcal{L}_m&=iN_f\overline{\psi}\Slash{D}\psi+iN_{\mathrm{ad}}\overline{\psi}^a\Slash{D}\psi^a,
\end{align}
where $M,N=0,1,2,3,y$ are the five-dimensional spacetime indices.
We use $\mu,\nu=0,1,2,3$ as the four-dimensional spacetime indices and $y$ as the fifth space index. 
We denote $a$ as a gauge index for $SU(2)$.
We introduce $N_f$ fermions $\psi$ in the fundamental representation and $N_{\mathrm{ad}}$ fermions $\psi^a$ in the adjoint representation.
Note that we do not introduce any scalar fields that acquire the VEV.
As in the later discussion, the $y$-component of the gauge field $A^a_y$ plays the role of the scalar field breaking the gauge symmetry.

We denote $x^\mu~(\mu=0,1,2,3)$ as the coordinates in Minkowski spacetime and $y\in [0,2\pi R)$ as the fifth-dimensional coordinate.
Since the $Z_2$ parity is imposed, two fixed points ($y=0$ and $y=\pi R$) appear.
Under the $Z_2$ transformation at $y=0$,
the bulk gauge fields $A_M(x^\mu,y)$ transform as
	\begin{align}
	A_\mu(x^\mu,-y)&=P_0A_\mu(x^\mu,y)P^{-1}_0\,, \label{y=0gauge}\\
	A_y(x^\mu,-y)&=-P_0A_y(x^\mu,y)P^{-1}_0 \label{y=0WL}\,,
	\end{align}
where $P_0=P^{-1}_0=P^\dag_0$ denotes the operation of $Z_2$ transformation at $y=0$.
Under the $Z_2$ transformation at $y=\pi R$, the bulk gauge fields transform as
	\begin{align}
	A_\mu(x^\mu,\pi R-y)&=P_1A_\mu(x^\mu,\pi R+y)P^{\dag}_1\,, \label{y=piRgauge} \\
	A_y(x^\mu,\pi R-y)&=-P_1A_y(x^\mu,\pi R+y)P^{\dag}_1\,,\label{y=piRWL}
	\end{align}
where $P_1=P^{-1}_1=P^\dag_1$ is the operation of $Z_2$ transformation at $y=\pi R$.
On the other hand, we must impose the periodic boundary condition on the bulk gauge fields as
	\begin{align}
	A_M(x^\mu,y+2\pi R)=U A_M(x^\mu,y) U^\dag\,, \label{pbc}
	\end{align}
where $U$ is a unitary matrix.
Note that Eq. \eqref{y=piRgauge} can be derived using Eqs. \eqref{y=0gauge} and \eqref{pbc}, and then we can obtain the relation $U=P_1P_0$.

Due to the orbifold boundary conditions $P_0$ and $P_1$, the gauge symmetry is explicitly broken.
If we start with a $SU(2)$ gauge group with the orbifold boundary conditions $P_0=P_1=\text{diag}(1, -1)$, 
$A_\mu$ and $A_y$ are decomposed as
	\begin{align}
	A_\mu=\left(\begin{array}{cc}
	(+,+) & (-,-) \\
	(-,-) & (+,+)
	\end{array}\right)\,,\quad
	A_y=\left(\begin{array}{cc}
	(-,-) & (+,+) \\
	(+,+) & (-,-)
	\end{array}\right)\,, \label{Amuy.decom}
	\end{align}
where $(\pm,\pm)$ is the $Z_2$ charges at $y=0$ (left) and $y=\pi R$ (right), respectively. 
The only bulk gauge field components with $(+,+)$ parity can have a four-dimensional massless zero mode. 
Therefore, only the $U(1)$ gauge symmetry, whose gauge field is $A^3_\mu$, remains in the Minkowski spacetime. 
Additionally, $A_y$ has zero modes that are proportional to $\sigma^1$ and $\sigma^2$, which are the Pauli matrices.
The zero modes of $A^{1,2}_y$ are identified as two real scalar fields in four dimension. 
We call them WL scalar fields.

While there are no potential for the WL scalar fields at the classical level, it may acquire a VEV due to radiative corrections.
For simplicity, a VEV of $A_y$ can be expressed as
\begin{align}
\braket{A_y}=\frac{\alpha}{g_4R}\frac{\sigma^1}{2}, 
\label{VEV}
\end{align}
with a dimensionless real parameter $\alpha\in[0,1]$ and a four-dimensional gauge coupling $g_4\equiv g_5/\sqrt{2\pi R}$. 
Here, $g_5$ is a five-dimensional gauge coupling.
If $A_y$ obtains a VEV except for $\alpha\ne0,1$, the residual gauge symmetry $U(1)$ is spontaneously broken .
\footnote{To see the pattern of the gauge symmetry breaking, we investigate the Wilson line phase.
See \cite{Kubo:2001zc}.}

\subsection{The effective potential}

To determine the effective potential, it is essential to clarify the mass spectrum of the particles in our model. 
Using periodic boundary condition \eqref{pbc} and $Z_2$ transformation at $y=0$ and $y=\pi R$, we can expand $A_\mu$ in terms of Kaluza-Klein (KK) modes as
	\begin{align}
	A_{\mu}\left(x^{\mu}, y\right)_{(+,+)}&=\frac{1}{\sqrt{2\pi R}}A^{(0)}_\mu(x^\mu)_{(+,+)}+\frac{1}{\sqrt{\pi R}}\sum_{k=1}^\infty A^{(k)}_\mu(x^\mu)_{(+,+)}\cos\left(\frac{ky}{R}\right)\,, \label{Amu(++)expand} \\
	A_{\mu}\left(x^{\mu}, y\right)_{(-,-)}&=\frac{1}{\sqrt{\pi R}}\sum_{k=1}^\infty A^{(k)}_\mu(x^\mu)_{(-,-)}\sin\left(\frac{ky}{R}\right)\,. \label{Amu(--)expand}
	\end{align}
The expansion of the extra-dimensional component $A_{y}\left(x^{\mu}, y\right)_{(+,+)}$ and $A_{y}\left(x^{\mu}, y\right)_{(-,-)}$ are the same expansion as in Eqs. \eqref{Amu(++)expand} and \eqref{Amu(--)expand}, respectively.
This is because the $Z_2$ assignments are the same.
Due to a VEV of $A_y$ in (\ref{VEV}), the squared KK mass eigenvalues of $A_\mu^{(k)}$ are derived as
	\begin{align}
	\frac{k^2}{R^2},~\frac{(k\pm \alpha )^2}{R^2}\quad (k\geq1) \,, \label{gaugeKKmass}
	\end{align}
and $\alpha^2/R^2$ for $A^{3(0)}_\mu$.
Based on the KK mass spectra in Eq. \eqref{gaugeKKmass}, the four-dimensional effective potential for $A_y$ are calculated. For a detail derivation, see~\cite{Haba:2004qf, Kubo:2001zc}. 
The effective potential derived from the contributions of  the gauge fields at one-loop level is
	\begin{align}
	V^{g}_\mathrm{eff}( \alpha )
	&=-3C\sum_{k=1}^{\infty}\frac{1}{k^5}\cos(2\pi k \alpha ) \,,
	\end{align}
where $C=3/(64\pi^6R^4)$.

In addition, we consider the contributions to the effective potential from the matter fields introducing in Eq. \eqref{matterLag}.
The $Z_2$ transformations of these matter fields are given by
	\begin{align}
	&\psi(x,-y)=\eta P_0 \gamma^{5} \psi(x, y) \quad, \quad \psi(x, \pi R-y)=\eta^{\prime} P_1\gamma^{5} \psi(x, \pi R+y)\,, \\
	&\psi^{a}(x,-y)=\eta P_0 \gamma^{5} \psi^{a}(x, y) P^{\dagger}_0 \quad, \quad \psi^{a}(x, \pi R-y)=\eta^{\prime} P_1\gamma^{5} \psi^{a}(x, \pi R+y) P^{\dagger}_1\,,
	\end{align}
with $\eta,\eta'=\pm$.
The fields with $\eta\eta'=+$ are the same expansion of Eqs. \eqref{Amu(++)expand} and \eqref{Amu(--)expand}.
On the other hand, the fields with $\eta\eta'=-$ are expanded as
	\begin{align}
	\Phi\left(x^{\mu}, y\right)_{(+,-)}&=\frac{1}{\sqrt{\pi R}}\sum_{k=1}^\infty \Phi^{(k)}(x^\mu)_{(+,-)}\cos\left(\frac{\left(k+\frac{1}{2}\right)y}{R}\right)\,, \\
	\Phi\left(x^{\mu}, y\right)_{(-,+)}&=\frac{1}{\sqrt{\pi R}}\sum_{k=1}^\infty \Phi^{(k)}(x^\mu)_{(-,+)}\sin\left(\frac{\left(k+\frac{1}{2}\right)y}{R}\right)\,,
	\end{align}
where $\Phi$ can be replaced for $\psi$, and $\psi^a$.
As in the derivation of Eq. \eqref{gaugeKKmass}, we can obtain the KK masses of the fields with $\eta\eta'=+$ or $\eta\eta'=-$ in the fundamental (adjoint) representation.
The effective potential, which arises from the contributions from matter fields, is given by
	\begin{align}
	V^{m}_\mathrm{eff}( \alpha )&=4N^{(+)}_f C\sum_{k=1}^{\infty}\frac{1}{k^5}\cos(\pi k \alpha )
	+4N^{(-)}_f C\sum_{k=1}^{\infty}\frac{1}{k^5}\cos(\pi k( \alpha -1)) \nonumber \\
	&\quad+4N^{(+)}_\mathrm{ad} C\sum_{k=1}^{\infty}\frac{1}{k^5}\cos(2\pi k \alpha )+4N^{(-)}_\mathrm{ad} C\sum_{k=1}^{\infty}\frac{1}{k^5}\cos\left(2\pi k\left( \alpha -1/2\right)\right),
	\end{align}
where $N^{(+)}_f$ and $N^{(-)}_f$ are the degree of freedom of the fundamental fermion $\psi$ with $\eta\eta'=+$ and $\eta\eta'=-$, and $N^{(+)}_{\mathrm{ad}}$ and $N^{(-)}_{\mathrm{ad}}$ are the degree of freedom of the adjoint fermion $\psi^a$ with $\eta\eta'=+$ and $\eta\eta'=-$, respectively.
Note that $N_f=N^{(+)}_f + N^{(-)}_f$ and $N_{\mathrm{ad}}=N^{(+)}_{\mathrm{ad}} + N^{(-)}_{\mathrm{ad}}$.
The total effective potential of $A_y$ is described as the combination of the contributions,
	\begin{align}
	\label{4DGHUpotential}
	V_\mathrm{eff}(\alpha)&\equiv V^g_\mathrm{eff}(\alpha)+V^m_\mathrm{eff}(\alpha) \nonumber \\
	&=(-3+4N^{(+)}_\mathrm{ad}) C\sum_{k=1}^{\infty}\frac{1}{k^5}\cos(2\pi k \alpha )+4N^{(-)}_\mathrm{ad} C\sum_{k=1}^{\infty}\frac{1}{k^5}\cos(2\pi k( \alpha -1/2)) \nonumber \\
	&\quad+4N^{(+)}_f C\sum_{k=1}^{\infty}\frac{1}{k^5}\cos(\pi k \alpha )
	+4N^{(-)}_f C\sum_{k=1}^{\infty}\frac{1}{k^5}\cos(\pi k( \alpha -1)).
	\end{align}
	\begin{figure}[htbp]
	  \begin{minipage}[b]{0.5\linewidth}
	    \centering
	    \includegraphics[keepaspectratio, scale=0.45]{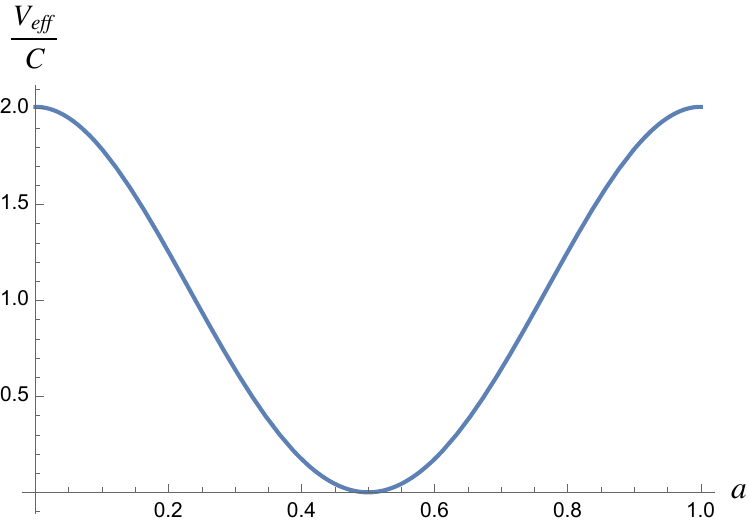}
	    \subcaption{$N^{(+)}_{ad}=1$ (Case 1).}
	  \end{minipage}
	  \begin{minipage}[b]{0.5\linewidth}
	    \centering
	    \includegraphics[keepaspectratio, scale=0.45]{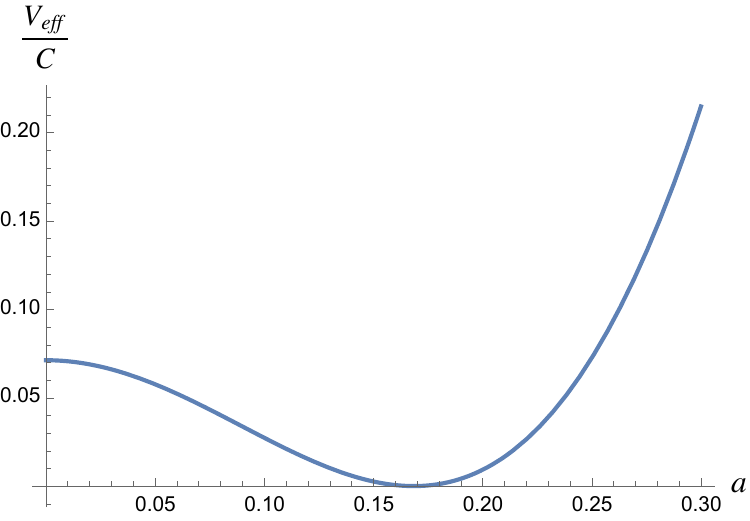}
	    \subcaption{$N^{(+)}_{ad}=1$, $N^{(-)}_{f}=1$ (Case 2).}
	  \end{minipage}
   \begin{minipage}[b]{0.5\linewidth}
	    \centering
	    \includegraphics[keepaspectratio, scale=0.45]{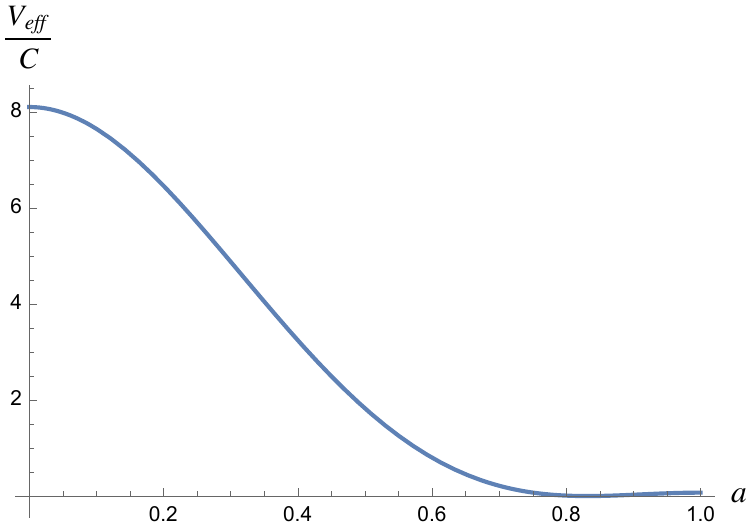}
	    \subcaption{$N^{(+)}_{ad}=1$, $N^{(+)}_{f}=1$ (Case 3).}
	  \end{minipage}
        \begin{minipage}[b]{0.5\linewidth}
	    \centering
	    \includegraphics[keepaspectratio, scale=0.45]{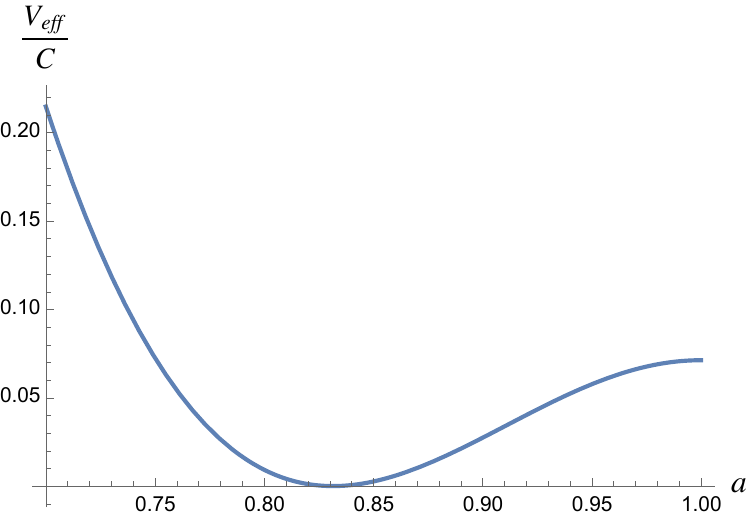}
	    \subcaption{Case 3, around the minimum.}
	  \end{minipage}
	  \caption{Some graphs of the effective potential in \eqref{4DGHUpotential}.}
	  \label{5DGHUpt(zero)}
	\end{figure}

\begin{table}[t]
    \centering
    \begin{tabular}{|c|c|c|c|} \hline
        & Case 1 & Case 2 & Case 3  \\ \hline
        $(N^{(+)}_\mathrm{ad}, N^{(-)}_\mathrm{ad}, N^{(+)}_{f}, N^{(-)}_{f})$  & $(1,0,0,0)$ & $(1,0,0,1)$ & $(1,0,1,0)$  \\ \hline
    \end{tabular}
    \caption{The fermion contents considered in this paper. }
    \label{Table_cases}
\end{table}

In this paper, we consider three cases, referred to as Case 1, Case 2, and Case 3, as summarized in Table \ref{Table_cases}. 
We plot their effective potentials \eqref{4DGHUpotential} in Fig. \ref{5DGHUpt(zero)}. 
We denote the minimum of $V_\mathrm{eff}(\alpha)$ as $\alpha_\mathrm{min}$.
Numerical values of $\alpha_\mathrm{min}$ are $\alpha_{\text{min}}=0.5$ in Case 1, $\alpha_{\text{min}}=0.169$ in Case 2, and $\alpha_{\text{min}}=0.831$ in Case 3.

We close this section with a discussion on the mass of $A_y^{(0)}$ in the four-dimensional theory. 
The squared mass of $A_y^{(0)}$ can be obtained by
\begin{align}
    \label{scalarmass}
    m^2_\phi&=\left.\frac{\partial^2 V_\mathrm{eff}(\alpha)}{\partial (A^1_y)^2}\right|_{\alpha=\alpha_\mathrm{min}}
    =\left.g^2_4 R^2\frac{\partial^2 V_\mathrm{eff}(\alpha)}{\partial \alpha^2}\right|_{\alpha=\alpha_\mathrm{min}} 
   =\frac{3g^2_4}{16\pi^4 R^2}M(\alpha_\mathrm{min}),
\end{align}
where $M(\alpha)$ is defined as
\begin{align}
    M(\alpha)&\equiv-\left[(-3+4N^{(+)}_\mathrm{ad}) \sum_{k=1}^{\infty}\frac{1}{k^3}\cos(2\pi k \alpha_\mathrm{min} )+4N^{(-)}_\mathrm{ad} \sum_{k=1}^{\infty}\frac{1}{k^3}\cos(2\pi k( \alpha_\mathrm{min} -1/2))\right. \nonumber \\
	&\hspace{7mm}\left.+N^{(+)}_f \sum_{k=1}^{\infty}\frac{1}{k^3}\cos(\pi k \alpha_\mathrm{min} )
	+N^{(-)}_f \sum_{k=1}^{\infty}\frac{1}{k^3}\cos(\pi k( \alpha_\mathrm{min} -1))\right].
\end{align}
In Case 1, Case 2, and Case 3, the masses are concretely given by
\begin{align}
    m^2_\phi|_{N^{(+)}_{\text{ad}}=1}&=\frac{9g^2_4\zeta(3)}{64\pi^4 R^2}, \\
    m^2_\phi|_{N^{(+)}_{\text{ad}}=1,N^{(-)}_f=1}&=-\frac{3g^2_4}{16\pi^4 R^2}\sum_{k=1}^{\infty}\frac{1}{k^3}\Big(\cos(2\pi k\alpha_\mathrm{min})+\cos(\pi k(\alpha_\mathrm{min}-1))\Big), \label{mass1b}\\
     m^2_\phi|_{N^{(+)}_{\text{ad}}=1,N^{(+)}_f=1}&=-\frac{3g^2_4}{16\pi^4 R^2}\sum_{k=1}^{\infty}\frac{1}{k^3}\Big(\cos(2\pi k\alpha_\mathrm{min})+\cos(\pi k\alpha_\mathrm{min})\Big). \label{mass1c}
\end{align}
We numerically estimate the sum for the KK modes in Eqs. \eqref{mass1b} and \eqref{mass1c}.
For instance, 
\begin{align}
   \sum_{k=1}^{50}\frac{1}{k^3}\left.\Big(\cos(2\pi k\alpha_\mathrm{min})+\cos(\pi k(\alpha_\mathrm{min}-1))\Big)\right|_{\alpha_\mathrm{min}=0.169}&\simeq-0.418876, \\
   \sum_{k=1}^{50}\frac{1}{k^3}\left.\Big(\cos(2\pi k\alpha_\mathrm{min})+\cos(\pi k\alpha_\mathrm{min})\Big)\right|_{\alpha_\mathrm{min}=0.831}&\simeq-0.418876.
\end{align}


\section{String in the higher-dimensional gauge theory} \label{Sec.1stringresult}

In this section, we consider a vortex string solution in the $SU(2)$ gauge theory on $M^4\times S^1/Z_2$. 
The string solution exists in the four-dimensional effective theory where $U(1)$ gauge symmetry is broken by a VEV of WL scalar fields. 

\subsection{String solution in the four-dimensional effective theory}

To clarify the discussion of the string solution, we first demonstrate that the four-dimensional effective theory from the $SU(2)$ gauge theory on $M^4\times S^1/Z_2$ can be treated as a $U(1)$ gauge theory with a complex scalar field. 
At the classical level, only the zero modes of the KK modes become massless in the four-dimensional spacetime. 
In this case, there are three zero modes, $A_\mu^{3(0)}$, $A_y^{1(0)}$ and $A_y^{2(0)}$. 
Extracting the four-dimensional effective Lagrangian consisting of the zero modes from Eq. \eqref{5Dlag}, we have
\begin{align}
\label{4dim_tree_lagrangian}
    \mathcal{L}_{4D}=-\frac{1}{4} F_{\mu\nu}^{(0)} F^{(0)\mu\nu} - \frac{1}{2}|D^{(0)}_\mu A_y^{a(0)}|^2 \qquad (a=1,2),
\end{align}
where $F_{\mu\nu}^{(0)} \equiv \partial_\mu A_\nu^{3(0)} - \partial_\nu A_\mu^{3(0)}$ and $D^{(0)}_\mu A_y^{a(0)} \equiv \left( \partial_\mu - ig_4 A^{3(0)}_\mu \right) A_y^{a(0)}$. 
As shown in Eq. \eqref{4dim_tree_lagrangian}, $A_\mu^{3(0)}$ is regarded as a $U(1)$ gauge field in four-dimensional effective theory. 
Under the $U(1)$ gauge transformation, $A_y^{1(0)}$ and $A_y^{2(0)}$ are transformed as the real and imaginary parts of a complex scalar field with $U(1)$ charge $-1$, respectively. 
Therefore, denoting
\begin{align}
    \phi \equiv \frac{A_y^{1(0)} - i A_y^{2(0)}}{\sqrt{2}} ,
\end{align}
we can treat $\phi$ as a charged complex scalar field with $U(1)$ charge $+1$. 

Taking account of the effective potential \eqref{4DGHUpotential}, $A_y$ obtains a VEV as in Eq. \eqref{VEV}. 
To rewrite $V_\mathrm{eff}(\alpha)$ as a function of $\phi$, we discuss the KK mass spectrum under a more general assumption. 
We assume that a VEV of $A_y$ takes a value as
\begin{align}
    \braket{A_y} = \frac{\alpha_1}{g_4R}\frac{\sigma^1}{2} + \frac{\alpha_2}{g_4R}\frac{\sigma^2}{2}, 
\end{align}
where $\alpha_1$ and $\alpha_2$ are dimensionless real parameters. 
In terms of $\phi$, this is equivalent to assuming 
\begin{align}
    \braket{\phi}=\frac{\alpha_1 - i \alpha_2}{\sqrt{2}g_4R} .
\end{align}
In this case, the squared KK mass spectra denoted in Eq. \eqref{gaugeKKmass} are rewritten as
\begin{align}
    \frac{k^2}{R^2},~\frac{\left(k\pm \sqrt{\alpha_1^2 + \alpha_2^2} \right)^2}{R^2} \,.
\end{align}
This implies that the discussion in Sec.\ref{5dgaugetheory} can be applied to this case by replacing $\alpha$ with $\sqrt{\alpha_1^2 + \alpha_2^2}$ or $\sqrt{2}g_4R|\braket{\phi}|$. 
Therefore, the four-dimensional effective Lagrangian is given as
\begin{align}
    \mathcal{L}_{4D}=-\frac{1}{4} F_{\mu\nu}^{(0)} F^{(0)\mu\nu} - |D^{(0)}_\mu \phi|^2 - V_{\mathrm{eff}}(\phi),
\end{align}
where $V_{\mathrm{eff}}(\phi)$ is obtained by replacing $\alpha$ in Eq. \eqref{4DGHUpotential} with $\sqrt{2}g_4R|\braket{\phi}|$. 
Hereafter, we denote the absolute value of the VEV of $\phi$ as $v$ for simplicity; hence, 
\begin{align}
    v = \frac{\sqrt{\alpha_1^2+\alpha_2^2}}{\sqrt{2}g_4R} .
\end{align}

If $\alpha_\mathrm{min}\neq 0$ (or $v\neq 0$), the moduli space of $\phi$ is homeomorphic to $S^1$. 
\footnote{Taking into account the case where $\alpha\geq1$, the moduli space becomes complex. We will discuss this in Sec.\ref{conclusion}.}
For the case where $0<\alpha_\mathrm{min}<1$, the degrees of freedom (dof) of the moduli space is equivalent to the dof of the spontaneously broken $U(1)$ gauge symmetry. 
Therefore, we can construct a string solution by utilizing this dof. 
The ansatz for a string solution is obtained from the ansatz in the Abelian-Higgs model, as shown in Eq. \eqref{NOstring}. 
Specifically, this means
\begin{align}
    \phi(x) = f(r) v e^{in\theta} , \quad
    A^{3(0)}_\theta(x) = \frac{n a(r)}{g_4 r} , \quad A^{3(0)}_r(x) = A^{3(0)}_z(x) = 0 \,,
    \label{stringansatz_ED}
\end{align}
where $n$ is the winding number. 
$f(r)$ and $a(r)$ are monotonically increasing functions which satisfy the boundary conditions \eqref{BC}. 
In terms of WL scalar fields, 
\begin{align}
    A^{1(0)}_y(x) = f(r)\frac{\alpha_\mathrm{min}}{g_4R}\cos (n\theta), \quad
    A^{2(0)}_y(x) = -f(r)\frac{\alpha_\mathrm{min}}{g_4R}\sin (n\theta) \,.
\end{align}

On the normalized radial coordinate $\rho=g_4 vr$, $f(\rho)$ and $a(\rho)$ satisfy Eqs. \eqref{eom_f_nmlzd} and \eqref{eom_a_nmlzd}, where $\widetilde{V}$ in Eq. \eqref{eom_f_nmlzd}
corresponds to a normalized effective potential $V_\mathrm{eff}/(g_4^2v^4)$ in this case.
We characterize the shape of the normalized effective potential by introducing a parameter.
In analogy with the Abelian-Higgs case, we define the parameter $\beta$ as the ratio of the squared mass of $\phi$ (Eq. \eqref{scalarmass}) to the squared mass of $A_\mu^{3(0)}$ after $U(1)$ symmetry breaking;
\begin{align}
    \beta&=\frac{3g^2_4}{16\pi^4\alpha^2_\mathrm{min}}M(\alpha_\mathrm{min}).
\end{align}
Consequently, the normalized effective potential can be expressed as   
\begin{align}
    \widetilde{V}(\phi) = \frac{\beta}{\pi^2\alpha^2_\mathrm{min}M(\alpha_\mathrm{min})} 
    &\left[(-3+4N^{(+)}_\mathrm{ad}) \sum_{k=1}^{\infty}\frac{1}{k^5}\cos\left(2\pi k \alpha_\mathrm{min}\tilde{\phi} \right) \right. \nonumber\\
    &\quad +4N^{(-)}_\mathrm{ad} \sum_{k=1}^{\infty}\frac{1}{k^5}\cos\left(2\pi k\left( \alpha_\mathrm{min}\tilde{\phi} -\frac{1}{2}\right)\right) \nonumber \\
    &\quad+4N^{(+)}_f \sum_{k=1}^{\infty} \frac{1}{k^5}\cos\left(\pi k \alpha_\mathrm{min} \tilde{\phi} \right) \nonumber\\
    &\left. \quad+4N^{(-)}_f\sum_{k=1}^{\infty}\frac{1}{k^5}\cos\left(\pi k\left( \alpha_\mathrm{min}\tilde{\phi} -1\right)\right)\right] \,.
    \label{normalized_Veff}
\end{align}
Note that we denote a normalized scalar field $\phi/v$ as $\tilde{\phi}$ in Eq. \eqref{normalized_Veff}.

\begin{figure}[t]
    \centering
    \includegraphics[width=0.5\linewidth]{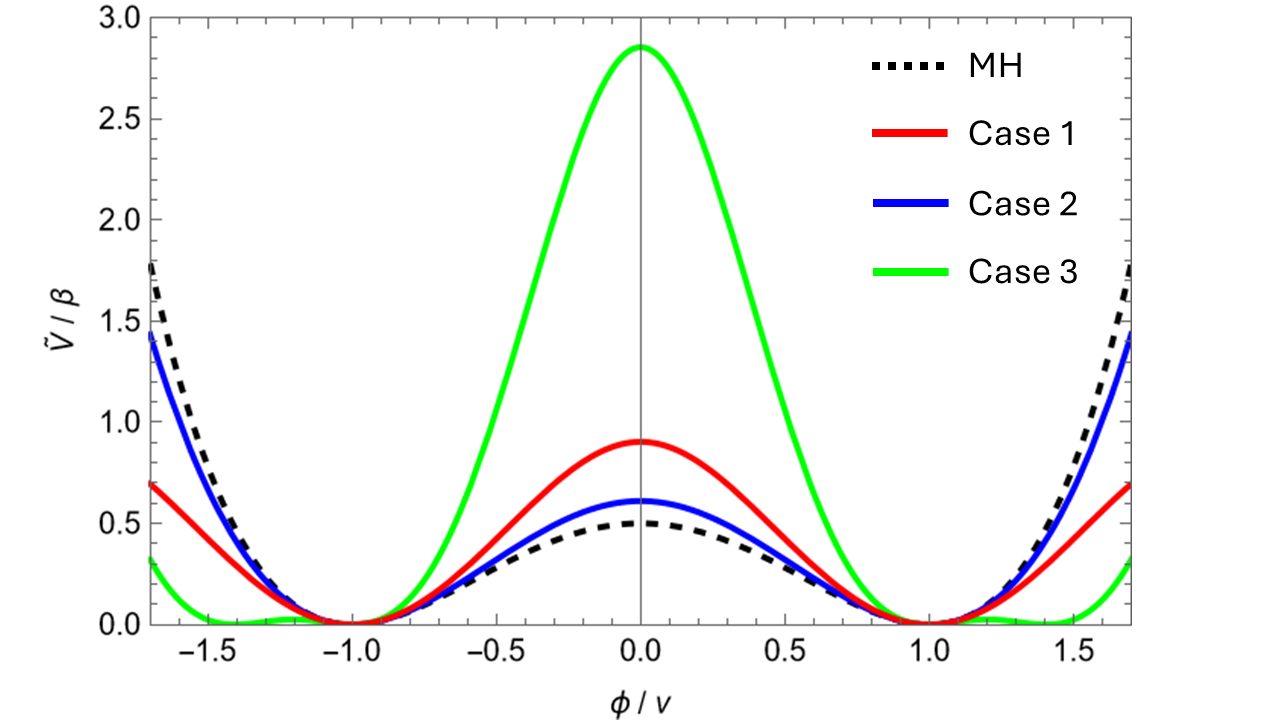}
    \caption{Comparison of the normalized potentials. The solid lines are $\widetilde{V}_\mathrm{eff}/\beta$ for Case 1 (red), Case 2 (blue), and Case 3 (green). The dashed black line is $\widetilde{V}/\beta$ for the Mexican hat potential.}
    \label{PotentialsGraph}
\end{figure}

It is important to emphasize the difference between the Mexican hat potential and the effective potentials considered in our study. 
As shown in Fig. \ref{PotentialsGraph}, the normalized effective potentials take larger values at $\phi/v=0$ than the normalized Mexican hat potential when $\beta$ is the same. 
This difference leads to significant variations in the interaction between two parallel strings, as discussed in Sec.\ref{2strings}.

We mention that the spontaneous breaking of $U(1)$ gauge symmetry in the four-dimensional effective theory can be understood within the framework of cosmological thermal phase transitions. 
At finite temperature, the effective potential deviates from Eq. \eqref{4DGHUpotential}. 
For detailed derivations of the thermal effective potential in higher-dimensional gauge theory, see \cite{Shiraishi:1986wu, Ho:1990xz, Panico:2005ft, Maru:2005jy}. 
In our cases, the thermal effective potential reaches a minimum at $\alpha=0,1,\dots$ when the temperature $T\gtrsim 0.2 \times R^{-1}$. 
This means that $U(1)$ symmetry is restored at a high temperature. 
Therefore, we apply the cosmological thermal phase transitions to our models and expect cosmic strings to form via the Kibble-Zurek mechanism.

In the rest of our paper, we mainly perform numerical calculations 
for $\beta$ in the range of $0.1\leq\beta\leq1.5$. 
There are two reasons for this. 
The primary reason is that one of the main motivations of this work is to understand how the interactions between the strings are affected when the scalar potential deviates from the Mexican hat potential. 
As we mentioned in Sec. 1, it has been found in \cite{Eto:2022hyt} that strings in the Abelian-Higgs model with CW potential exhibit different interactions compared to the ANO string with the Mexican hat potential. 
The effective potential considered in this study also differ from the Mexican hat potential, and its characteristic contrasts with those studied in \cite{Eto:2022hyt}; 
the normalized CW potential takes a lower value at $\phi/v=0$ than the normalized Mexican hat potential. 
Hence, we consider that it is important to study the string solutions in the Abelian-Higgs model with the potential given in Eq. \eqref{normalized_Veff} for any $\beta$. 
Indeed, as we discuss later, the significant differences occur in the range of $0.1\leq\beta\leq1.5$.

The second reason is that it is difficult to calculate 
the string solution for very small $\beta$ with our current computational resources. 
As $\beta$ decreases, the width of the excited region of the scalar field is increased. 
Hence, we must solve differential equations, \textit{e.g.} Eqs. \eqref{eom_f_nmlzd} and \eqref{eom_a_nmlzd}, numerically on a large region for very small $\beta$. 
This computational difficulty significantly increases the time required for obtaining accurate results for very small values of $\beta$.

Note that the value of $\beta$ is determined by $g_4$ and $\alpha_\mathrm{min}$ in our models.
Actually, $\beta$ tends to take very small values since the mass of $\phi$ is derived from the one-loop effective potential. 
For instance, in the cases considered in this paper, $\beta$ takes values as shown in Table \ref{beta_values}.
For $\beta>0.1$, the coupling $g_4$ is larger than one and the perturbative expansion would fail normally. 
However, our motivation is in the behavior of interaction with the cosmic strings not only in higher-dimensional models but also in Abelian-Higgs model with various scalar potentials. 
Therefore, we deal with $\beta$ or $g_4$ such as a free parameter in later section. 
On the other hand, since cosmic strings in higher-dimensional models remains a subject of interest, we briefly comment on the very small $\beta$ in Sec.\ref{Sec.smallbeta}.

\begin{table}[t]
    \centering
    \begin{tabular}{|c|c|c|} \hline
        Case 1 & Case 2 & Case 3  \\ \hline
        $\beta\simeq g_4^2\times0.0075$  & $\beta\simeq g_4^2\times0.0257$ & $\beta\simeq g_4^2\times0.0011$  \\ \hline
    \end{tabular}
    \caption{The values of $\beta$ for the each cases. }
    \label{beta_values}
\end{table}


\subsection{Numerical results} \label{Sec.Numerical_1string}
We show some numerical results and discuss differences between the strings within the Mexican hat potential and those within the effective potentials shown in Fig. \ref{PotentialsGraph}.
First, we show the behavior of the strings which are obtained by solving Eqs. \eqref{eom_f_nmlzd} and \eqref{eom_a_nmlzd} numerically. 
To solve them, we use the gradient flow method. 
In detail, see Appendix \ref{GFM}.
Hereafter, we refer to the case with the Mexican hat potential as the MH case for simplicity. 

\begin{figure}[t]
    \begin{minipage}[b]{0.5\linewidth}
        \centering
        \includegraphics[keepaspectratio, width=.8\columnwidth]{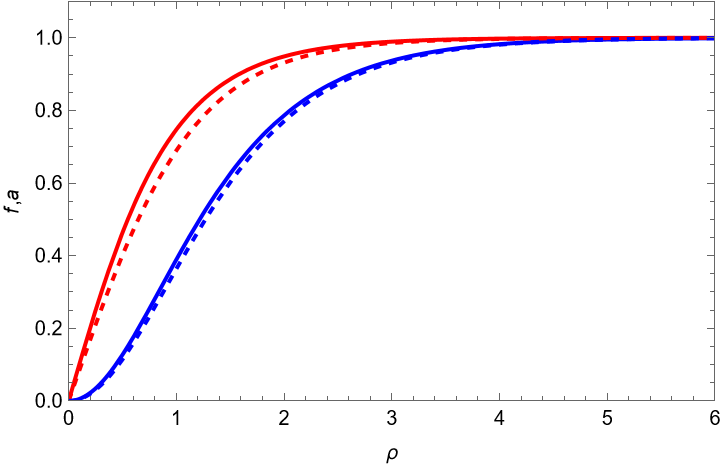}
        \subcaption{$\beta=1$}
    \end{minipage}
    \begin{minipage}[b]{0.5\linewidth}
        \centering
        \includegraphics[keepaspectratio, width=.8\columnwidth]{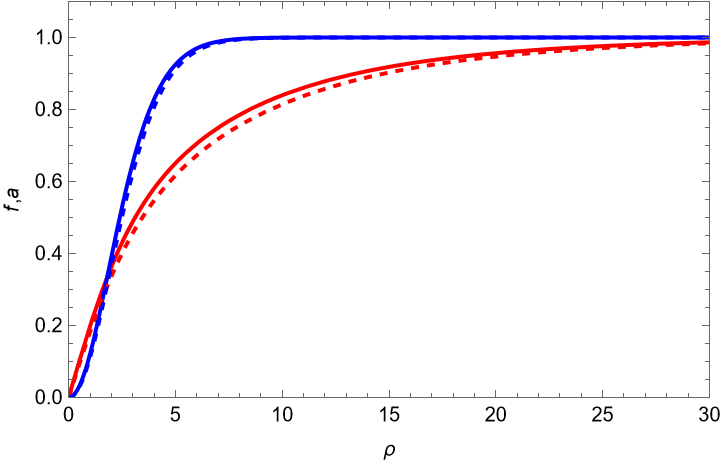}
        \subcaption{$\beta=0.005$}
    \end{minipage}
    \caption{
    The behaviors of $f(\rho)$ (red) and $a(\rho)$ (blue). The solid lines represent the results in Case 1. The dashed lines correspond to those in the MH case.
    }
    \label{fa_nad1}
\end{figure}

\begin{figure}[t]
    \begin{minipage}[b]{0.5\linewidth}
        \centering
        \includegraphics[keepaspectratio, width=.8\columnwidth]{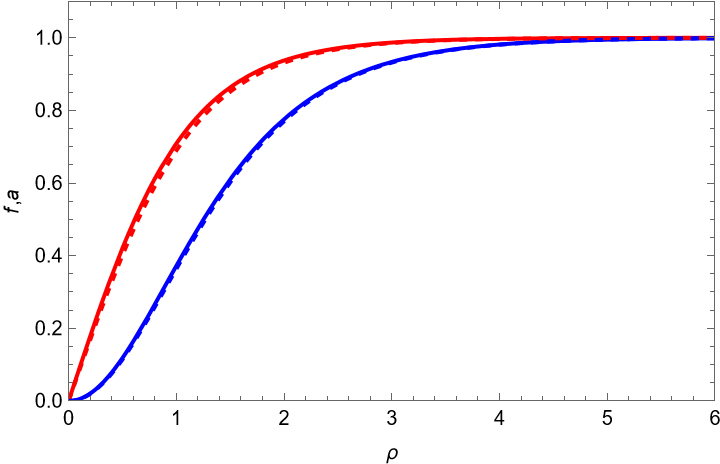}
        \subcaption{$\beta=1$}
    \end{minipage}
    \begin{minipage}[b]{0.5\linewidth}
        \centering
        \includegraphics[keepaspectratio, width=.8\columnwidth]{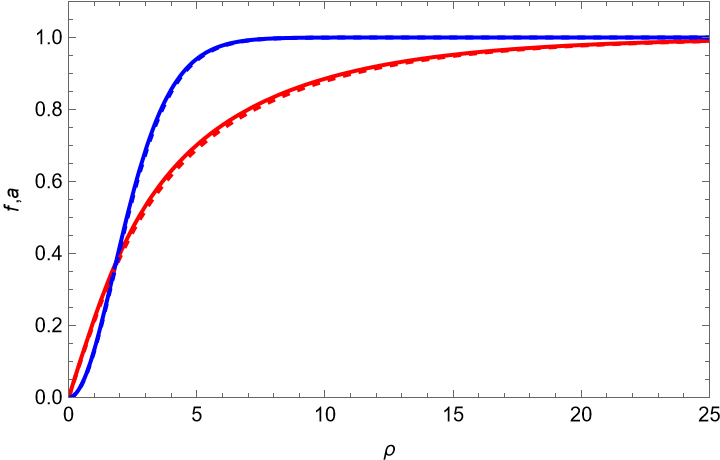}
        \subcaption{$\beta=0.01$}
    \end{minipage}
    \begin{minipage}[b]{0.5\linewidth}
        \centering
        \includegraphics[keepaspectratio, width=.8\columnwidth]{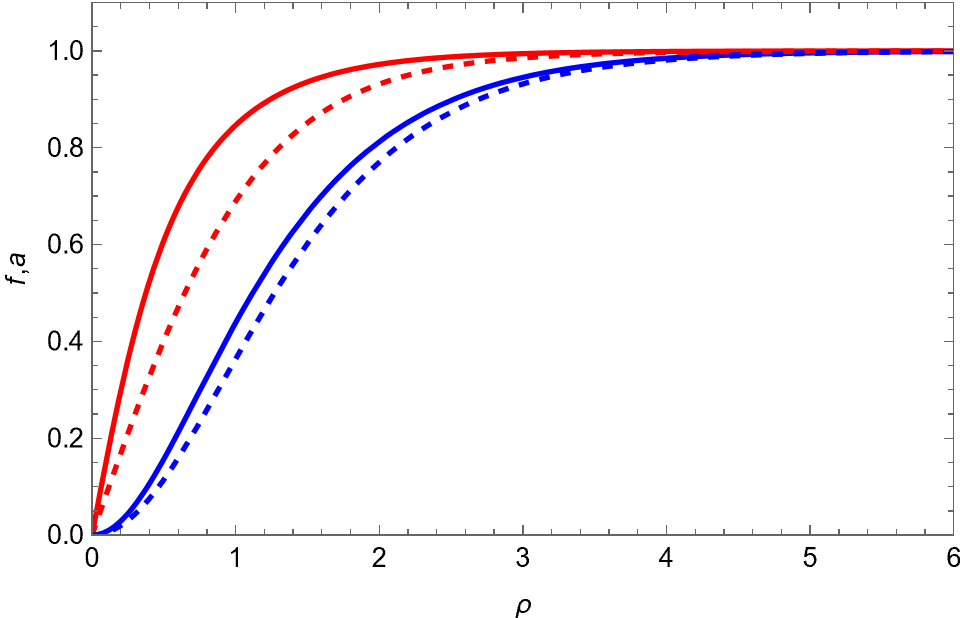}
        \subcaption{$\beta=1$}
    \end{minipage}
    \begin{minipage}[b]{0.5\linewidth}
        \centering
        \includegraphics[keepaspectratio, width=.8\columnwidth]{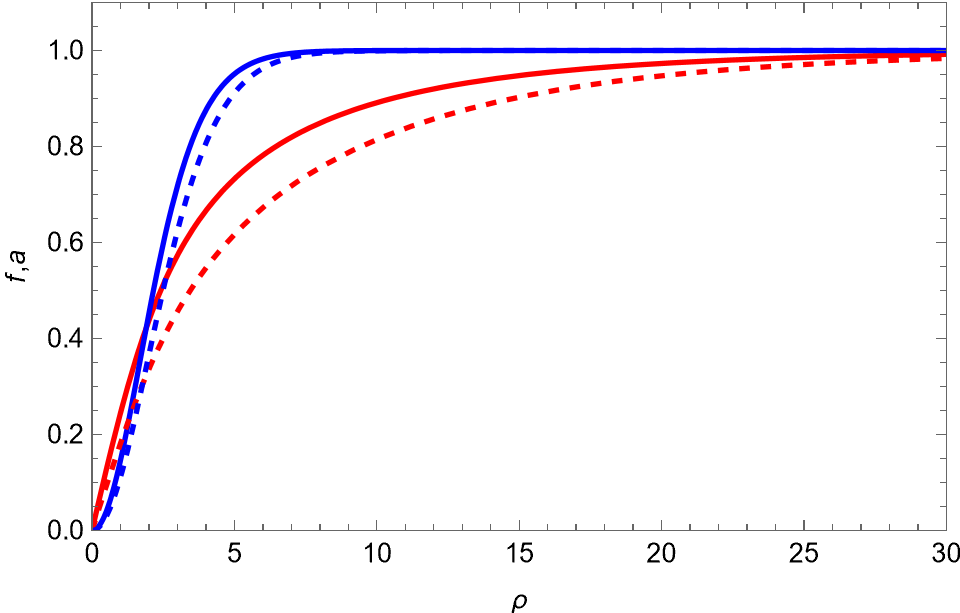}
        \subcaption{$\beta=0.005$}
    \end{minipage}
    \caption{
    The behaviors of $f(\rho)$ (red) and $a(\rho)$ (blue). The solid lines in the top panels represent Case 2, while those in the bottom panels represent Case 3. The dashed lines correspond to conventional ANO strings with the same $\beta$ values in each case.
    }
    \label{fa_nad1nf1}
\end{figure}

The results of $f(\rho)$ and $a(\rho)$ for Case 1 are summarized in Fig. \ref{fa_nad1}. 
We plot $f(\rho)$ and $a(\rho)$ for $\beta=1.0$ (left panel) and $0.005$ (right panel).
Note that $\beta=0.005$ is a natural value from the viewpoint of the extra-dimensional model as seen in Table \ref{beta_values}. 
These results indicate that the behavior of $f(\rho)$ and $a(\rho)$ of Case 1 is thinner than in the MH case.
This variation seems more significant for $f(\rho)$ than $a(\rho)$. 
We also show our results of Case 2 and Case 3 in Fig. \ref{fa_nad1nf1}.

We consider that the values of the potential around $\phi/v\sim0$ are related to the behavior of $f(\rho)$ and $a(\rho)$. 
Even though this may not be the case, let's assume for the sake of argument that we impose the behavior of $f(\rho)$ and $a(\rho)$ in the MH case on our case. 
In that scenario, the string in our case would have larger tension from the potential energy than the string within a Mexican hat potential, but the tension from the kinetic energy would be the same. 
Hence, it is considered that $f(\rho)$ and $a(\rho)$ in our case becomes thinner to strike a balance between the tension from the potential energy and the kinetic energy. 
This effect becomes more pronounced in the order of Case 3, Case 1, and Case 2, as the contribution from the potential energy increases.

\begin{figure}[t]
    \begin{minipage}[b]{0.5\linewidth}
        \centering
        \includegraphics[keepaspectratio, width=.9\columnwidth]{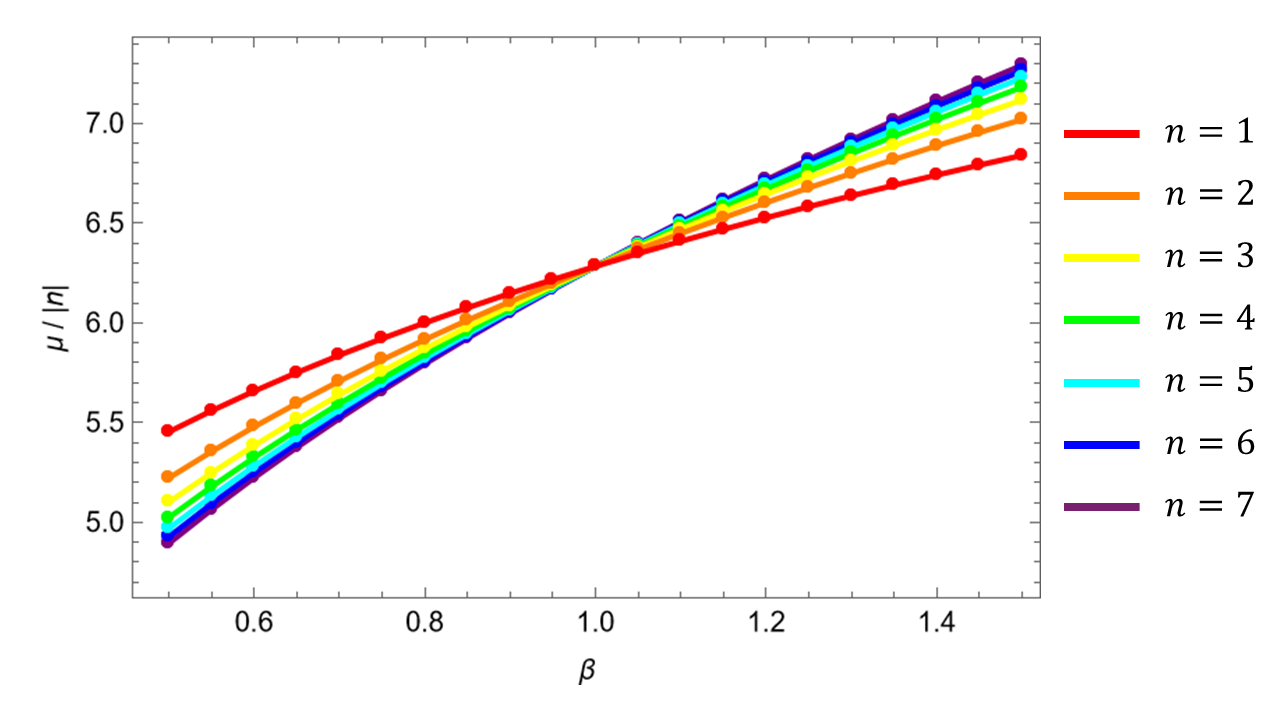}
        \subcaption{$0.5\leq\beta\leq1.5$}
    \end{minipage}
    \begin{minipage}[b]{0.5\linewidth}
        \centering
        \includegraphics[keepaspectratio, width=.9\columnwidth]{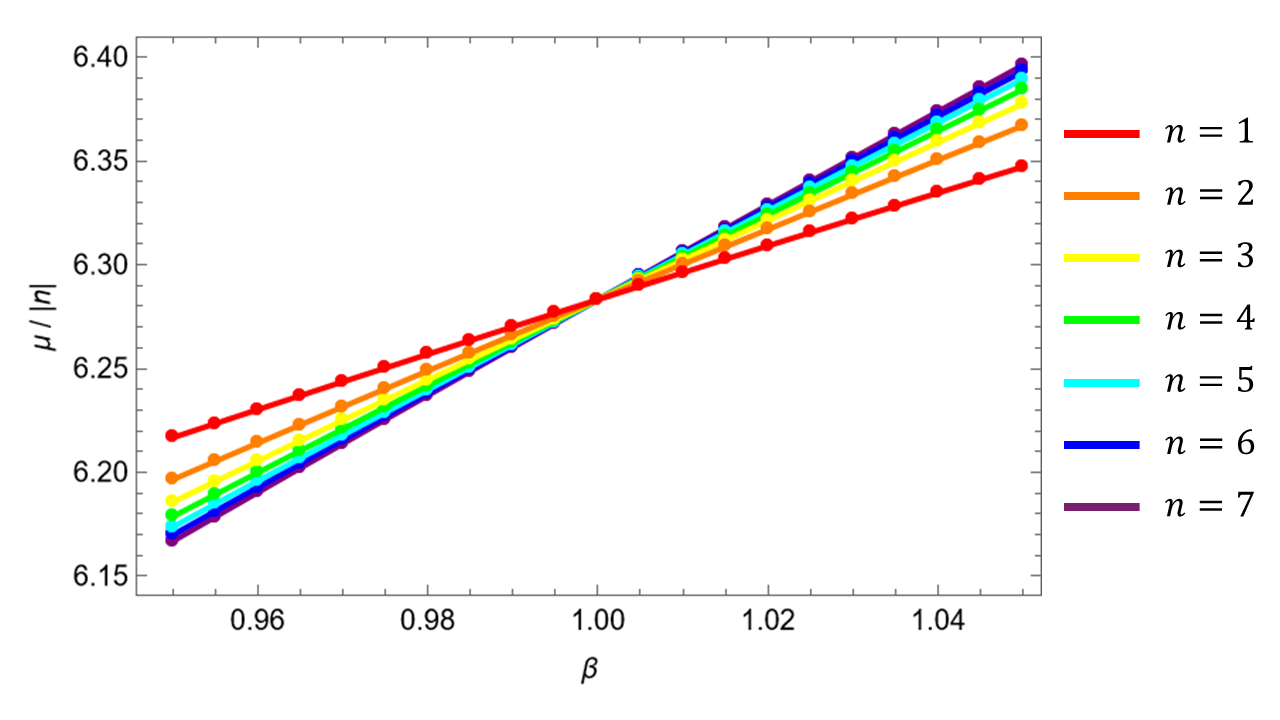}
        \subcaption{$0.95\leq\beta\leq1.05$}
    \end{minipage}
    \begin{minipage}[b]{0.5\linewidth}
        \centering
        \includegraphics[keepaspectratio, width=.9\columnwidth]{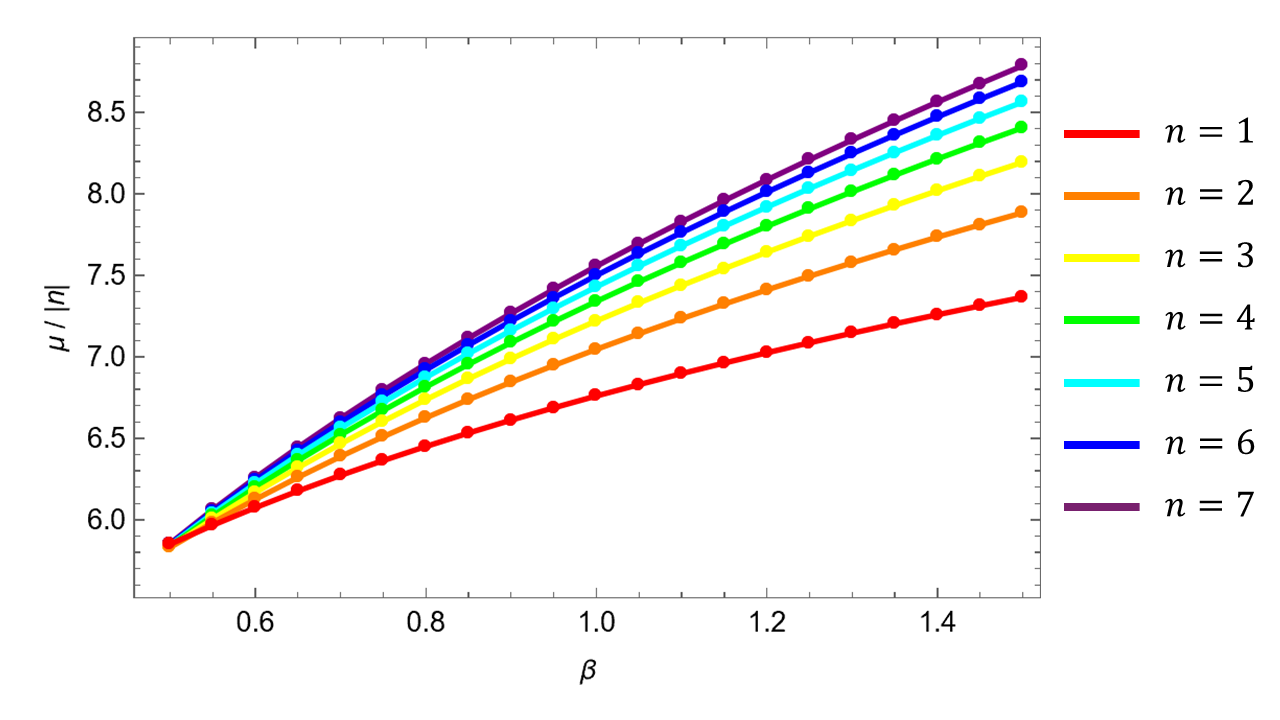}
        \subcaption{$0.5\leq\beta\leq1.5$}
    \end{minipage}
    \begin{minipage}[b]{0.5\linewidth}
        \centering
        \includegraphics[keepaspectratio, width=.9\columnwidth]{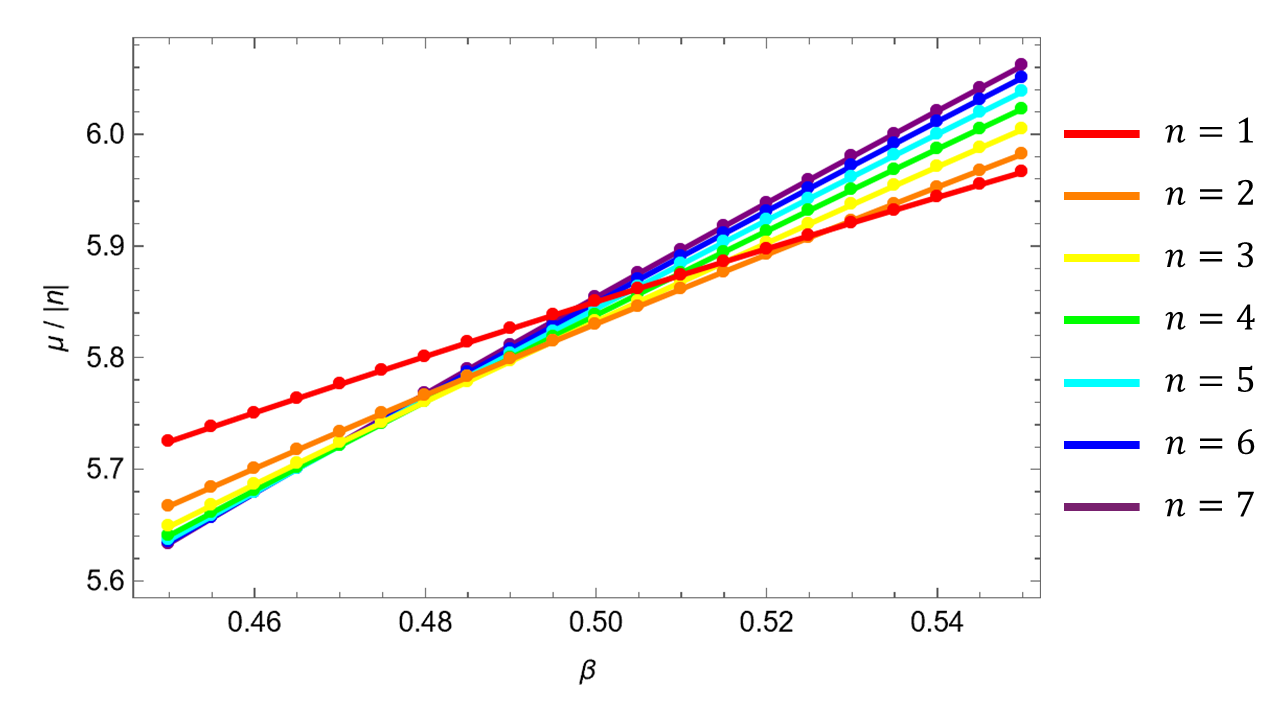}
        \subcaption{$0.45\leq\beta\leq0.55$}
    \end{minipage}
    \caption{
    Energy per unit length of the strings $\mu$ divided by the absolute value of its winding number $n$. 
    Each line corresponds to a different winding number.
    The top panels corresponds to the MH case, and the bottom panels corresponds to Case 1. 
    }
    \label{Bogomolnyi_nad1}
\end{figure}

Second, we investigate how the tension of a string $\mu$ per the winding number $n$ depends on $\beta$. 
\footnote{We set $v=1$ because the value of $v$ is irrelevant to qualitative features of the results. Hereafter, we hold this setting in our numerical calculations.}
The results in Case 1 and those in the MH case are summarized in Fig. \ref{Bogomolnyi_nad1}. 
In the MH case, all the lines intersect at $\beta=1$ due to the BPS state. 
On the other hand, we find that all lines do not intersect at a certain point in Case 1. 
This indicates that there is no state corresponding to the BPS state in Case 1.
This implication is consistent with the fact that the transformation shown in Eq. (\ref{Bogomolnyi completion}) cannot be applied to the potential considered in Case 1.

Furthermore, this result indicates that the winding number $n$ of the most stable string configuration depends on $\beta$ in Case 1. 
For example, if we take $\beta=0.5$, the most stable string configuration corresponds to $|n|=2$, as shown in Fig. \ref{Bogomolnyi_nad1}.
On the other hand, if we take $\beta=0.48$, the most stable string configuration corresponds to $|n|=3$. 
This interesting feature does not appear in the case of the Mexican hat potential, and it may affect the interaction between strings with different winding numbers from each other. 
However, analysis of this topic is beyond the scope of the present paper.

\begin{figure}[t]
    \begin{minipage}[b]{0.5\linewidth}
        \centering
        \includegraphics[keepaspectratio, width=.9\columnwidth]{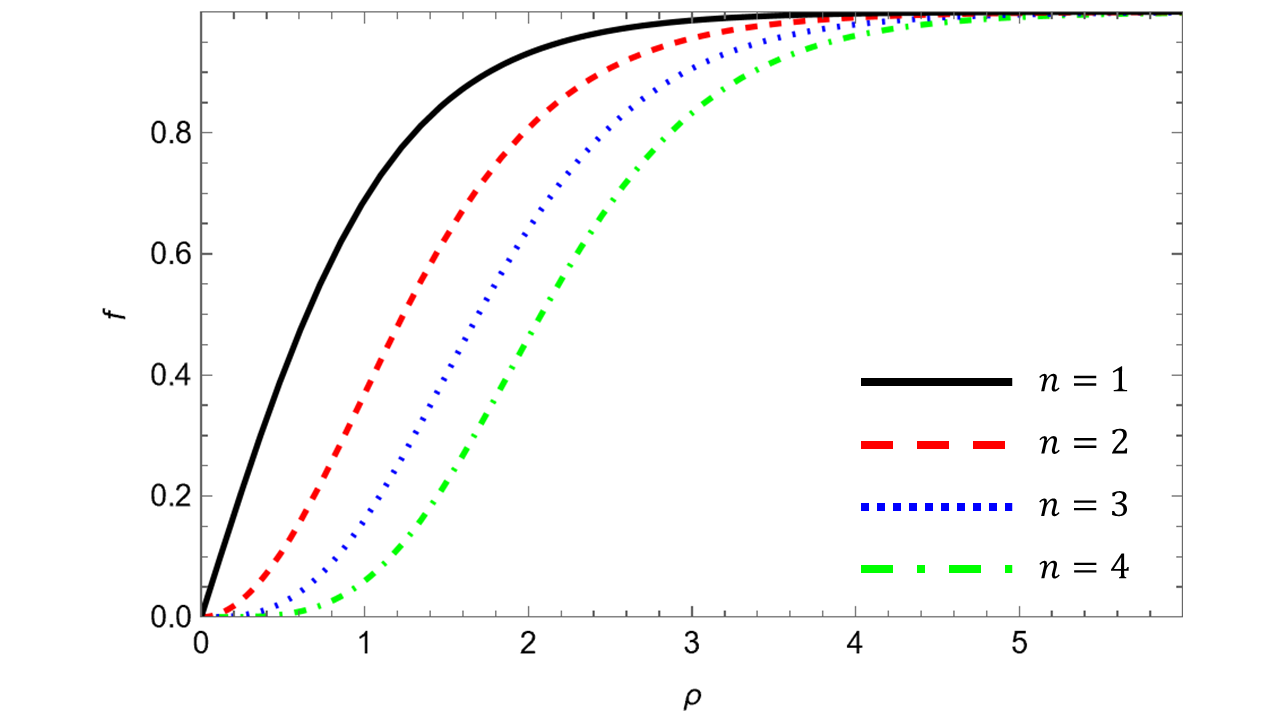}
        \subcaption{The MH case}
    \end{minipage}
    \begin{minipage}[b]{0.5\linewidth}
        \centering
        \includegraphics[keepaspectratio, width=.9\columnwidth]{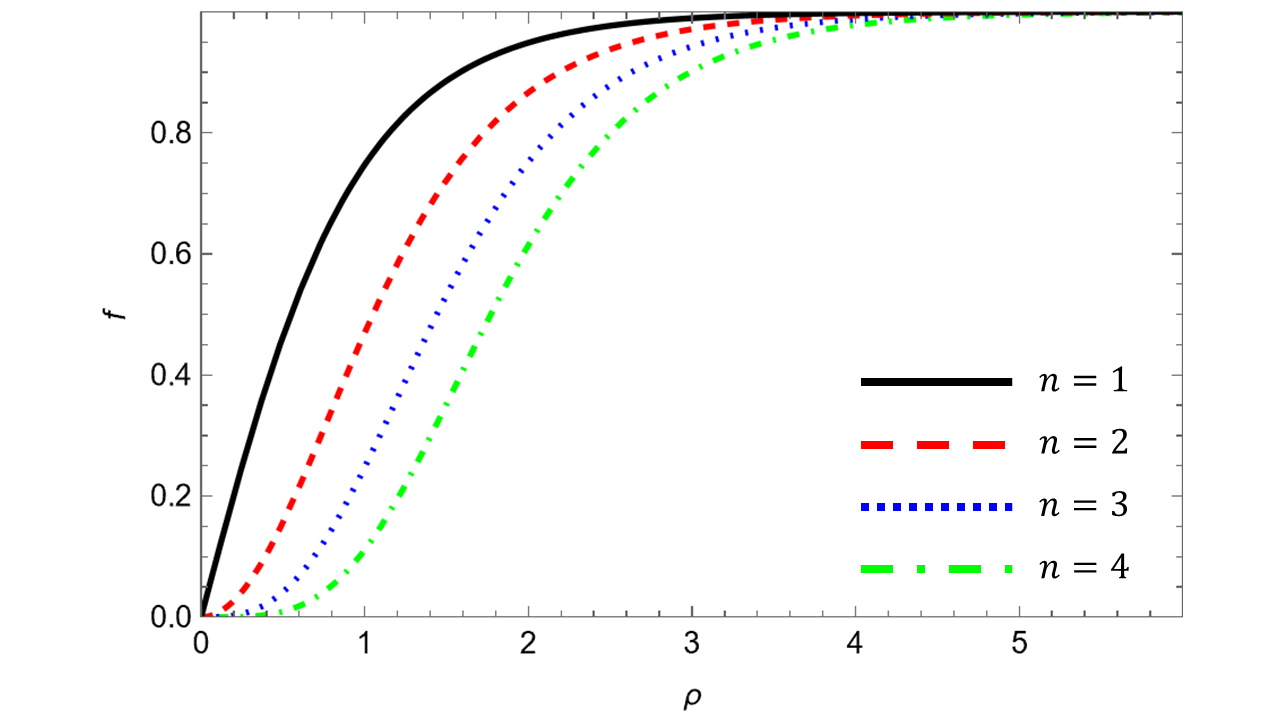}
        \subcaption{Case 1}
    \end{minipage}
    \caption{The behaviors of $f(\rho)$ for different winding numbers. The left panel shows the MH case, while the right panel shows Case 1. Each line corresponds to a different winding number.}
    \label{fa_winding}
\end{figure}

We consider the difference between the two cases shown in Fig. \ref{Bogomolnyi_nad1} by focusing on the potentials. 
In the first place, the value of $\mu/|n|$ in Case 1 is larger than that in the MH case for the same $n$ and $\beta$. 
We infer that this difference is due to the variation in the potential values around $\phi/v\sim0$. 
In addition, the excited region of the scalar field becomes large when $n$ increase as shown in Fig. \ref{fa_winding}. 
Hence, the contribution from the potential energy to the tension becomes more significant in strings with larger $n$, and the deviation of the value of $\mu/|n|$ from the MH case also increases. 
Compared to the MH case, the line of $\mu/|n|$ corresponding to a larger $n$ in Fig. \ref{fa_winding} moves upwards by a greater amount, causing the intersection points to shift to the side with smaller $\beta$ as in Fig. \ref{Bogomolnyi_nad1}.

\begin{figure}[t]
    \begin{minipage}[b]{0.5\linewidth}
        \centering
        \includegraphics[keepaspectratio, width=.9\columnwidth]{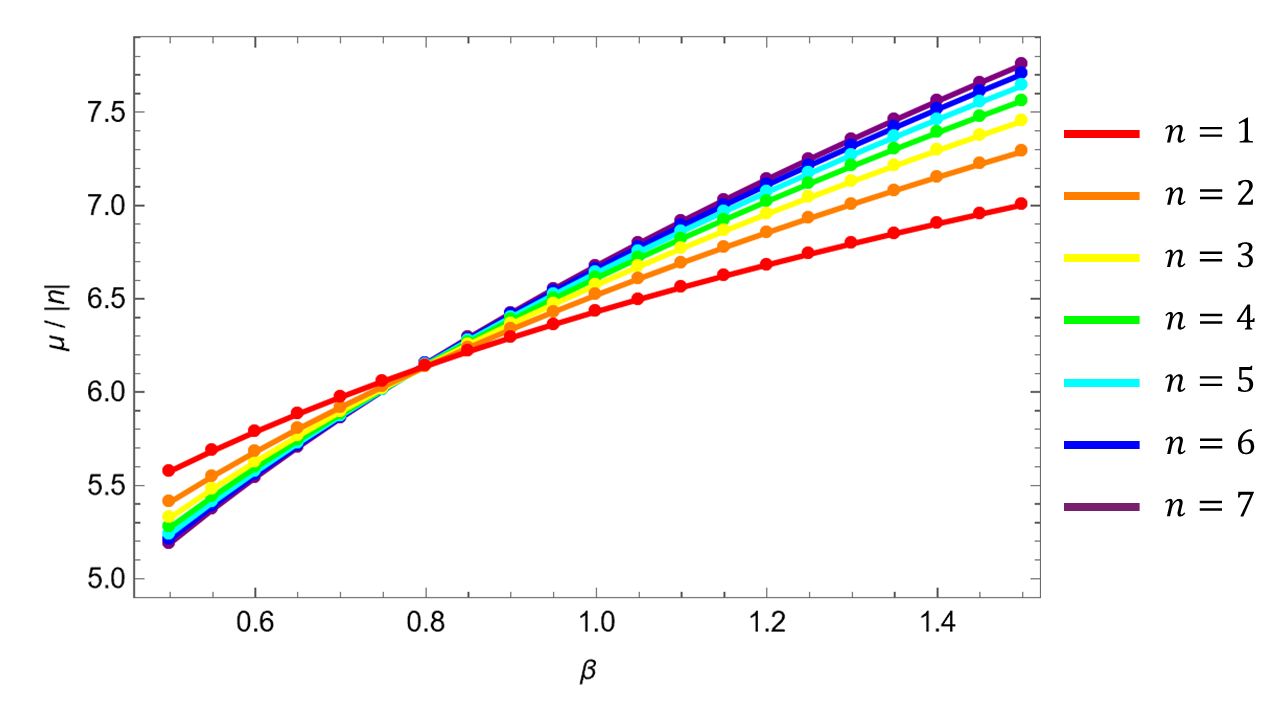}
        \subcaption{$0.5\leq\beta\leq1.5$}
    \end{minipage}
    \begin{minipage}[b]{0.5\linewidth}
        \centering
        \includegraphics[keepaspectratio, width=.9\columnwidth]{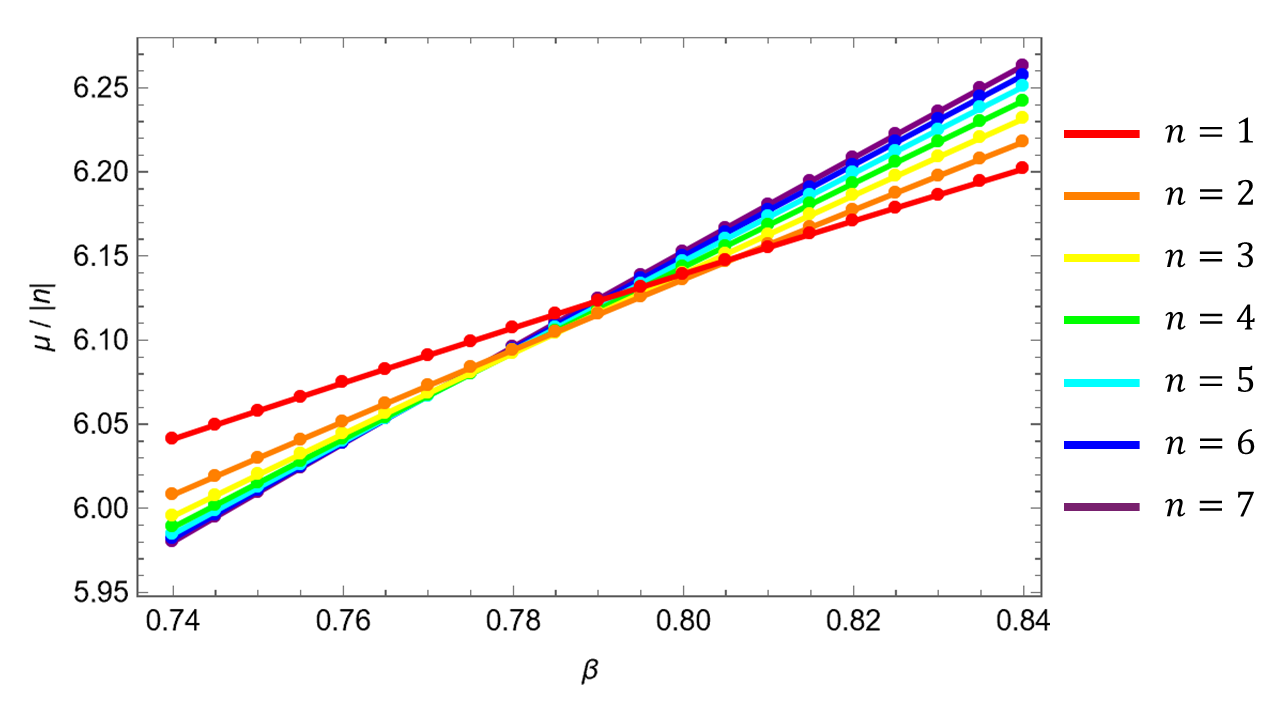}
        \subcaption{$0.74\leq\beta\leq0.84$}
    \end{minipage}
    \begin{minipage}[b]{0.5\linewidth}
        \centering
        \includegraphics[keepaspectratio, width=.9\columnwidth]{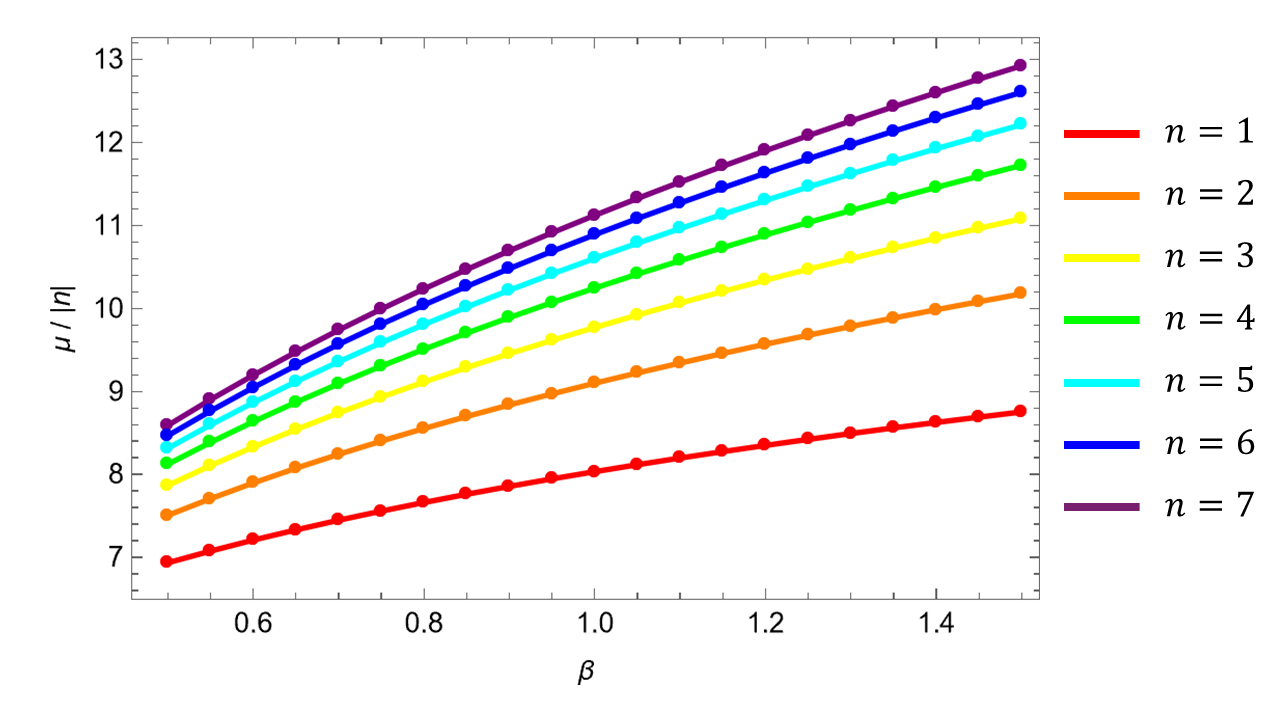}
        \subcaption{$0.5\leq\beta\leq1.5$}
    \end{minipage}
    \begin{minipage}[b]{0.5\linewidth}
        \centering
        \includegraphics[keepaspectratio, width=.9\columnwidth]{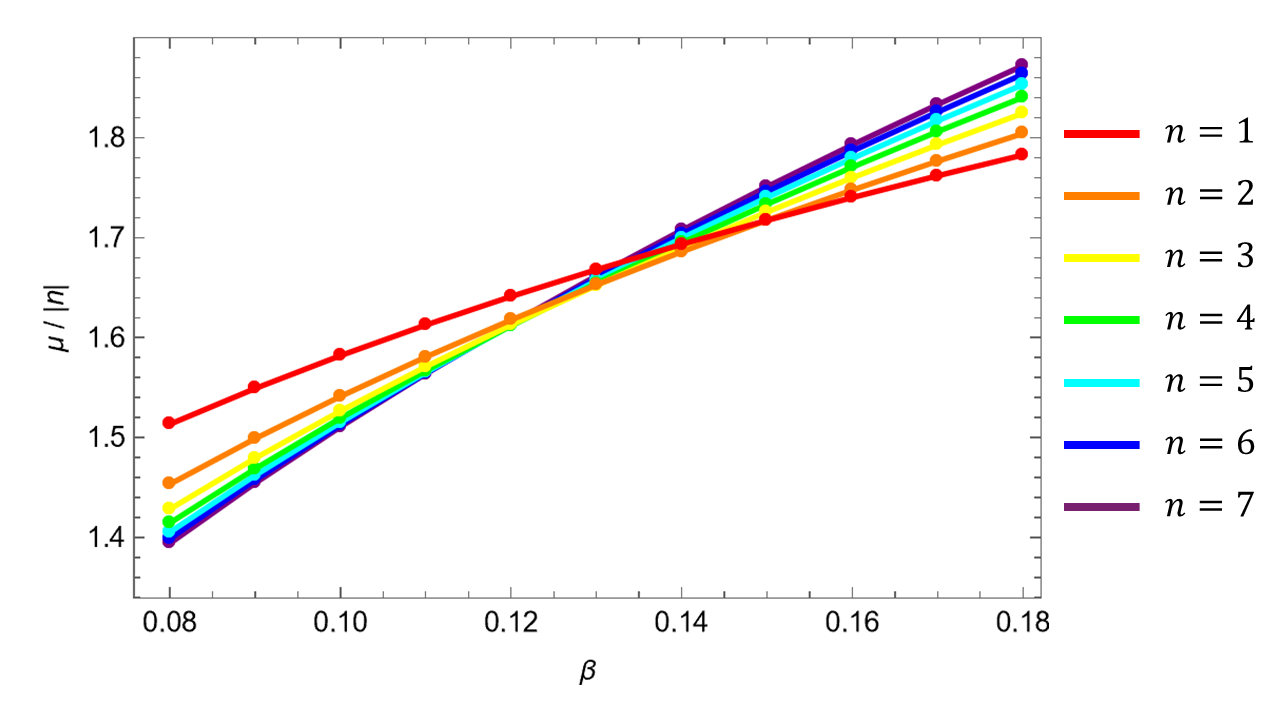}
        \subcaption{$0.08\leq\beta\leq0.18$}
    \end{minipage}
    \caption{
    Energy per unit length of the strings $\mu$ divided by the absolute value of its winding number $n$. 
    Each line corresponds to a different winding number. 
    The top panels corresponds to Case 2, and the bottom panels corresponds to Case 3. 
    }
    \label{Bogomolnyi_nad1nf1}
\end{figure}

This inference can be applied to other cases. 
In the case of the CW potential, the intersection points of lines of $\mu/|n|$ shift to around $\beta\sim2$ as shown in Fig. 5 of \cite{Eto:2022hyt}. 
Since this deviation is opposite to the result in Case 1, we can interpret it as being due to the CW potential having a lower value at $\phi/v \sim 0$ compared to the Mexican hat potential. 
Moreover, we show the results of Case 2 and Case 3 in Fig. \ref{Bogomolnyi_nad1nf1}. 
The intersection points of the lines are in the region of $\beta<1$ in the both cases, but we can find that the detailed behavior differs in Case 1, Case 2, and Case 3. 
In Fig. \ref{Bogomolnyi_nad1nf1}, the intersection points are around $\beta\sim0.8$ for Case 2, but those for Case 3 are around $\beta\sim0.1$. 
In line with our inference in the previous paragraph, we can understand that this fact is caused by the differences in the deviations of the potentials from the MH potential. 

We finally note that interaction between two strings is inferred from Figs. \ref{Bogomolnyi_nad1} and \ref{Bogomolnyi_nad1nf1}. 
Comparing the line with $|n|=1$ and $|n|=2$, it is possible to understand which is more stable: a state with two strings that are sufficiently far apart with $n=1$ or one string with $n=2$. 
Therefore, we naively infer the interaction is repulsive for $\beta>0.5$ and attractive for $\beta<0.5$ in Case 1. 
However, the dependence of a distance between strings is nontrivial as discussed in \cite{Eto:2022hyt}. 
We investigate this in detail by using some numerical methods in the next section.

\section{Interaction between two cosmic strings}\label{2strings}

In this section, we calculate the interaction energy of two parallel strings separated by a distance $d$. 
Specifically, we derive the field configurations in such a two string system, and calculate its energy. 
Due to the translational symmetry along the strings, we only consider the energy in the plane perpendicular to the strings. 

\subsection{Calculation method}

Before explaining the calculation method that we use, we summarize the normalization of the coordinates and the fields. 
In Sec. \ref{Sec.ANOstring}, we defined the normalized polar coordinates $(\rho, \theta)$ where $\rho=g_4vr$. 
In this section, we mainly use the normalized cartesian coordinates as
\begin{align}
    \Bar{x}^1\equiv \rho \cos\theta, \quad \Bar{x}^2\equiv \rho \sin\theta \,.
    \label{cartesian_normalization}
\end{align}
For the scalar field, we have already defined how to normalize in Eq. (\ref{normalized_Veff}) as $\tilde{\phi}\equiv\phi/v$. 
We also define a normalization of the gauge field as 
\begin{align}
    \tilde{A}_\mu \equiv \frac{A_\mu}{v} \,.
\end{align}
Note that $\Bar{x}^i, \tilde{\phi}$ and $\tilde{A}_\mu$ becomes dimensionless quantities, and it is useful to perform numerical calculations. 

The energy per unit length of the two string system $\mu$ is rewritten under the above normalization as
\begin{align}
    \label{mu_normalized_general}
    \mu = v^2 \int d\Bar{x}^1d\Bar{x}^2 \left[ \frac{1}{4} \left(\partial_i\tilde{A}_j-\partial_j\tilde{A}_i\right)^2 + \left| \left(\partial_i-i\tilde{A}_i\right) \tilde{\phi} \right|^2 + \widetilde{V}(\tilde{\phi}) + \frac{1}{2}\left(\partial_i\tilde{A}_i\right)^2 \right] \,,
\end{align}
where $i,j=1,2$ and $\partial_i\equiv\frac{\partial}{\partial \Bar{x}^i}$. 
The normalized potential $\widetilde{V}(\tilde{\phi})$ is 
\begin{align}
    \widetilde{V}(\tilde{\phi}) = \frac{\beta}{2} \left(|\tilde{\phi}|^2 - 1\right)^2
\end{align}
for the MH case, and shown in Eq. (\ref{normalized_Veff}) for Case 1, Case 2, and Case 3. 
The last term of the integrand in Eq. (\ref{mu_normalized_general}) is a gauge fixing term based on the Coulomb gauge. 
This term helps us to avoid technical problems in numerical calculations such as convergence. 
We perform numerical calculations by using the energy density in Eq. (\ref{mu_normalized_general}). 

We explain how to derive the field configurations corresponding to the two-string system. 
The numerical methods used in this section are primarily based on the approaches described in Refs. \cite{Eto:2022hyt, Fujikura:2023lil}.
First, we prepare a two-dimensional lattice plane with a box size $L=30$ and a grid size $\Delta L=0.05$. 
We then assume that the cores of the two strings are fixed at $(\Bar{x}^1,\Bar{x}^2)=(\pm d/2,0)$, respectively. 
We derive the field configuration by using the gradient flow method. 
This method is useful for finding a configuration that minimizes the energy.
We apply it using the energy per unit length shown in Eq. (\ref{mu_normalized_general}) for each case. 
A detailed explanation is provided in Appendix \ref{GFM}. 
The interaction energy is obtained by substituting the field configuration derived using the gradient flow method into Eq. (\ref{mu_normalized_general}).

\subsection{Numerical results}

In this subsection, we present the numerical results for the MH case, Case 1, Case 2, and Case 3. 
For simplicity, we focus on the case where each string has a winding number $n=1$. 
We mainly show the results for $\beta$ in the range of $0.1 \leq \beta \leq 1.5$.
However, we will perform our calculations for $\beta=0.01$ in Sec. \ref{Sec.smallbeta} to investigate the interaction for small values of $\beta$, motivated by the higher-dimensional gauge theory. 

\subsubsection{Field configurations}

\begin{figure}[t]
    \begin{minipage}[b]{0.5\linewidth}
        \centering
        \includegraphics[keepaspectratio, width=.9\columnwidth]{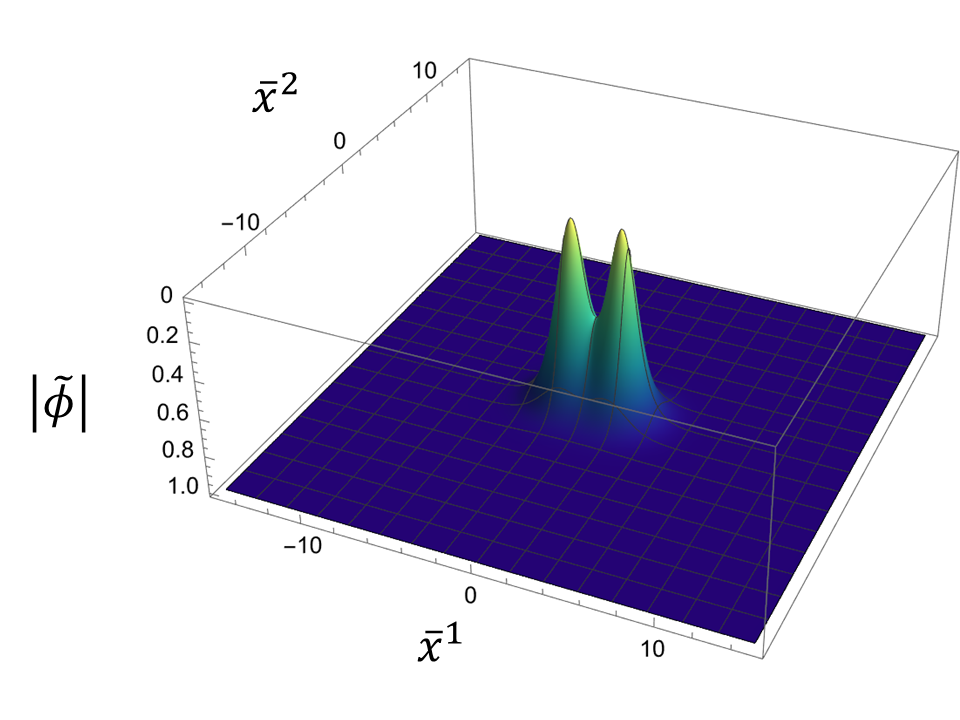}
        \subcaption{The MH case}
    \end{minipage}
    \begin{minipage}[b]{0.5\linewidth}
        \centering
        \includegraphics[keepaspectratio, width=.9\columnwidth]{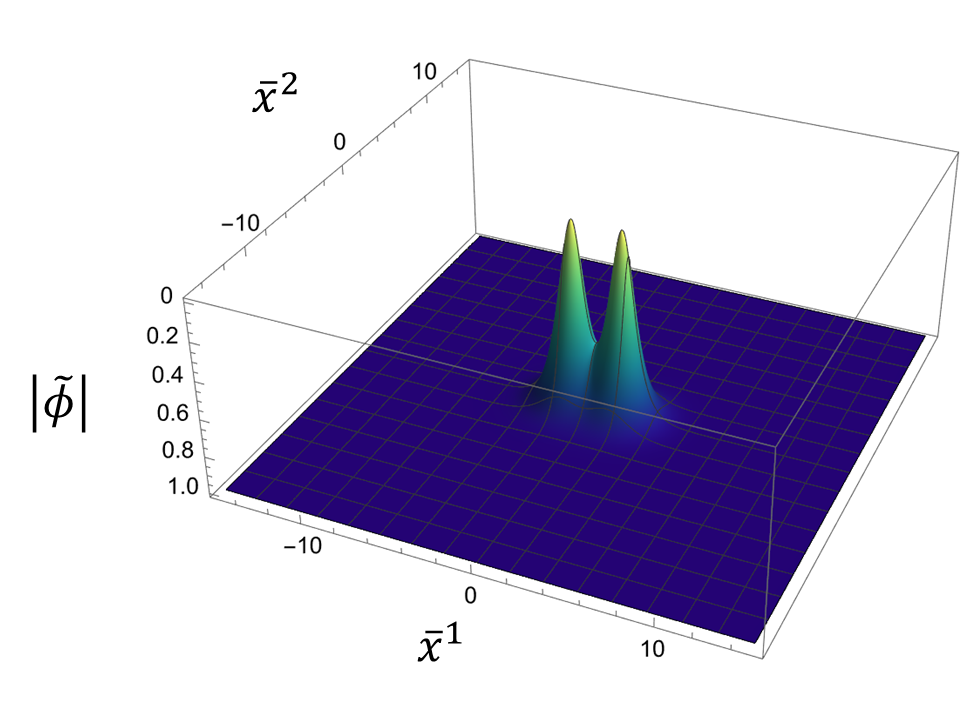}
        \subcaption{Case 1}
    \end{minipage}
    \begin{minipage}[b]{0.5\linewidth}
        \centering
        \includegraphics[keepaspectratio, width=.9\columnwidth]{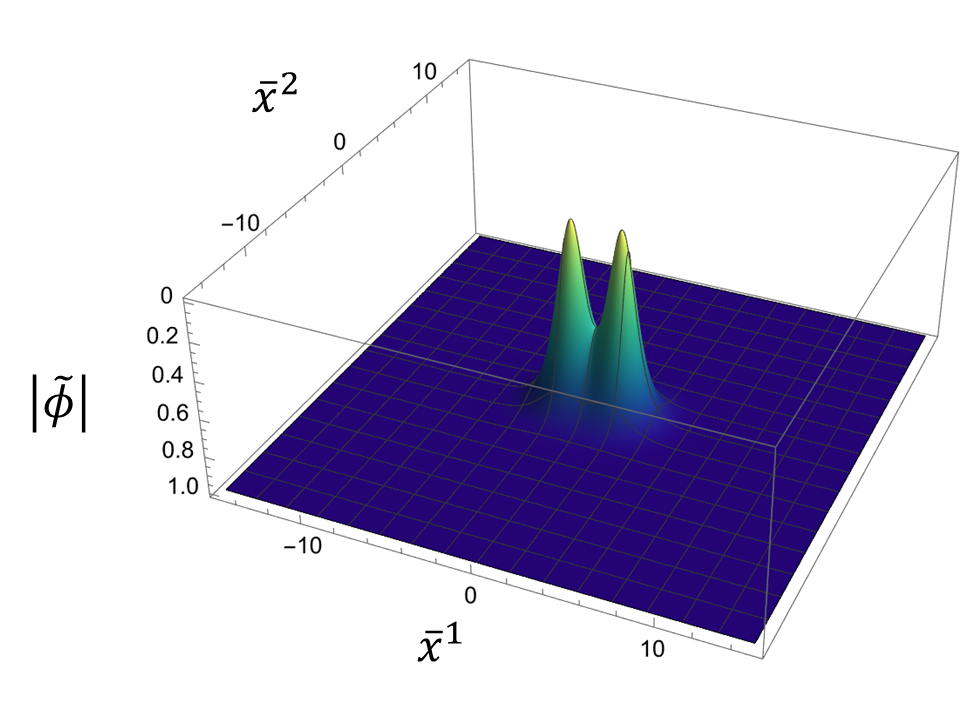}
        \subcaption{Case 2}
    \end{minipage}
    \begin{minipage}[b]{0.5\linewidth}
        \centering
        \includegraphics[keepaspectratio, width=.9\columnwidth]{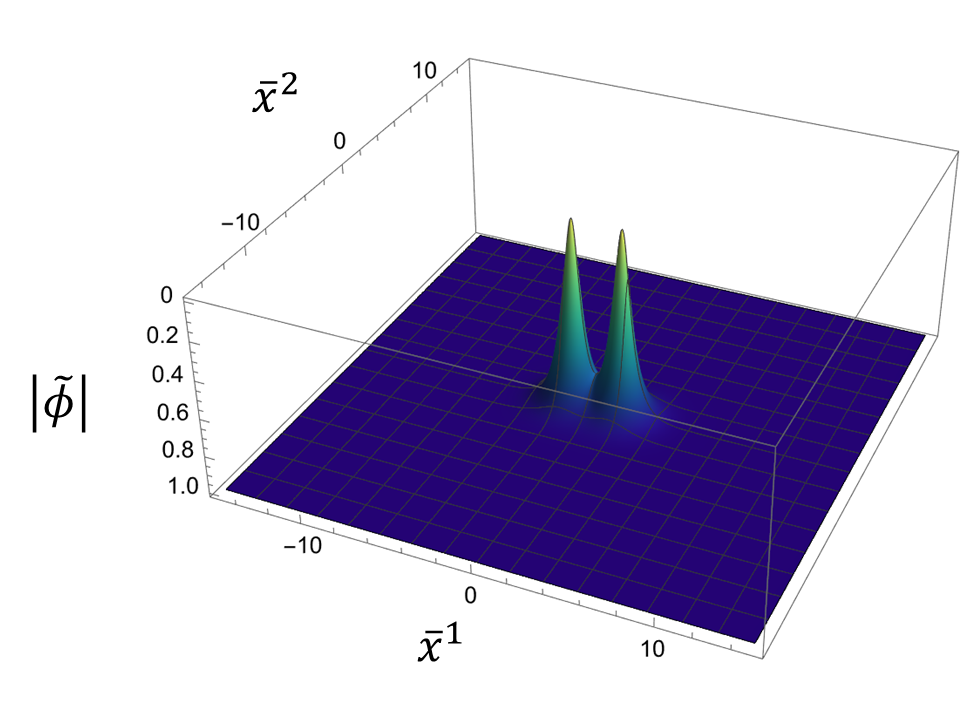}
        \subcaption{Case 3}
    \end{minipage}
    \caption{Field configurations of $|\tilde{\phi}|$ for $\beta=1.0$ and $d=3.0$ on a two-dimensional plane. 
    Each panel corresponds to different cases: The MH case (top-left), Case 1 (top-right), Case 2 (bottom-left), and Case 3 (bottom-right).
    }
    \label{2string_f}
\end{figure}

\begin{figure}[t]
    \centering
    \includegraphics[width=0.7\linewidth]{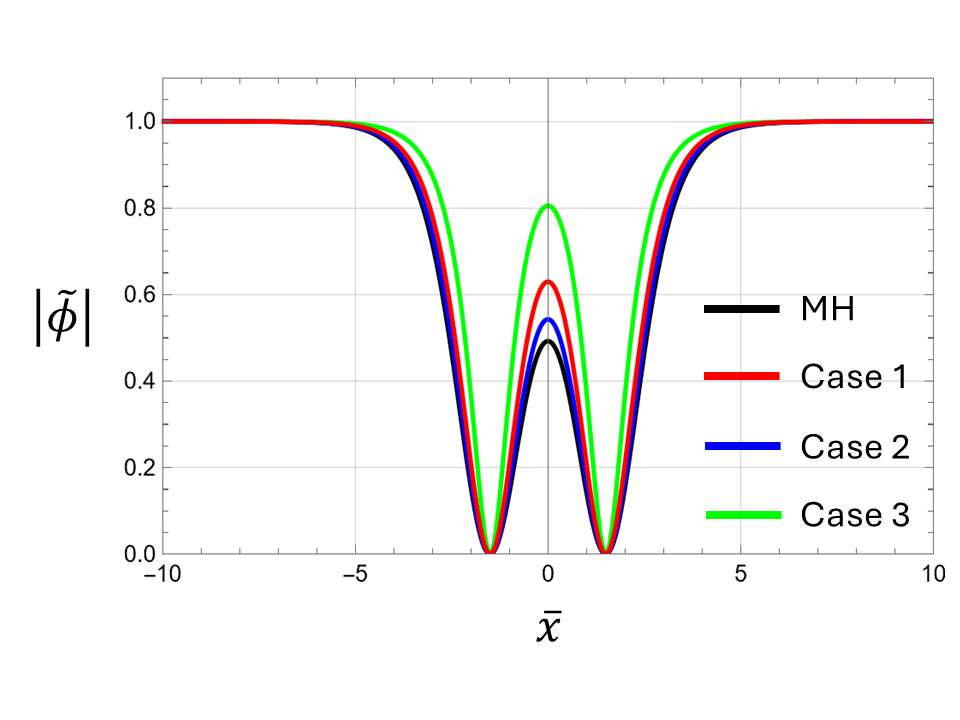}
    \caption{Comparison of configurations of $|\tilde{\phi}|$ along $\bar{x}^2=0$. 
    Each line corresponds to different cases: The MH case (black), Case 1 (red), Case 2 (blue), and Case 3 (green).
    }
    \label{2string_f_cut}
\end{figure}

First, we present the results for the field configurations in the two-string systems. 
In Fig. \ref{2string_f}, we show the absolute value of the scalar field $|\tilde{\phi}|$ on the two-dimensional space for $\beta=1.0$ and $d=3.0$ for the each case. 
The two peaks in these graphs correspond to the points where the cores of the two strings are located. 
Note that we invert the vertical axis for clarity in presenting the results. 
The slices of these graphs along $\bar{x}^2=0$ are shown in Fig. \ref{2string_f_cut}. 

We find that excited regions of the scalar field in our cases are thinner than those of conventional ANO strings. 
The sharpness of the scalar field configurations increases in the order of Case 3, Case 1, and Case 2. 
These features are consistent with our consideration regarding the width of a single string, as discussed in Sec. \ref{Sec.Numerical_1string}.

\begin{figure}[t]
    \begin{minipage}[b]{0.5\linewidth}
        \centering
        \includegraphics[keepaspectratio, width=.9\columnwidth]{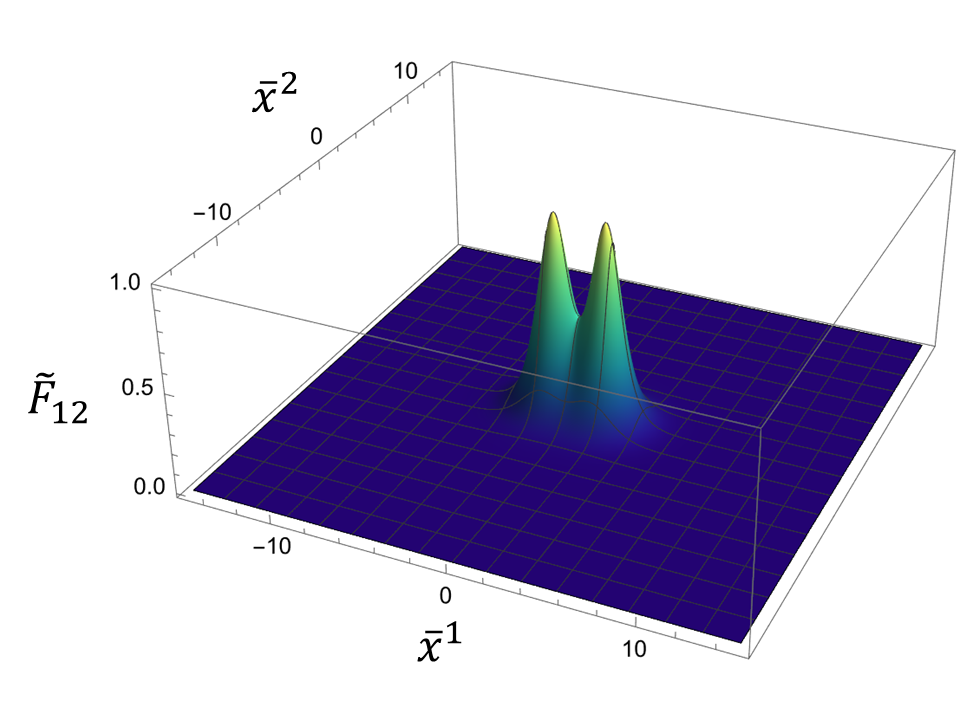}
        \subcaption{The MH case}
    \end{minipage}
    \begin{minipage}[b]{0.5\linewidth}
        \centering
        \includegraphics[keepaspectratio, width=.9\columnwidth]{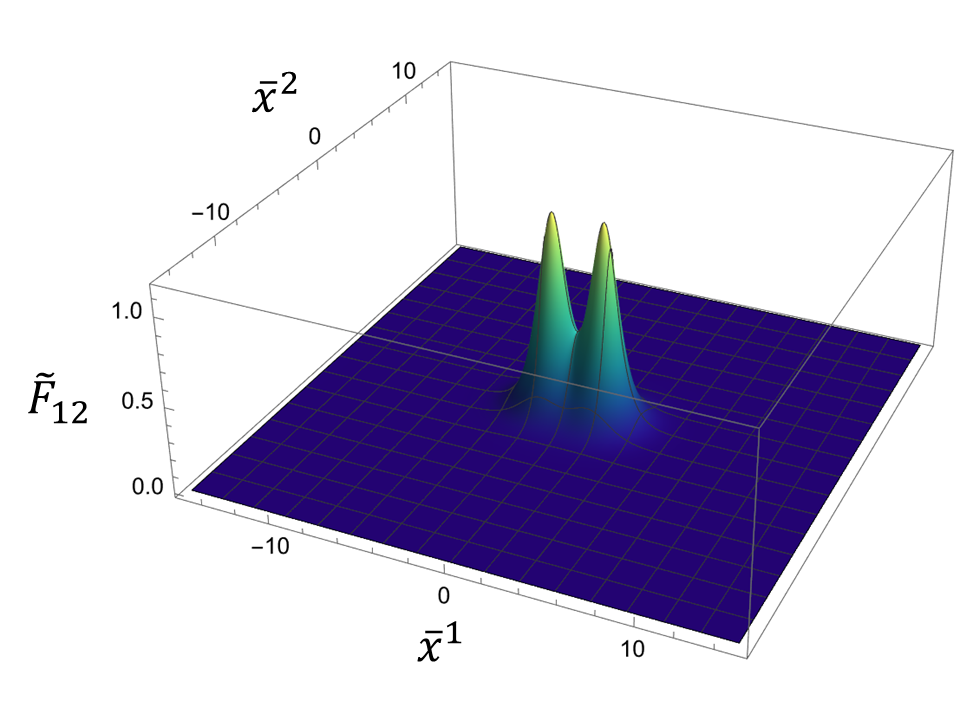}
        \subcaption{Case 1}
    \end{minipage}
    \begin{minipage}[b]{0.5\linewidth}
        \centering
        \includegraphics[keepaspectratio, width=.9\columnwidth]{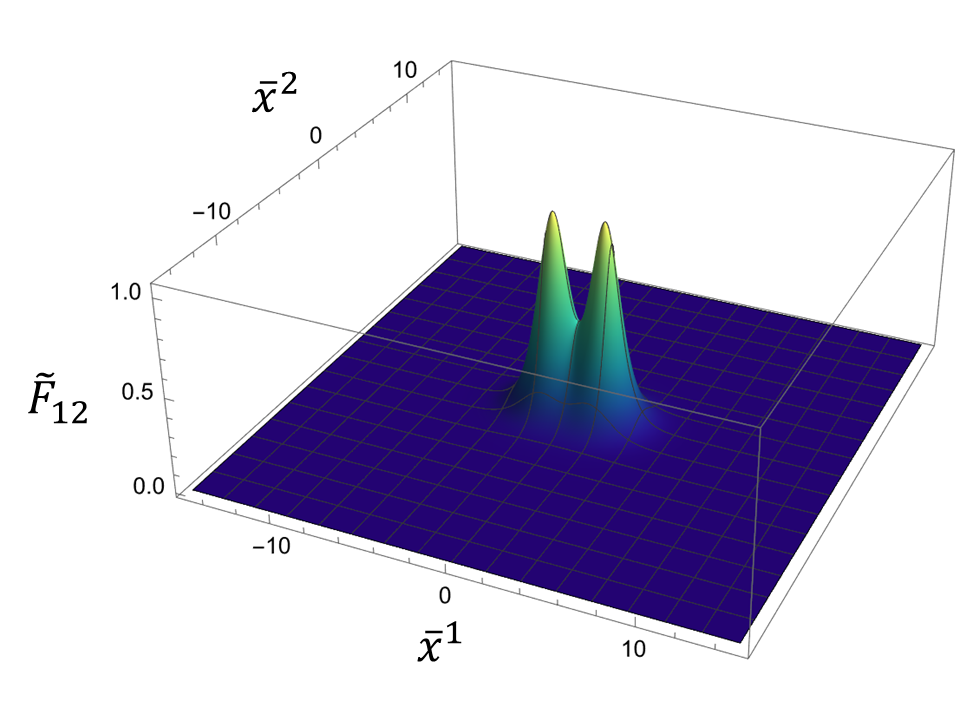}
        \subcaption{Case 2}
    \end{minipage}
    \begin{minipage}[b]{0.5\linewidth}
        \centering
        \includegraphics[keepaspectratio, width=.9\columnwidth]{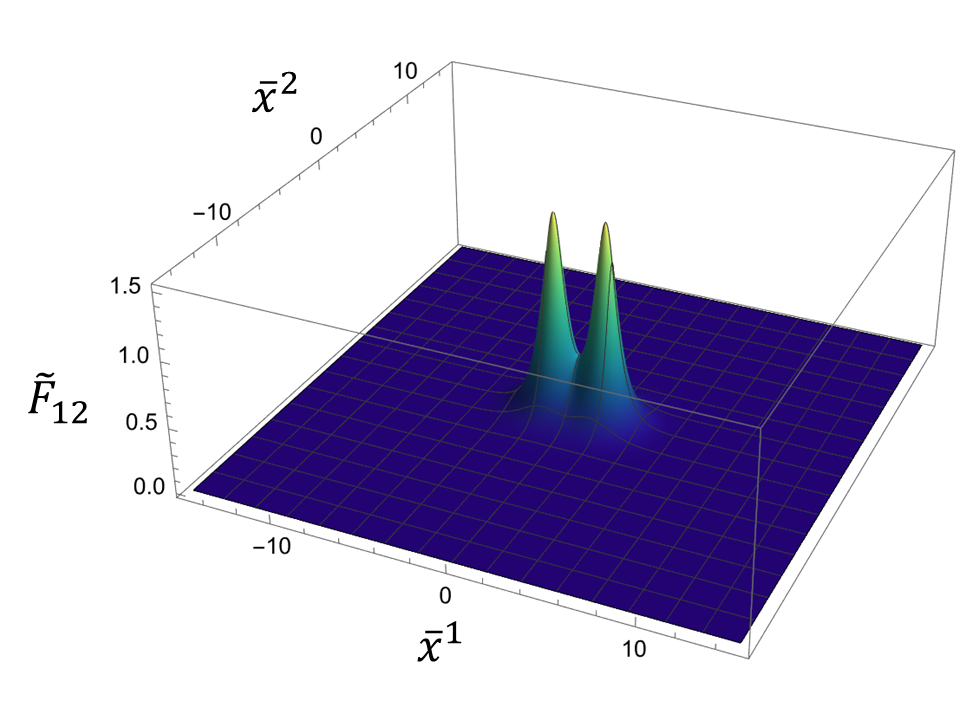}
        \subcaption{Case 3}
    \end{minipage}
    \caption{Field configurations of the flux $\tilde{F}_{12}$ for $\beta=1.0$ and $d=3.0$ on a two-dimensional plane. 
    Each panel corresponds to different cases: The MH case (top-left), Case 1 (top-right), Case 2 (bottom-left), and Case 3 (bottom-right).
    }
    \label{2string_F12}
\end{figure}

\begin{figure}[t]
    \begin{minipage}[b]{0.5\linewidth}
        \centering
        \includegraphics[keepaspectratio, width=.9\columnwidth]{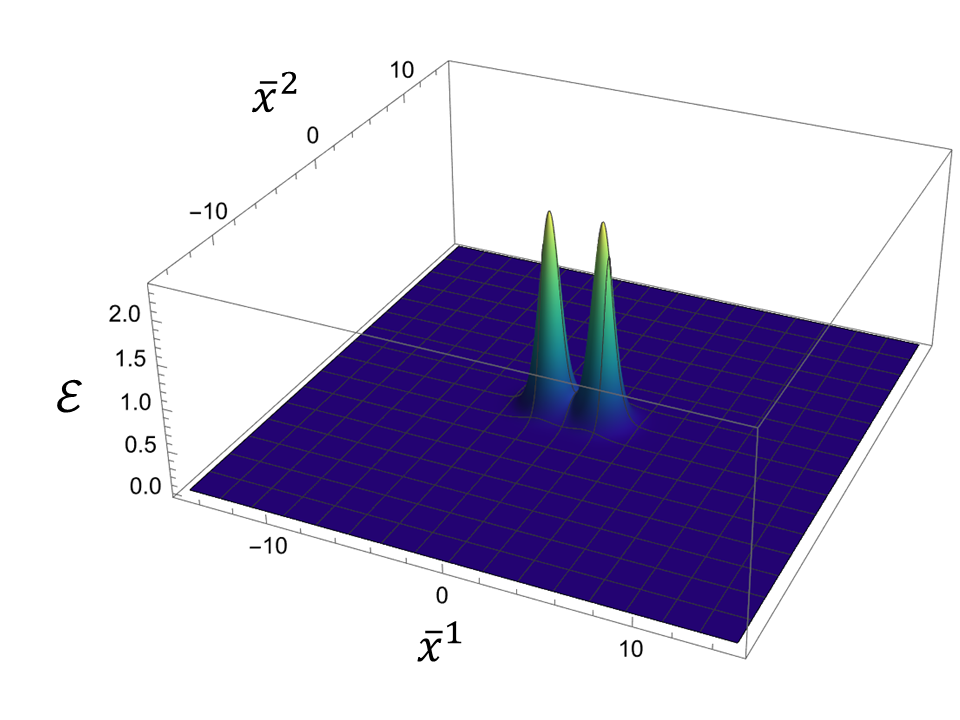}
        \subcaption{The MH case}
    \end{minipage}
    \begin{minipage}[b]{0.5\linewidth}
        \centering
        \includegraphics[keepaspectratio, width=.9\columnwidth]{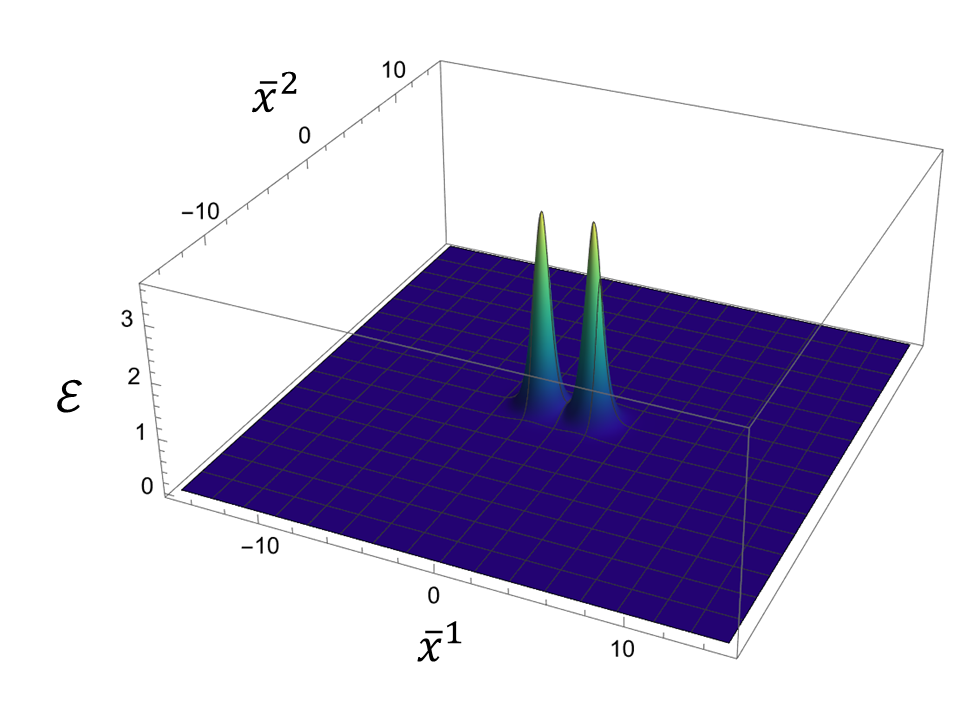}
        \subcaption{Case 1}
    \end{minipage}
    \begin{minipage}[b]{0.5\linewidth}
        \centering
        \includegraphics[keepaspectratio, width=.9\columnwidth]{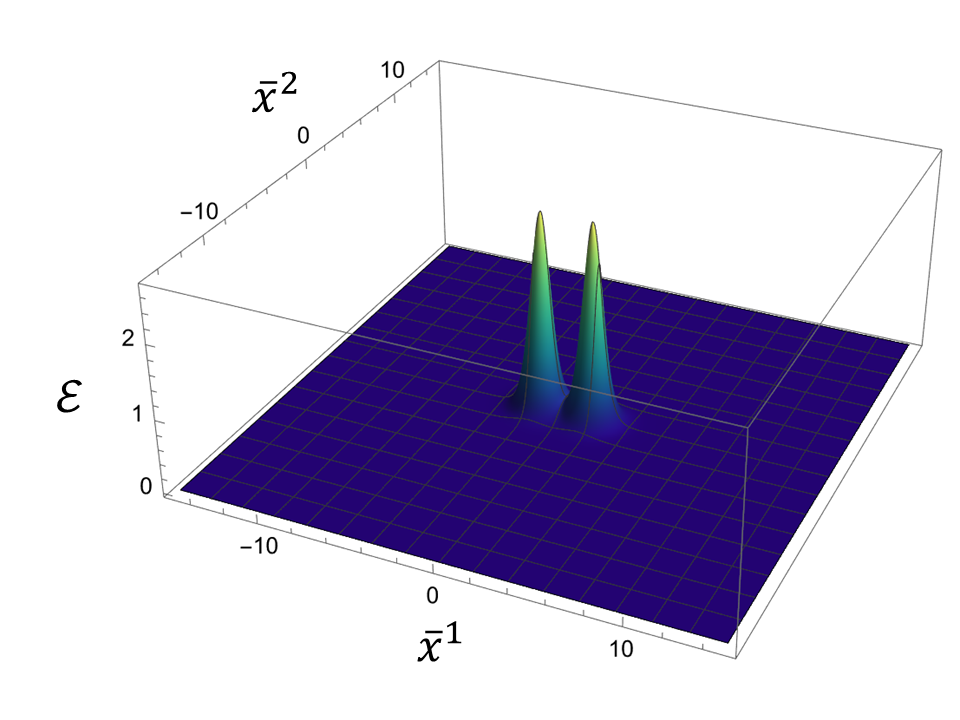}
        \subcaption{Case 2}
    \end{minipage}
    \begin{minipage}[b]{0.5\linewidth}
        \centering
        \includegraphics[keepaspectratio, width=.9\columnwidth]{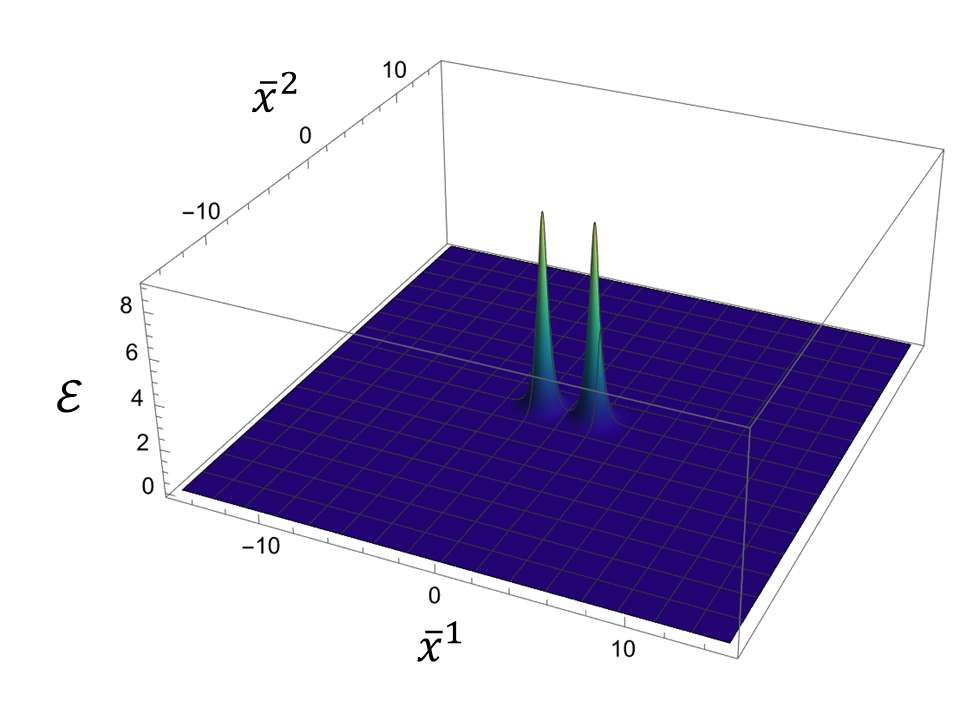}
        \subcaption{Case 3}
    \end{minipage}
    \caption{Field configurations of the energy density $\mathcal{E}$ for $\beta=1.0$ and $d=3.0$ on a two-dimensional plane. 
    Each panel corresponds to different cases: The MH case (top-left), Case 1 (top-right), Case 2 (bottom-left), and Case 3 (bottom-right).
    }
    \label{2string_E}
\end{figure}

\begin{figure}[t]
    \begin{minipage}[b]{0.5\linewidth}
        \centering
        \includegraphics[keepaspectratio, width=.95\columnwidth]{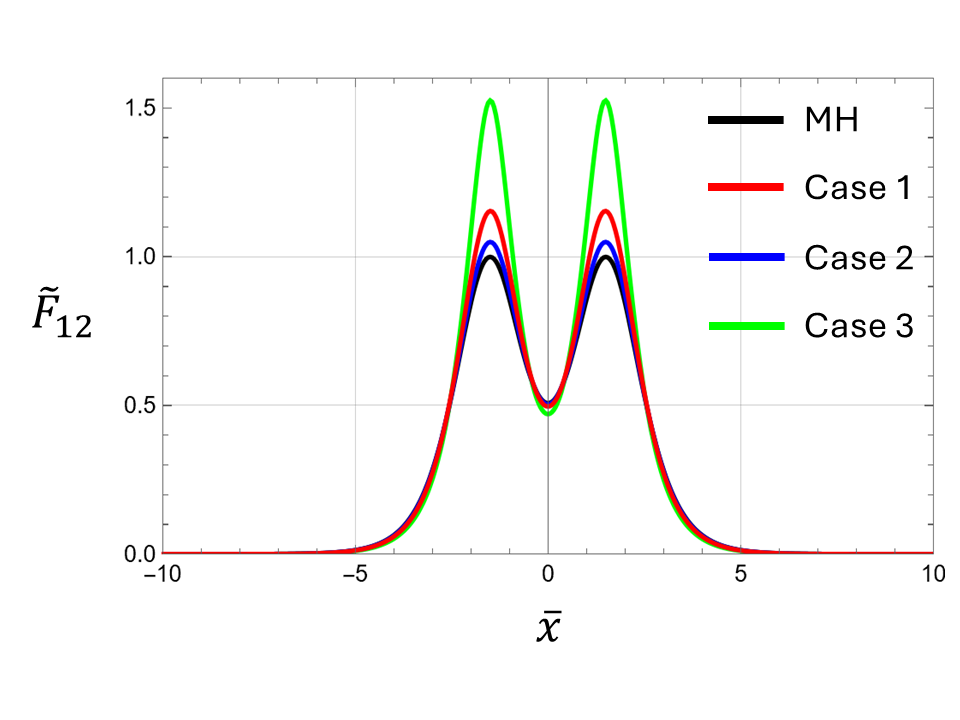}
        \subcaption{Configurations of $\tilde{F}_{12}$}
    \end{minipage}
    \begin{minipage}[b]{0.5\linewidth}
        \centering
        \includegraphics[keepaspectratio, width=.95\columnwidth]{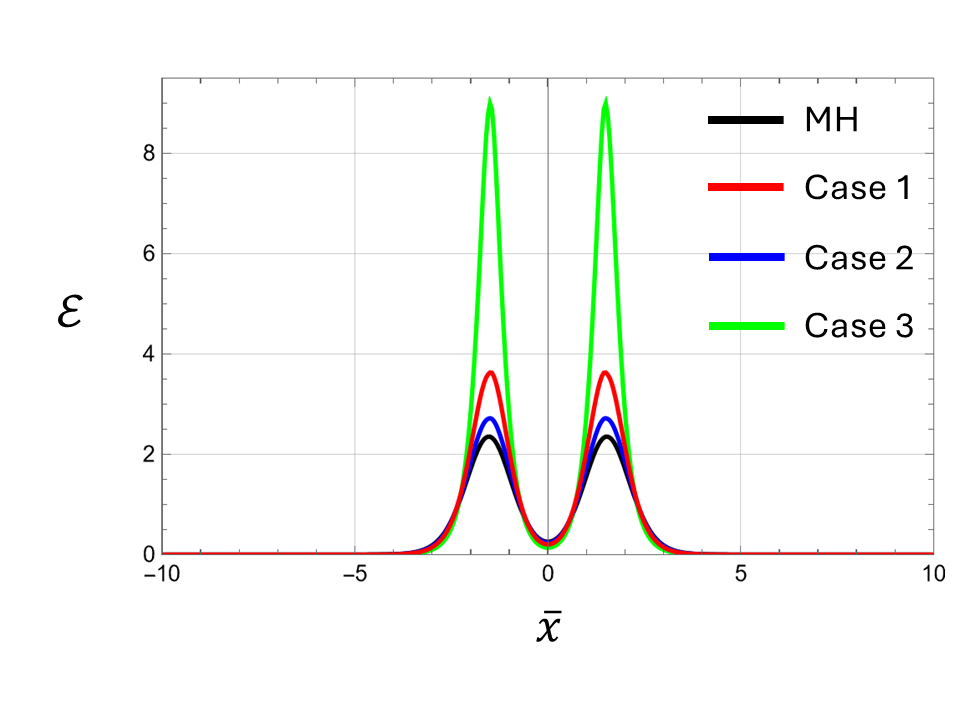}
        \subcaption{Configurations of $\mathcal{E}$}
    \end{minipage}
    \caption{
    Comparison of field configurations along $\bar{x}^2=0$. 
    The left panel shows the flux $\tilde{F}_{12}$, and the right panel shows the energy density $\mathcal{E}$. 
    Each line corresponds to different cases: The MH case (black), Case 1 (red), Case 2 (blue), and Case 3 (green).
    }
    \label{2string_FE_cut}
\end{figure}

We also present the two-dimensional field configurations for the flux $\tilde{F}_{12}$ and the energy density $\mathcal{E}$ for the same values of $\beta$ and $d$ in Figs. \ref{2string_F12} and \ref{2string_E}, respectively. 
Their slices along $\bar{x}^2=0$ are summarized in Fig. \ref{2string_FE_cut}. 
Similar to the results for $|\tilde{\phi}|$, the excited regions in our cases are thinner than those in the MH case. 
In addition, we observe an increase in the flux values at the centers of the strings in our cases, as illustrated in the left panel of Fig. \ref{2string_FE_cut}. 
This would be due to the differences in the gradients of the scalar field configurations. 
For the energy density $\mathcal{E}$, the values at the centers of the strings significantly rise in our cases. 
We attribute this to the combined increase in both the potential energy and the flux.

\subsubsection{Interaction between two strings}

Performing above numerical calculations for several values of $\beta$ and $d$, we calculate the interaction energy for each case. 
To investigate whether the interaction force is attractive or repulsive, we perform numerical calculations for $d$ ranging from $0.0$ to $6.0$ in increments of $0.1$. 

\begin{figure}[t]
    \centering
    \includegraphics[width=0.7\linewidth]{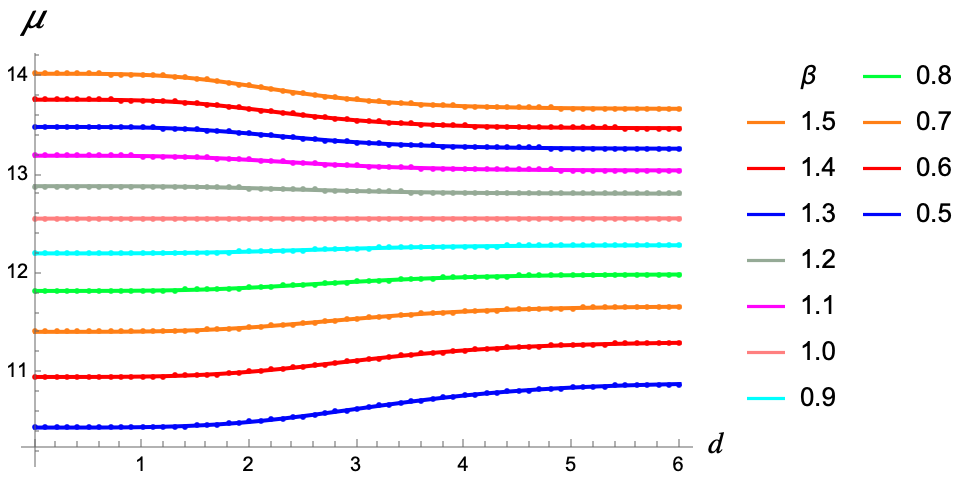}
    \caption{Distance dependence of the energy per unit length in the two-strings system for the MH case.}
    \label{WB_energy_vs_d}
\end{figure}

First, we present the results for the MH case in Fig. \ref{WB_energy_vs_d}. 
We find that the value of $\mu$ increases (decreases) as $d$ approaches 0 for $\beta>1$ ($\beta<1$). 
This result indicates that the interaction force between the two strings is repulsive (attractive) for $\beta>1$ ($\beta<1$) in the MH case. 
We also find that the line for $\beta=1$ in Fig. \ref{WB_energy_vs_d} is flat, indicating that the BPS state is realized. 
These results are consistent with conventional studies on the ANO string, including Refs. \cite{Eto:2022hyt, Fujikura:2023lil}. 

\begin{figure}[t]
    \begin{minipage}[b]{0.5\linewidth}
        \centering
        \includegraphics[keepaspectratio, scale=0.47]{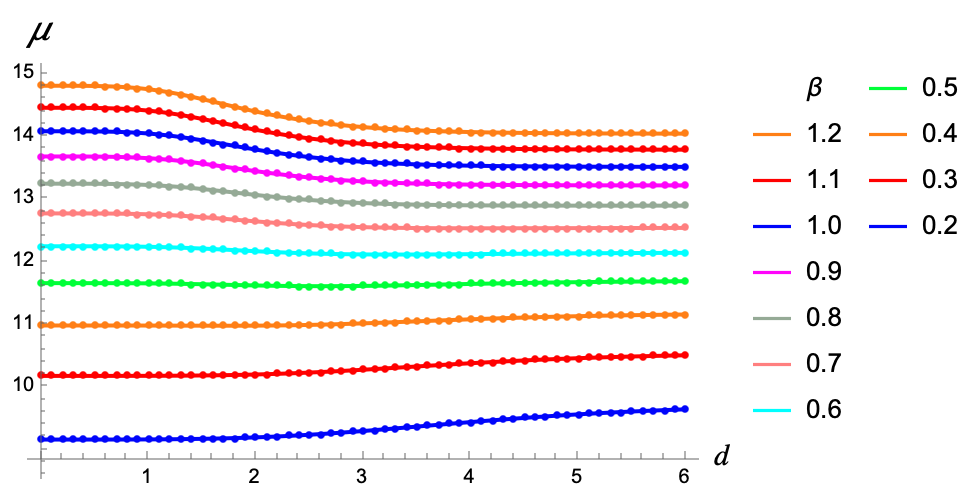}
        \subcaption{$0.2\le\beta\le1.2$}
    \end{minipage}
    \begin{minipage}[b]{0.5\linewidth}
        \centering
        \includegraphics[keepaspectratio, scale=0.47]{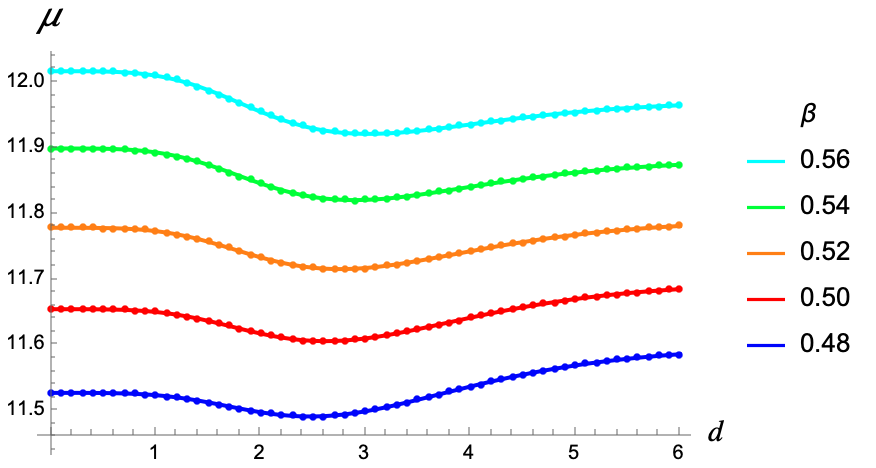}
        \subcaption{$0.48\le\beta\le0.56$}
    \end{minipage}
    \caption{
    Distance dependence of the energy per unit length in the two-strings system for Case 1.
    }
    \label{ED1_energy_vs_d}
\end{figure}

The interaction properties described above change in higher-dimensional scenarios. 
Fig. \ref{ED1_energy_vs_d} shows the results for Case 1. 
We find that the $\beta$ parameter space can be classified into three regions based on the $d$ dependence of $\mu$. 
Two of these regions correspond to the type-I and type-II regimes, similar to those observed in the MH case. 
For $\beta\geq0.9$, $\mu$ increase as $d\rightarrow0$ within the range $0\leq d\leq6$, indicating that the interaction force is always repulsive (type-II) in our calculations. 
In contrast, $\mu$ decreases as $d\rightarrow0$ for $\beta\leq0.3$, indicating that the interaction force is attractive (type-I). 
Between these two regions, \textit{i.e.} $0.4\leq\beta\leq0.8$, $\mu$ reaches a minimum at a certain value of $d$, denoted $d_c$. 
Note that $d_c\neq0$ and $d_c\neq\infty$. 
In this region, the interaction force is repulsive for $d<d_c$, but attractive for $d>d_c$. 
This behavior arises in neither the MH case nor the CW case.

We present the results for values around $\beta=0.52$ in the right panel in Fig. \ref{ED1_energy_vs_d}. 
In this region of $\beta$, the values of $\mu$ at $d=0.0$ and $d=6.0$ are close, allowing the $d$ dependence described above to be observed more clearly. 
We can see that $d_c$ is between $2$ and $3$, and the slopes of lines switch at $d=d_c$.

We mention that similar non-trivial $d$-dependent interactions of vortex solutions have been explored in condensed matter physics.  
In Refs. \cite{Babaev:2004hk, PhysRevLett.102.117001}, superconductors classified as ``type-1.5'' have been studied to investigate multi-component superconductors. 
In the type-1.5 superconductors, vortices formed in the background magnetic field exhibit an interaction potential with long-range attractive component and short-range repulsive component.  
This characteristic of vortices is analogous to the behavior of strings in our study. 
Therefore, we refer to the region of $\beta$ where the string interaction is repulsive (attractive) at small (large) $d$ as the type-1.5 regime, following Refs. \cite{Babaev:2004hk, PhysRevLett.102.117001}. 
We call a string in this region as the type-1.5 string. 

However, we emphasize that the models considered in Refs. \cite{Babaev:2004hk, PhysRevLett.102.117001} and our study are quite different. 
In Refs. \cite{Babaev:2004hk, PhysRevLett.102.117001}, the authors considered a model having two scalar fields. 
Each scalar field has a different Mexican hat potential: one corresponds to a type-I superconductor, while the other corresponds to a type-II superconductor. 
Vortex solutions are formed by these two scalar fields and a gauge field. 
As a result,  they exhibit interactions characteristic of the type-1.5 regime. 
On the other hand, in our study, we consider a model with only one scalar field and vary the shape of its potential away from the Mexican hat type. 
Thus, we investigate how changes in the potential affect the interaction properties of cosmic strings. 
The appearance of the type-1.5 regime in our results occurs only by coincidence. 

\begin{figure}[t]
    \begin{minipage}[b]{0.5\linewidth}
        \centering
        \includegraphics[keepaspectratio, scale=0.47]{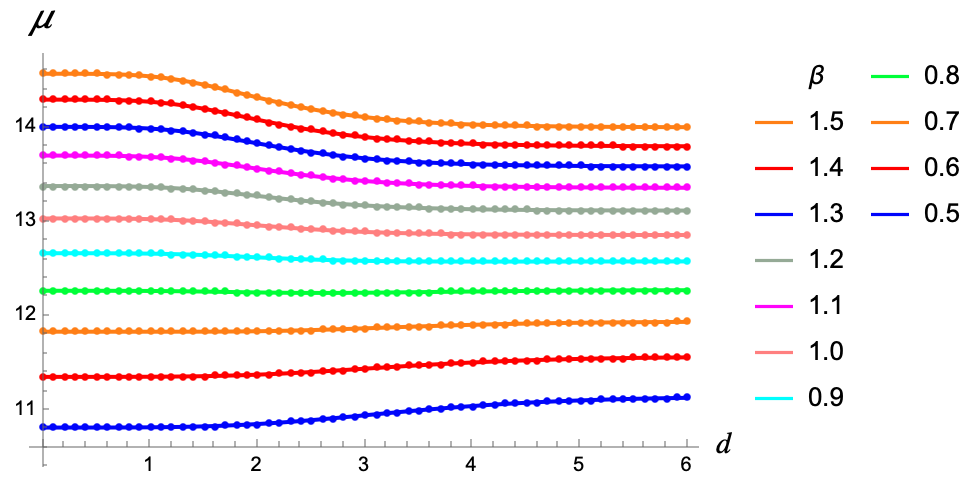}
        \subcaption{$0.5\le\beta\le1.5$}
    \end{minipage}
    \begin{minipage}[b]{0.5\linewidth}
        \centering
        \includegraphics[keepaspectratio, scale=0.47]{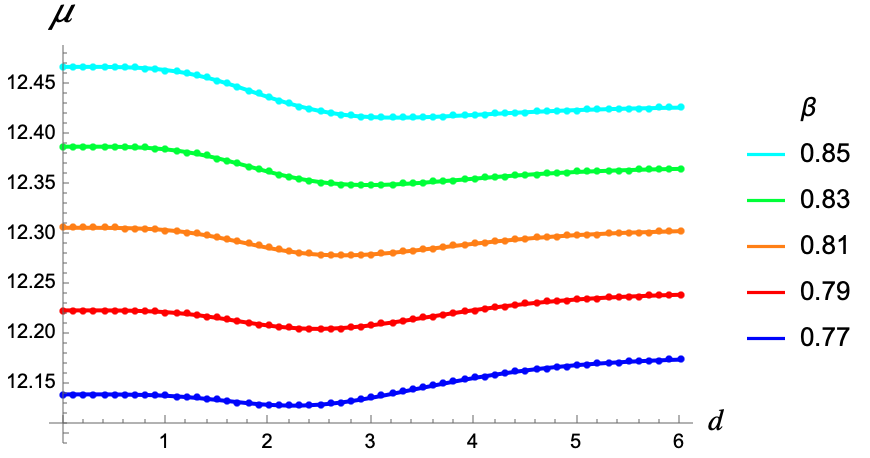}
        \subcaption{$0.77\le\beta\le0.85$}
    \end{minipage}
    \begin{minipage}[b]{0.5\linewidth}
        \centering
        \includegraphics[keepaspectratio, scale=0.47]{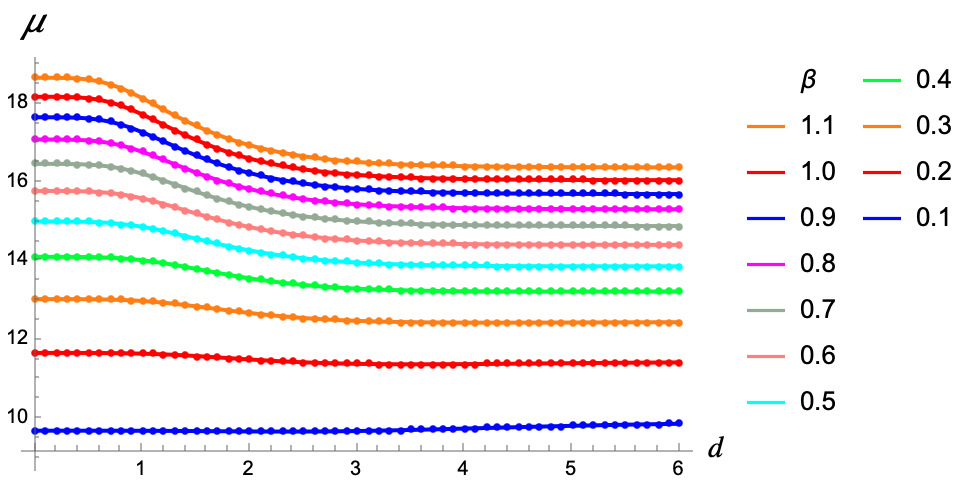}
        \subcaption{$0.1\le\beta\le1.1$}
    \end{minipage}
    \begin{minipage}[b]{0.5\linewidth}
        \centering
        \includegraphics[keepaspectratio, scale=0.47]{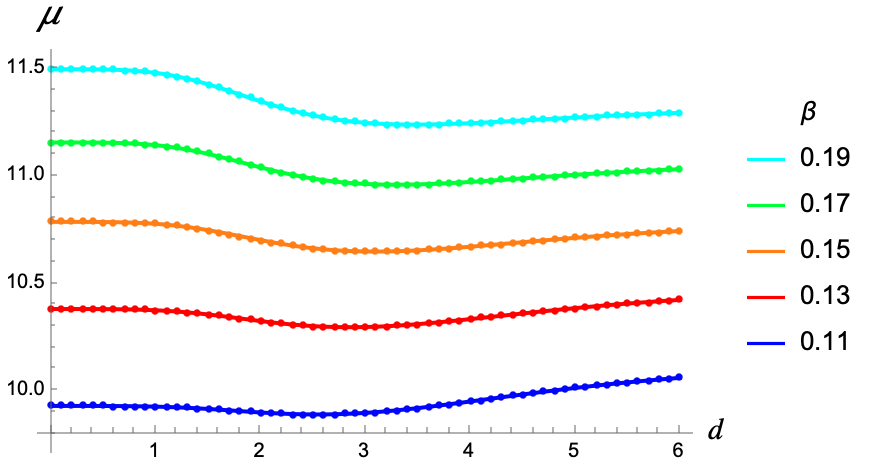}
        \subcaption{$0.11\le\beta\le0.19$}
    \end{minipage}
    \caption{
    Distance dependence of the energy per unit length in the two-strings system. 
    The top panels correspond to Case 2, and the bottom panels correspond to Case 3.
    }
    \label{ED23_energy_vs_d}
\end{figure}

The type-1.5 regime can be observed in both Case 2 and Case 3, as shown in Fig. \ref{ED23_energy_vs_d}. 
For Case 2, the type-1.5 regime is in the range $0.7\leq\beta\leq0.9$. 
In Case 3, the type-I regime does not appear within the range $0.1\leq\beta\leq1.1$. 
Here, the type-1.5 regime occurs for $0.1\leq\beta\leq0.5$, while the type-II regime begins at $\beta\geq0.6$. 

Comparing the results in Figs. \ref{ED1_energy_vs_d} and \ref{ED23_energy_vs_d}, we find that the values of $\beta$ belonging to the type-1.5 regime suggest a correlation with the shape of the scalar potentials. 
In all cases, it is common that the type-1.5 regime only appears for $\beta<1$. 
However, the specific region of $\beta$ at which the type-1.5 string occurs differ between each case. 
The boundary value of $\beta$ between the type-I and type-1.5 regimes is largest in Case 2, decreases to a lower value in Case 1, and is even lower in Case 3. 
In addition, the size of the type-1.5 regime seems to become small in the order of Case 3, Case 1, and Case 2. 
This order is the same as that in which the values of the scalar potential around the origin approach those of the Mexican hat potential, as shown in Fig. \ref{PotentialsGraph}. 
These observations suggest that the proximity of the scalar potential to the Mexican hat potential influences the extent and position of the type-1.5 regime. 
We discuss this in detail in Sec. \ref{Sec.Analysis}.

\subsubsection{For small $\beta$ cases} \label{Sec.smallbeta}

Within the framework of higher-dimensional gauge theory, the value of $\beta$ is very small, as shown in Table \ref{beta_values}. 
While our above results for $\beta\sim\mathcal{O}(0.1)$ help us to understand the relationship between string interactions and the scalar potentials, we are also interested in the behavior of interactions for such small $\beta$. 
If the interactions for small $\beta$ differ from that of conventional ANO strings, it suggests that we can distinguish between $U(1)$ breakings in higher-dimensional gauge theory and in the four-dimensional Abelian-Higgs model with the Mexican hat potential. 

However, Figs. \ref{ED1_energy_vs_d} and \ref{ED23_energy_vs_d} imply that the interactions between the strings are attractive (type-I) for small $\beta$. 
Hence, we infer that the interactions for the values of $\beta$ shown in Table \ref{beta_values} is attractive at any interstring distance. 
Unfortunately, it is difficult to perform numerical calculations for $\beta\sim10^{-3}$ with our computational resources. 
Alternatively, we estimate the interaction for $\beta=0.01$, which is the lowest value we can calculate numerically. 

\begin{figure}
    \centering
    \includegraphics[keepaspectratio, scale=0.8]{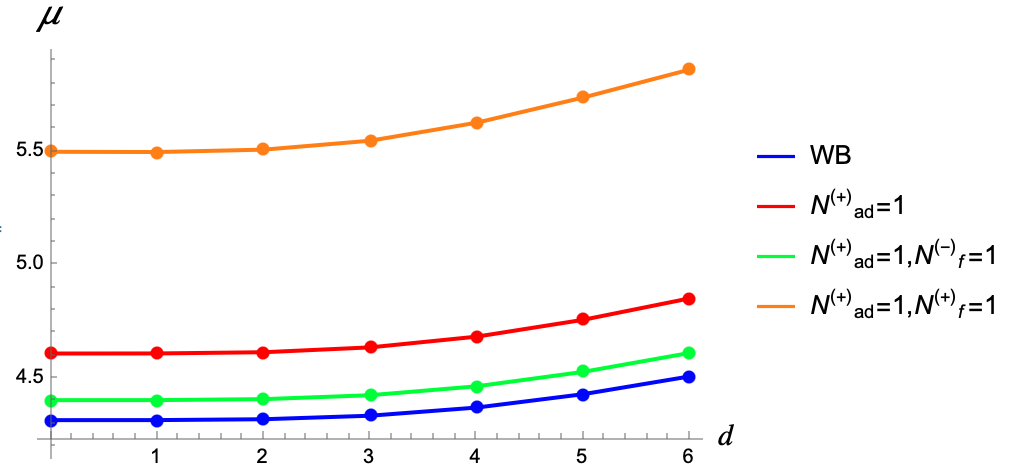}
    \caption{Dependence of the energy of two-strings system for $\beta=0.01$ in our cases.
   }
    \label{beta0.01compare}
\end{figure}

The $\mu$-$d$ graph for each case with $\beta=0.01$ is shown in Fig. \ref{beta0.01compare}. 
We find that the behavior of the lines is not significantly different from each other, and all of them indicate attractive interactions. 
Thus, we can conclude that it is difficult to distinguish them based on the string interactions. 
In other words, if cosmic string with very small $\beta$ are found, it may suggest the presence of extra-dimensional space, as considered in this paper. 

Note that our model represents a simple setup within the framework of higher-dimensional gauge theory. 
We do not entirely rule out the possibility of probing extra dimensions through cosmic strings. 
If a higher-dimensional gauge theory predicting a large $\beta$ and a scalar potential that takes large values around the origin exists, interactions in the type-1.5 regime may arise, potentially affecting the evolution of string networks.

\subsection{Analysis of the type-1.5 regime} \label{Sec.Analysis}

In this subsection, we explore the reason behind the distance-dependent interaction observed in the type-1.5 regime, based on our numerical results. 
Since this type of the interaction does not occur with the Mexican hat potential, it must be attributed to the differences in the scalar potentials. 
Therefore, our goal in this subsection is to explain the distance dependence of interaction in relation to the scalar potentials. 

The basic idea for our analysis is to consider situations for large and small distances between the strings separately. 
The key characteristic of the interaction in the type-1.5 regime is that, as the two strings approach each other, an attractive force transitions to a repulsive one at a certain distance. 
If the leading factors determining the interaction properties at large and small distances differ, this could explain the switching of the interaction force. 
In the following paragraphs, we discuss what controls the interaction properties at large and small distances individually. 

At large distances, we consider that the interaction can be described in terms of the competition between a repulsive force due to the gauge field and an attractive force due to the scalar field. 
As we have reviewed in Sec. \ref{Sec.Rev_Int_ANO}, the magnetic fluxes along the strings cause a repulsive interaction, while the force to minimize the potential energy of the scalar field leads to an attractive interaction. 
This description for the string interaction is based on which fields interact first when the two strings approach each other. 
Therefore, we believe that this framework is suitable for explaining the interaction at long distances. 
The point source formalism aligns well with this framework because it calculates the interaction energy based on the description of the superposition of the string configurations. 
We use this formalism to evaluate the string interaction at large distances in our cases.

\begin{table}[t]
    \centering
    \begin{tabular}{c|cccc} \hline
         $\beta$ & Case 1 & Case 2 & Case 3 & The MH case \\ \hline
         $1.5$ & $(2.1046, 2.0176)$ & $(2.591, 2.1082)$ & $(1.1453, 1.8002)$ & $(2.8943, 2.1563)$  \\ \hline
         $1.4$ & $(2.033, 2.0484)$ & $(2.5057, 2.0764)$ & $(1.0987, 1.8191)$ & $(2.798, 2.1945)$  \\ \hline
         $1.3$ & $(1.9627, 2.0833)$ & $(2.4206, 2.1844)$ & $(1.0533, 1.8405)$ & $(2.7060, 2.2378)$  \\ \hline
         $1.2$ & $(1.8924, 2.1224)$ & $(2.3352, 2.2309)$ & $(1.0092, 1.8649)$ & $(2.6059, 2.2874)$  \\ \hline
         $1.1$ & $(1.8221, 2.1684)$ & $(2.2492, 2.2847)$ & $(0.9662, 1.8931)$ & $(2.5092, 2.3465)$  \\ \hline
         $1.0$ & $(1.7584, 2.2242)$ & $(2.1757, 2.3481)$ & $(0.924, 1.9263)$ & $(2.4152, 2.4152)$  \\ \hline
         $0.9$ & $(1.6856, 2.2894)$ & $(2.0856, 2.4247)$ & $(0.8806, 1.9656)$ & $(2.3155, 2.4972)$  \\ \hline
         $0.8$ & $(1.6117, 2.369)$ & $(1.9941, 2.5178)$ & $(0.8379, 2.0134)$ & $(2.2134, 2.5976)$  \\ \hline
         $0.7$ & $(1.536, 2.4689)$ & $(1.9002, 2.635)$ & $(0.795, 2.073)$ & $(2.1076, 2.7241)$  \\ \hline
         $0.6$ & $(1.4576, 2.5988)$ & $(1.8017, 2.7877)$ & $(0.7522, 2.1498)$ & $(1.9968, 2.8891)$  \\ \hline
         $0.5$ & $(1.3749, 2.77559)$ & $(1.697, 2.9967)$ & $(0.7083, 2.2535)$ & $(1.8788, 3.1154)$  \\ \hline
         $0.4$ & $(1.2859, 3.0347)$ & $(1.5833, 3.3036)$ & $(0.6628, 2.4027)$ & $(1.7504, 3.4484)$  \\ \hline
         $0.3$ & $(1.1865, 3.456)$ & $(1.4559, 3.8068)$ & $(0.614, 2.6405)$ & $(1.606, 3.9965)$  \\ \hline
         $0.2$ & $(1.0692, 4.2892)$ & $(1.3041, 4.8136)$ & $(0.5586, 3.0947)$ & $(1.4341, 5.0993)$  \\ \hline
         $0.1$ & $(0.9119, 6.91)$ & $(1.1005, 8.0538)$ & $(0.4873, 4.4198)$ & $(1.2029, 8.69)$  \\ \hline
    \end{tabular}
    \caption{The values of $(c_\phi, c_A)$ for different values of $\beta$ in each case. }
    \label{PSFcharge_values}
\end{table}

We estimate charges $c_\phi$ and $c_A$, as defined in Eq. (\ref{2strings_energy}) for the strings with different values of $\beta$ in our cases. 
These charges are derived from the behavior of $f(\rho)$ and $a(\rho)$ for a single string. 
The results are summarized in Table \ref{PSFcharge_values}. 
For comparison, we also estimate $c_\phi$ and $c_A$ for the conventional ANO strings. 
We confirm that $c_\phi=c_A$ for $\beta=1.0$, which is consistent with the existence of the BPS state. 

We calculate the interaction energy $E_{int}$ as a function of the interstring distance $d$ by substituting these charges into Eq. (\ref{2strings_energy}). 
\footnote{Note that $d$ in Eq. (\ref{2strings_energy}) is not normalized as in Eq. (\ref{cartesian_normalization}). }
By taking the derivative of the energy per unit length along the strings $\mu$ with respect to $d$, we find that
\begin{align}
    \frac{d\mu}{dd} \propto \left( c_\phi^2\sqrt{\beta} K_1\left(\sqrt{2\beta}d\right) - c_A^2K_1\left(\sqrt{2}d\right) \right) \,.
    \label{deriv_PSF}
\end{align}
The sign of $d\mu/dd$ indicates whether the interaction force is attractive or repulsive. 
If $d\mu/dd$ is positive (negative), it corresponds to an attractive force (repulsive force).

Substituting charges in Table \ref{PSFcharge_values} into Eq. \eqref{deriv_PSF}, we then evaluate the interaction force at $d=6.0$ in our cases. 
In Case 1, the interaction is repulsive for $\beta\geq0.9$, while it becomes attractive for $\beta\leq0.8$.
Similarly, in Case 2, the interaction switches from repulsive to attractive between $\beta=1.0$ and $\beta=0.9$, while in Case 3, this transition occurs between $\beta=0.6$ and $\beta=0.5$. 
The values of $\beta$ at which the interaction switches coincide with the boundary between the type-II and type-1.5 regimes in each case. 
Therefore, we conclude that the long-range interaction of the strings can be described by the competition between the scalar and gauge fields, and it is evaluated by using the point source formalism. 

At small distances, the explanation using the point source formalism cannot be applied for describing the interaction. 
Hence, we need to explore another factor determining the interaction in this regime. 
In this situation, the non-linear effects arising from internal structure of the strings that influence the interaction cannot be ignored. 
Although it is difficult to evaluate these effects analytically, 
we infer that the change of field configuration must be related to the behavior of interaction energy.

\begin{figure}[t]
    \begin{minipage}[b]{0.5\linewidth}
        \centering
        \includegraphics[keepaspectratio, width=.95\columnwidth]{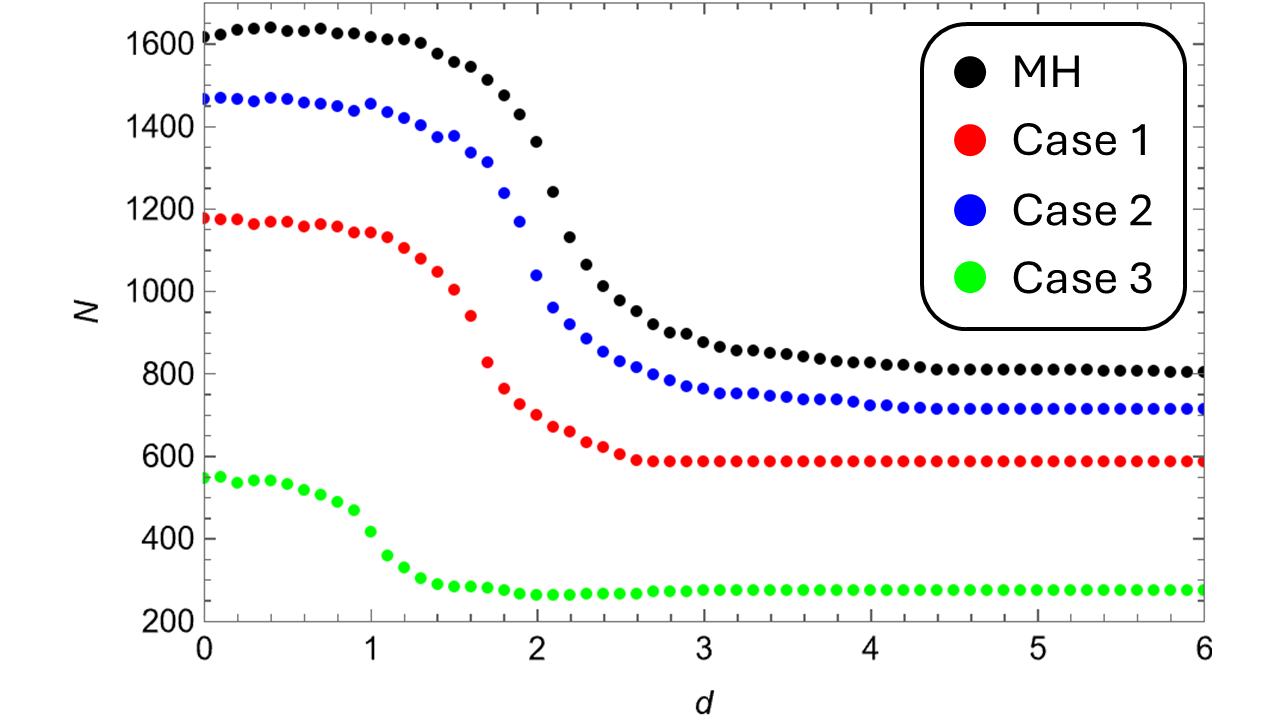}
        \subcaption{$\beta=1.0$}
    \end{minipage}
    \begin{minipage}[b]{0.5\linewidth}
        \centering
        \includegraphics[keepaspectratio, width=.95\columnwidth]{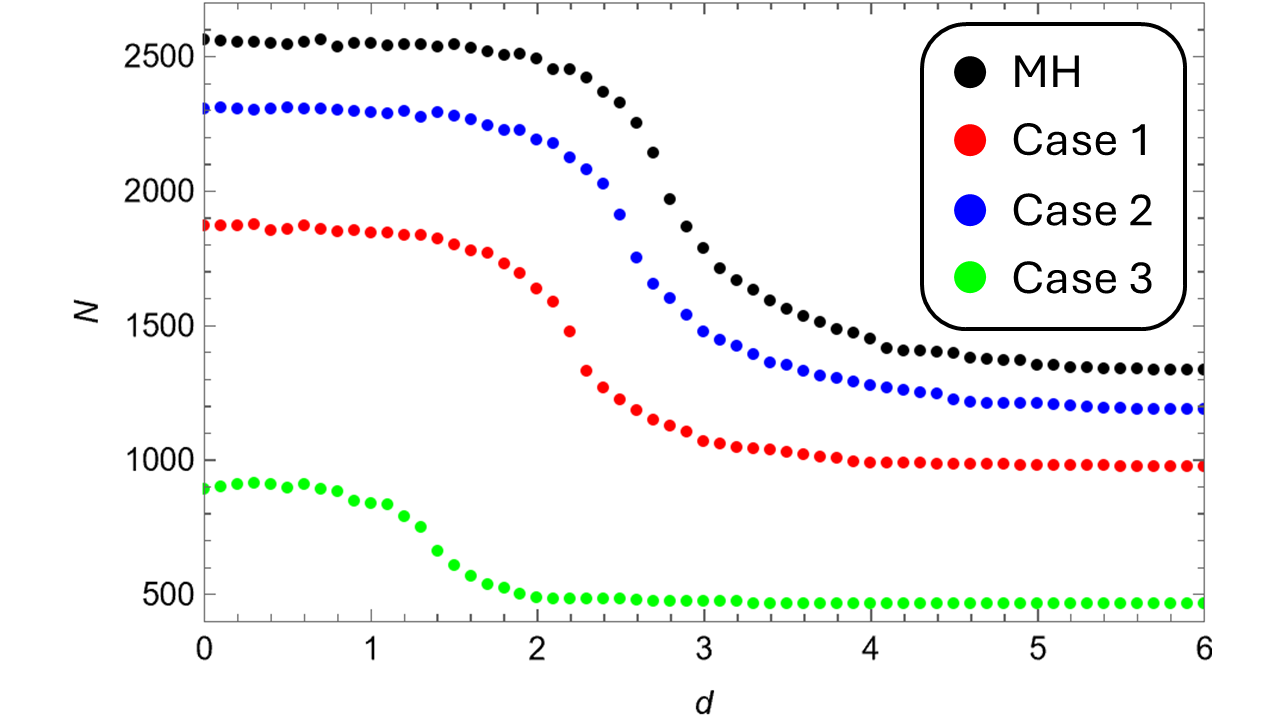}
        \subcaption{$\beta=0.5$}
    \end{minipage}
    \caption{
    Number of lattice sites where $|\tilde{\phi}|\leq0.2$ for each $d$. The left panel corresponds to $\beta=1.0$, while the right one corresponds to $\beta=0.5$. 
    }
    \label{excite_N}
\end{figure}

When the two strings are approaching to each other, the excited region of the scalar field increases. 
We show the number of lattice sites where the absolute value of the scalar field is less than or equal to 0.2 for each $d$ in each case, as depicted in Fig. \ref{excite_N}. 
We find that the region where $|\tilde{\phi}|\leq0.2$ increase around $d\sim2.0$ in all the cases we consider. 
This is a common feature for any $\beta$ that we examine, although the detailed behavior differs for each case. 

\begin{figure}[t]
    \centering
    \includegraphics[width=0.7\linewidth]{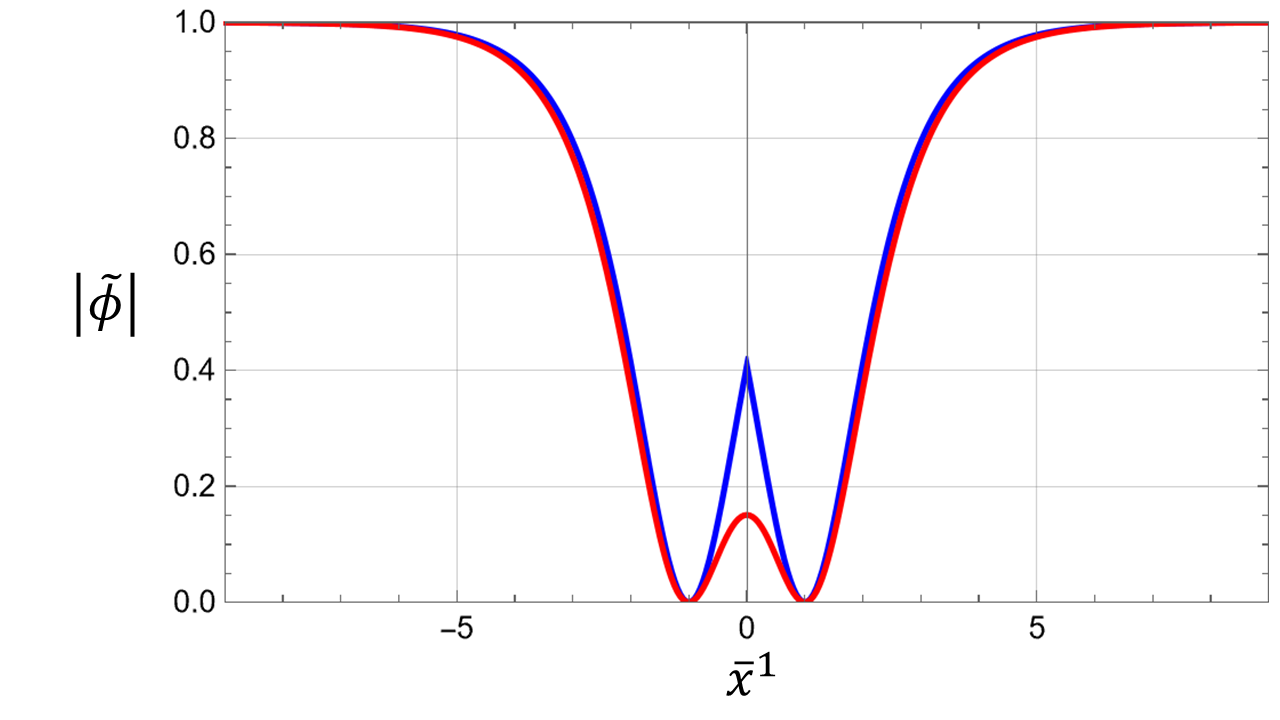}
    \caption{Comparison of the scalar field configurations along $\bar{x}^2=0$ with $\beta=0.5$ in Case 1. 
    The red line corresponds to the configuration for $d=2.0$. 
    The blue line corresponds to the configuration where the two strings are moved from $d=6.0$ to $d=2.0$ without changing their shapes.}
    \label{excite_compare}
\end{figure}

To explore which region of the scalar field is excited, we compare the scalar field configurations with the same $\beta$ but different $d$. 
In Fig. \ref{excite_compare}, we compare two configurations along $\bar{x}^2=0$ with $\beta=0.5$ in Case 1. 
One configuration is for $d=2.0$, while the other is for the scenario where the two strings are moved from $d=6.0$ to $d=2.0$ without changing their shapes. 
If the two strings were to approach each other without changing their shapes, the resulting configuration would resemble the blue line in Fig. \ref{excite_compare}. 
However, the actual field configuration is represented by the red line in Fig. \ref{excite_compare}. 
The significant difference between them is observed in the region between the cores of the two strings. 
Therefore, we conclude that the scalar fields located between the two strings significantly excited as the strings approach each other. 

\begin{figure}[t]
    \begin{minipage}[b]{0.5\linewidth}
        \centering
        \includegraphics[keepaspectratio, width=.95\columnwidth]{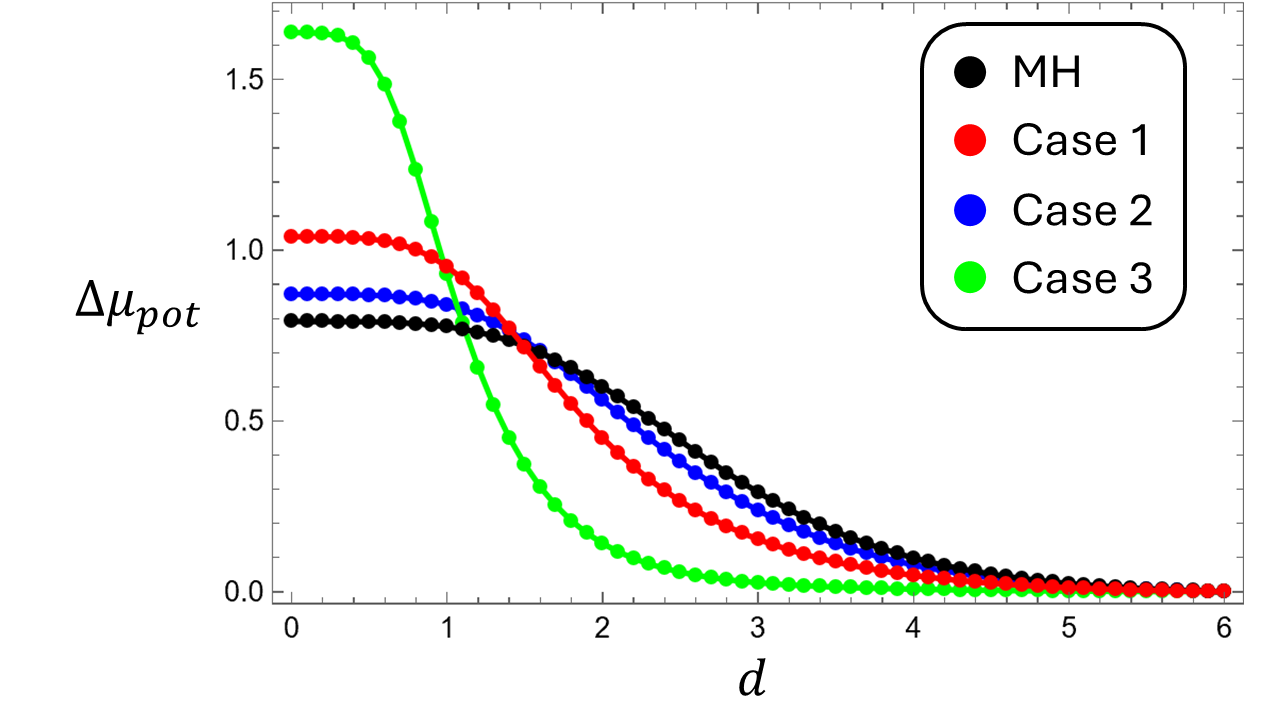}
        \subcaption{$\beta=1.0$}
    \end{minipage}
    \begin{minipage}[b]{0.5\linewidth}
        \centering
        \includegraphics[keepaspectratio, width=.95\columnwidth]{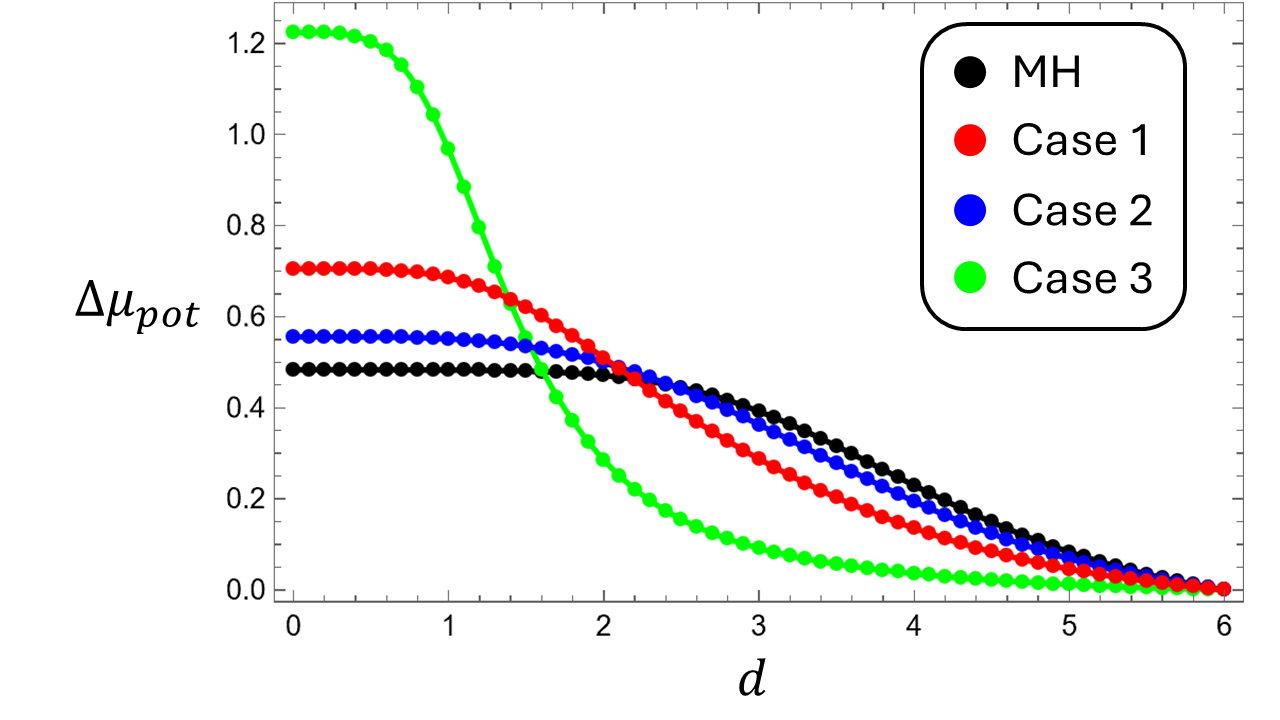}
        \subcaption{$\beta=0.5$}
    \end{minipage}
    \caption{
    Distance dependence of the potential energy $\mu_{pot}$ for each case. $\Delta\mu_{pot}(d)\equiv\mu_{pot}(d)-\mu_{pot}(6.0)$. The left panel shows the result for $\beta=1.0$, while the right panel shows that for $\beta=0.5$.
    }
    \label{potentialenergy_vs_d}
\end{figure}

The increase in the excited region of the scalar fields suggests that the contribution from the potential energy also increases. 
Specifically, the contribution from the potential energy is anticipated to be larger when the values of the scalar potential around the origin are larger. 
To confirm this, we investigate how the potential energy depends on $d$ in each case. 
Here, we denote the contribution from the potential energy to the energy per unit length as $\mu_{pot}$. 
We examine $\mu_{pot}$ over the range $d=0.0$ to $6.0$, and calculate the differences from $\mu_{pot}$ at $d=6.0$, which we denote as $\Delta\mu_{pot}(d)\equiv\mu_{pot}(d)-\mu_{pot}(6.0)$. 
The graph of $\Delta\mu_{pot}(d)$ for $\beta=1.0$ and $\beta=0.5$ are shown in Fig. \ref{potentialenergy_vs_d}. 
We find that $\Delta\mu_{pot}$ in our cases increases more rapidly as $d$ decreases than in the MH case at $d\lesssim3$. 
Since an increase in the interaction energy as $d$ decreases corresponds to a repulsive interaction between the two strings, we conclude that the behavior of $\mu_{pot}$ in our cases causes the repulsive force at small $d$ in the type-1.5 regime. 
Unfortunately, it is difficult for us to quantitatively evaluate the interaction energy at small $d$ based on this framework. 
Developing a method to do so is beyond the scope of this paper. 

Summarizing the above discussions, the interaction at large and small distances between the strings are determined by different reasons. 
The interaction at large distances is determined by the competition between the scalar and gauge fields. 
In contrast, at small distances, the interaction is strongly influenced by the behavior of the potential energy, which is affected by changes in the excited region of the scalar field. 

A common feature of these factors is that they cause a repulsive force between the strings for large $\beta$ but an attractive force for small $\beta$. 
We can easily confirm this at large $d$ by using Eq. (\ref{deriv_PSF}). 
At small distances, Fig. \ref{potentialenergy_vs_d} shows that the increase of $\Delta\mu_{pot}$ as $d$ decreases becomes less significant as $\beta$ decreases. 
This suggests that the contribution from the potential energy to the interaction energy is not important at small $\beta$. 
Alternatively, the decrease of the gradient energy as $d$ decreases is what makes the interaction attractive for such small $\beta$. 
Importantly, the critical values of $\beta$ where the switch between repulsion and attraction at large and small distances do NOT match in general.
We find that this discrepancy causes the $d$-dependent interaction between the strings.
Based on this idea, we reinterpret the results of several cases in the following paragraphs.

For the MH case, the values of $\beta$ at which the switch between repulsion and attraction occurs are the same at any distances. 
This consistency is guaranteed by the existence of the BPS state. 
When $\beta=1.0$, the interaction energy does not depend on $d$, resulting in no interaction force between the strings. 
From our view point, this suggests that the critical values of $\beta$ align to 1.0 at large and small distances. 
Consequently, the interaction becomes repulsive for $\beta>1$ at any distances, and vice versa. 

On the other hand, the situations in Case 1, Case 2, and Case 3 are different: the critical values of $\beta$ deviate between large and small distances.
Through the analysis using the point source formalism, we find that the values of $\beta$ at which the switch between repulsion and attraction occurs shifts from $\beta=1.0$ at large distances. 
The more the value of the potential around the origin deviates from the Mexican hat potential, the greater the decrease in $\beta$, as we discussed above. 
The boundary value of $\beta$ also shifts at small distances.
However, the magnitude of this shift differs from that observed at large distances. 
This is because the contribution from the potential energy, which arises from the difference in scalar potentials, significantly affects the interaction energy. 
Consequently, the boundary value of $\beta$ decreases further, leading to a region of $\beta$ where the interaction becomes attractive at large distances but becomes repulsive at small distances. 
That is precisely the type-1.5 regime. 
Based on this interpretation, the type-1.5 regime emerges between the type-I and type-II regimes. 
This new perspective is consistent with our results. 

\begin{figure}[t]
    \centering
    \includegraphics[width=0.7\linewidth]{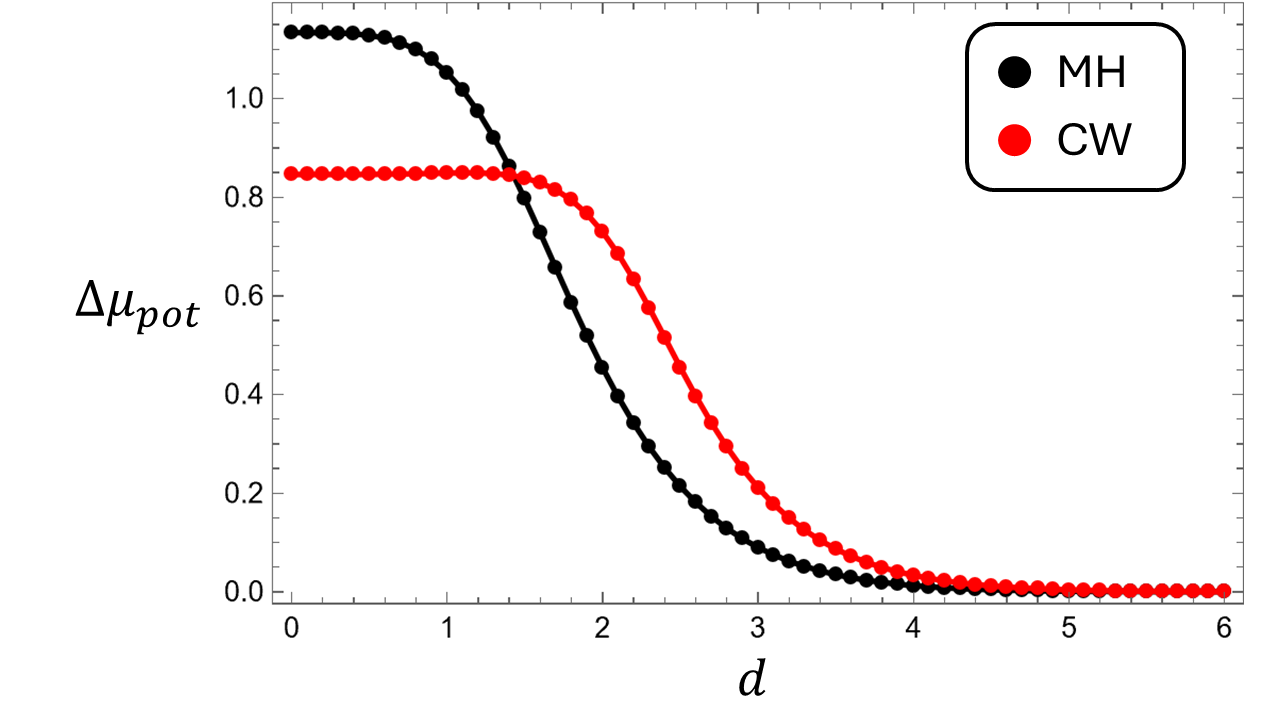}
    \caption{Distance dependence of the potential energy $\Delta\mu_{pot}$ for $\beta=2.0$. 
    The red line represents the case with the CW potential, while the black line represents the case with the Mexican hat potential. }
    \label{cw_potential_graph}
\end{figure}

Furthermore, our interpretation can be applied to the case with the CW potential, as investigated in Ref. \cite{Eto:2022hyt}. 
In this previous study, it has been observed that the interaction force is repulsive at large distances but attractive at small distances for a certain range of $\beta$, as shown in Fig 9 of Ref. \cite{Eto:2022hyt}. 
This range of $\beta$ is referred to as the type-$\bar{1.5}$ . 
We reinterpret their results based on our approach. 
Given that the type-$\bar{1.5}$ regime appears around $\beta \sim 2.0$, we select $\beta=2.0$ as a benchmark for our analysis.
First, we evaluate the interaction energy of the two CW-ANO string system using the point source formalism. 
We estimate charges as $(c_\phi, c_A)=(4.6377, 2.1496)$ based on a calculation of a single CW-ANO string. This result indicates that the interaction is repulsive at $d=6.0$, and it is consistent with Ref. \cite{Eto:2022hyt}.
Second, we investigate the $d$ dependence of the potential energy. 
The result is shown in Fig. \ref{cw_potential_graph}. 
We find that the potential energy with the CW potential does not change much at small distances compared to that with the Mexican hat potential.

Based on the above results, we can infer the following explanation for the emergence of the type-$\bar{1.5}$ regime in the case with the CW potential. 
In general, the values of $\beta$ at which the switch between repulsion and attraction occurs shift to the region where $\beta>1.0$. 
However, details are different between large and small distances. 
Since the shape of the string does not change significantly, the picture of the competition between the scalar and gauge fields does not deviate significantly from that in the MH case. 
Thus, the boundary value of $\beta$ slightly shift from $\beta=1.0$ at large distances. 
In contrast, the difference in the scalar potential strongly affects the potential energy, causing a significant deviation in the interaction energy behavior at small distances. 
Due to this, the boundary value of $\beta$ shifts significantly from $\beta=1.0$ at small distances. 
Consequently, a region where the interaction is repulsive at large distances but attractive at small distances emerges somewhere for $\beta>1.0$. 
That is precisely the type-$\bar{1.5}$ regime. 

As discussed above, our approach successfully describes both the type-1.5 regime in our cases and type-$\bar{1.5}$ regime observed in Ref. \cite{Eto:2022hyt}. 
We anticipate that this approach can also be applied to various types of scalar potentials. 
One of the challenges we face is the analytical evaluation of the interaction energy at small distances. 
Overcoming this challenge would enable a deeper understanding of the interaction between strings with various scalar potentials.

\section{Conclusion and discussion}\label{conclusion}

We have investigated ANO strings with the one-loop effective potentials induce by the higher-dimensional gauge theory and its interaction between strings. 
In our study, One of our motivations is to deeply understand the relationship between the scalar potentials and the properties of strings, as pointed out in Ref. \cite{Eto:2022hyt}.
We have considered a five-dimensional $SU(2)$ gauge theory with an extra-dimensional space $S^1/Z_2$. 
Due to the orbifold, $SU(2)$ gauge symmetry is explicitly broken to $U(1)$ symmetry in the four-dimensional effective theory. 
Furthermore, $U(1)$ symmetry is spontaneously broken by developing the VEV of the WL scalar field (Hosotani mechanism), which is why we considered this above model. 
Since the shape of the effective potential differs in what kind of fermions are introduced, we analyze three cases. 
In all cases, the height of the normalized effective potential around the origin is higher than that of the Mexican hat potential, though these heights differ from each other. 

Before investigating the interactions, we have numerically examined single string solutions in our models.  
We have found that the shape of the strings are slightly narrower than the conventional ANO string for the same $\beta$. 
We have also calculated the $\beta$ dependence of the tension of a string $\mu$ per the winding number $n$. 
We have found that there is no critical value of $\beta$ corresponding to the BPS state. 
The values of $\beta$ where the lines in Fig. \ref{Bogomolnyi_nad1} intersect are smaller than those in the MH case. 
Based on the results in Ref. \cite{Eto:2022hyt} and our results, we have inferred that the deviations from the MH case are due to variations in the scalar potential values around the origin. 

We have estimated the interaction energy of the two-string systems as a function of the interstring distance $d$ in our models.
We have found a region of $\beta$ where the interaction energy is minimized at a certain distance $d_c(\neq0)$. 
In this region, the interaction force switches from attraction to repulsion as two strings approach each other. 
We call this region of $\beta$ as the type-1.5 regime, since similar $d$-dependent interaction has known as the type-1.5 superconductivity in condensed matter physics \cite{Babaev:2004hk,PhysRevLett.102.117001}. 
Furthermore, an intriguing aspect is that the $d$ dependence of the string interactions differs from that of the CW-ANO string: 
the attractive/repulsive relation for the CW-ANO strings is opposite to what we observed in our cases. 

To interpret the emergence of the type-1.5 regime in our models, we have considered the leading factors that determine the interaction at both large and small distances. 
We have confirmed that the interaction at large distances is well described by using the point source formalism, and that it is determined by the competition between the scalar and gauge fields. 
Since it is difficult to apply the point source formalism at small distances, we numerically have analyzed the interaction in this regime by focusing on the changes of field configurations of the scalar field. 
As a result, we have found that the potential energy in our models changes more significantly at small distances compared to the MH case. 
We have inferred this significant change in potential energy contributes to the emergence of the repulsive force in the type-1.5 regime. 
Both factors cause a repulsive (attractive) force for large (small) values of $\beta$. 
The important things is that the critical values of $\beta$ where the switch between repulsion and attraction at large and small distances do NOT match in general, because the leading factors are different from each other. 
Building on this idea, we can interpret the emergence of the type-1.5 regime in our models as a result of the deviation in the critical $\beta$ at large and small distances.  
Furthermore, this interpretation can be applied to the results in the CW case. 
It is also worth noting that this interpretation is consistent with the MH case, as the BPS state guarantees that the critical values $\beta$ are the same at both large and small distances. 

We conjecture that the values around the origin of the normalized scalar potentials are related to the emergence of $d$-dependent interactions: if the value is larger (smaller) than that of the Mexican hat potential, the type-1.5 (type-$\bar{1.5}$) regime emerges. 
To prove this conjecture, we need to estimate the interaction energy at small distances quantitatively and analytically, similar to the point source formalism used for the interaction energy at large distances. 
However, the analysis presented in this paper is not sufficient to develop such a method. 
We believe that further investigation into the interaction for the scalar potentials with various shapes is required, but this is beyond the scope of this paper.

Finally, we have investigated the interaction energy for $\beta=0.01$, because a very small value of $\beta$ is predicted in the higher-dimensional gauge theory, as shown in Table \ref{beta_values}. 
We have found that the $d$ dependence of the interaction disappear as $\beta$ decreases, and that the interaction is attractive (type-I) for small $\beta$. 
Indeed, for $\beta=0.01$, the interaction energies in our models exhibit behaviors that closely resemble the interaction energy observed in the MH case. 
Therefore, we conclude that it is difficult to distinguish between the Abelian-Higgs model with the Mexican hat potential and our models based on the properties of cosmic strings.


On the other hand, investigating the dynamics of the strings studied here is crucial. 
While we have analyzed the interaction energy of a system with two parallel strings, the reconnection dynamics and the evolution of networks of them remain as open questions. 
Since cosmic strings are one of the targets of various cosmological observations, including the gravitational waves observations, it is important to understand the dynamics of them.

Our study may also be applicable to condensed matter physics. 
As mentioned earlier, the type-1.5 string was first identified in researches for the superconductors \cite{Babaev:2004hk, PhysRevLett.102.117001}. 
While two scalar fields were introduced in those studies, we have shown that the type-1.5 string can emerge in a model with a single scalar field and the scalar potential different from the Mexican hat type. 
Our findings suggest that this approach may be applicable to describing superconducting properties, opening up new possibilities for research in condensed matter physics.

In this paper, we have focused on the string solution derived from the ansatz in Eq. (\ref{stringansatz_ED}) to investigate the relation between the scalar potentials and the string interactions. 
However, it is likely that cosmic strings in higher-dimensional theories possess more complex structures than what we have pointed out. 
For example, fermions which we introduced to trigger the Hosotani mechanism form Yukawa couplings with the scalar field, and this Yukawa couplings induce fermionic zero modes on the string \cite{Witten:1984eb,Jackiw:1981ee}. 
This type of string is known as a fermionic superconducting string, and its dynamics differ from those of the ANO string \cite{Ostriker:1986xc, Davis:1988jp, Davis:1988jq, Ganoulis:1989hz}. 
In addition, it should be noted that the situation becomes complex when the periodicity of effective potentials is considered. 
Taking this periodicity into account, the full moduli space is homeomorphic not to a single $S^1$, but to a direct sum of multiple $S^1$ spaces. 
As a result, domain wall-like defects or other composite defects may also exist. 
Investigating them and their potential applications to physical phenomena is left for future work.


\section*{Acknowledgments}
We would like to thank Nobuhito Maru and Takashi Hiramatsu for their valuable discussions. 
We also would like to thank Naoya Kitajima for techniques of the numerical calculations. 
This work was supported by JSPS KAKENHI Grant No. JP24KJ1257 (Y.K.).

\appendix
\section{Point Source Formalism}\label{PSF}
In this appendix, we review the point source formalism (see \cite{Speight:1996px,Fujikura:2023lil}).
First, we find the external sources that represent the presence of the ANO string with the Abelian-Higgs potential \eqref{AHpotential}.
We take $\phi(x)$ and $A_\mu(x)$ as
\begin{align}
    \label{phi_Amu:sigma_Umu}
    \phi(x) = \left(v + \frac{\sigma(x)}{\sqrt{2}}\right) e^{i\pi(x)},\quad
    A_\mu(x) = U_\mu + \frac{1}{g}\partial_\mu\pi(x),
\end{align}
where $\sigma(x)$, $\pi(x)$ and $U_\mu(x)$ are the radial component of the scalar, the would-be NG boson and the massive $U(1)$ gauge fields, respectively.
Comparing Eq. \eqref{NOstring} and Eqs. \eqref{fa_asympt} with Eq. \eqref{phi_Amu:sigma_Umu}, we read $\sigma(x),\pi(x)$ and $\vec{U}(x)$ as
\begin{align}
    \label{concrete_configurations}
    \sigma(x)=-c_\phi vK_0(m_\phi r),\quad
    \pi(x)=n\theta,\quad
    \vec{U}(x)=\frac{n c_A v}{m_A} \Vec{e}_z \times \nabla K_0(m_Ar).
\end{align}

Defining the masses of $U_\mu, \sigma$ as $m^2_A=2g^2 v^2, m^2_\phi=4\lambda v^2$ respectively, the quadratic terms in the Lagrangian \eqref{AHmodel} with Abelian-Higgs potential is given by
\begin{align}
    \label{Lagrangian_PointSourceFormalism}
    \mathcal{L} = -\frac{1}{4}F_{\mu\nu}F^{\mu\nu} - \frac{1}{2}\partial_\mu\sigma\partial^\mu\sigma - \frac{1}{2} m_A^2 U_\mu U^\mu - \frac{1}{2}m_\phi^2\sigma^2 - J_\sigma\sigma - j^\mu U_\mu.
\end{align}
Here, we introduced the source terms $J_\sigma$ and $j^\mu$.
From Eq. \eqref{Lagrangian_PointSourceFormalism}, the equations of motion are expressed as
\begin{align}
    \label{eom_Jsigma}
    &\left(\partial_\mu\partial^\mu - m_\phi^2\right) \sigma = J_\sigma, \\
    \label{eom_JUmu}
    &\left(\partial_\nu\partial^\nu - m_A^2\right) U_\mu = j_\mu + \frac{1}{m_A^2}\partial_\mu\left(\partial^\nu j_\nu\right).
\end{align}
Substituting Eq. \eqref{concrete_configurations} into Eqs. \eqref{eom_Jsigma} and \eqref{eom_JUmu}, $J_\sigma$ and $j_i$ are represented by
\begin{align}
    J_\sigma=2\pi c_\phi v\delta^{(2)}(\bm{x}),\quad
    j_i=\frac{2\pi\phi v c_A v}{m_A}\epsilon_{i j}\partial_{j}\delta^{(2)}(x).
\end{align}

We evaluate the interaction between two strings by the changes of energy dependent on the distance of two strings.
We assume that two strings are parallel each other.
We introduce the two-dimensional plane with perpendicular to the string.
This two-dimensional Cartesian coordinate is described as $\bm{x} = (x_1,x_2)$.
When the superposition of two strings occurs, the source terms are the sum of the source term of each string like
\begin{align}
    \label{source_superposition_sigma}
    J_\sigma(x) &= 2\pi c_{\phi1} v \delta^{(2)}(x-x_1) + 2\pi c_{\phi2} v \delta^{(2)}(x-x_2), \\
    \label{source_superposition_U}
    j_i(x) &= \frac{2\pi n_1 c_{A1} v}{m_A} \epsilon_{ij} \partial_j \delta^{(2)}(x-x_1) + \frac{2\pi n_2 c_{A2} v}{m_A} \epsilon_{ij} \partial_j \delta^{(2)}(x-x_2),
\end{align}
where $c_{\phi 1}(c_{\phi2})$ and $c_{A 1}(c_{A2})$ are the constants of the first (second) string.

Next, we derive the energy of this system.
Using Eqs. \eqref{eom_Jsigma}, \eqref{eom_JUmu} and the propagator in the two-dimensional plane
\begin{align}
    \label{2Dpropagator}
    D(x,m)\equiv-\int d^2p\frac{e^{ipx}}{p^2+m^2}=-\frac{1}{2\pi}K_{0}(m|\bm{x}|),
\end{align}
$\sigma(x)$ and $U_i(x)$ are expressed as
\begin{align}
    \sigma &= \int d^2x' D(x-x',m_\phi) J_\sigma(x'), \\
    U_i &= \int d^2x' D(x-x',m_A) j_i(x').
\end{align}
Thus, the Lagrangian \eqref{Lagrangian_PointSourceFormalism} is summarized as
\begin{align}
    \mathcal{L}=-\frac{1}{2}J_\sigma \sigma-\frac{1}{2}j_i U_i.
\end{align}
Integrating in the three-dimensional space, the energy of the system has
\begin{align}
    E &= -\int d^3x \mathcal{L} \nonumber\\
    &= \frac{1}{2} \int dz \int d^2x\int d^2x' \left[ J_\sigma(x) D(x-x',m_\phi) J_\sigma(x') + j_i(x) D(x-x',m_A) j_i(x') \right].
\end{align}
Using Eq. \eqref{2Dpropagator}, we conclude the result of the energy of this system as
\begin{align}
    \label{Energy_PSF}
    E = 2\pi v^2 \int dz \left[ n_1n_2 c_{A1} c_{A2}  K_0(m_Ad) -  c_{\phi1} c_{\phi2} K_0(m_\phi d) \right],
\end{align}
where $d\equiv|x_1-x_2|$ is the distance of two strings.

\section{Gradient flow method}\label{GFM}

In this appendix, we describe how we solved the equations of motion considered in this paper. 
We employed the relaxation method, also referred to as the gradient flow method \cite{Eto:2022hyt,Fujikura:2023lil}. 

First, we review the basic idea of this method. 
Let us consider a function $p(x)$ and its associated functional $Q[p]$. 
The gradient flow method is useful for finding a configuration of $p(x)$ that minimizes $Q[p]$ if such a configuration exists. 
We introduce $\tau$, referred to as a flow time, as an additional variable of $p(x)$, \textit{i.e.} $p(x,\tau)$. 
Starting from a given configuration at a certain $\tau$, we evolve $p(x,\tau)$ according to the following differential equation: 
\begin{align}
    \frac{\partial p(x,\tau)}{\partial\tau} = - \frac{\delta Q}{\delta p} \,.
\end{align}
This equation is called as the flow equation. 
Since $p(x,\tau)$ evolves to decrease $Q$, we obtain the configuration that minimizes $Q[p]$ at a sufficiently large $\tau$. 

Since stable classical solutions minimize the energy of the system, we numerically derive them by employing the gradient flow method. 
In the following subsections, we explain the cases for a single string (Eqs. \eqref{eom_f_nmlzd} and \eqref{eom_a_nmlzd}) and a two-string system. 
Several of the techniques we used are primarily based on the approaches described in Refs. \cite{Eto:2022hyt, Fujikura:2023lil}.

\subsection{Single string solution}

As we mentioned,  $f(\rho)$ and $a(\rho)$ in Eq. \eqref{NOstring} determine the shape of string. 
While they are the solution of the equations of motion \eqref{eom_f_nmlzd} and \eqref{eom_a_nmlzd}, we now derive them numerically by employing the gradient flow method.
The energy per unit length of a single string $\mu$ is shown in Eq. \eqref{mu_normalized_general_single}. 
Since $f(\rho)$ and $a(\rho)$ minimize $\mu$, the flow equations are given by
\begin{align}
    \frac{\partial f}{\partial\tau}&=-\frac{\delta \mu}{\delta f} = 2\rho f'' + 2f' - \frac{2n^2(1-a)^2}{\rho}f - \rho \frac{\partial \widetilde{V}}{\partial f} \,,\\
    \frac{\partial a}{\partial\tau}&=-\frac{\delta \mu}{\delta a} = \frac{n^2}{\rho}a'' - \frac{n^2}{\rho}a' + \frac{n^2f^2}{\rho}(1-a) \,,
\end{align}
where we introduce a flow time $\tau$ as $f(\rho,\tau)$ and $a(\rho,\tau)$. 
We set the initial configuration satisfying the boundary condition \eqref{BC} as
\begin{align}
    f(\rho,0)= \tanh\left(\sqrt{\beta}\rho\right) \,,\quad a(\rho,0)= \tanh{\rho^2} \,.
\end{align}
We evolve $f(\rho,\tau)$ and $a(\rho,\tau)$ from $\tau=0$ to $\tau=500$ by \textit{Mathematica} \cite{Mathematica}, and obtain the classical solutions.

\subsection{Field configuration of a two-string system}

We derive the field configurations of $\tilde{\phi}$ and $\tilde{A}_\mu$ in a two-string system by employing the point source formalism. 
We use the temporal gauge $\tilde{A}_0=0$ and assume $\tilde{A}_3=0$. 
Since the energy per unit length of the system is given in Eq. \eqref{mu_normalized_general}, we create the flow equations to minimize this. 
For simplicity, we write the real and imaginary parts of $\tilde{\phi}$ as $\tilde{\phi}_1$ and $\tilde{\phi}_2$, respectively. 
Hence, the flow equations are obtained as
\begin{align}
    \frac{\partial\tilde{\phi}_1}{\partial\tau} &= \partial_i^2\tilde{\phi}_1 - \tilde{A}_i^2\tilde{\phi}_1 + 2 \tilde{A}_i\partial_i\tilde{\phi}_2 - \frac{1}{2}\frac{\partial \widetilde{V}}{\partial\tilde{\phi}_1} \,, \\
    \frac{\partial\tilde{\phi}_2}{\partial\tau} &= \partial_i^2\tilde{\phi}_2 - \tilde{A}_i^2\phi_2 - 2 \tilde{A}_i\partial_i\tilde{\phi}_1 - \frac{1}{2}\frac{\partial \widetilde{V}}{\partial\tilde{\phi}_2} \,, \\
    \frac{\partial\tilde{A}_1}{\partial\tau} &= \partial_i^2 \tilde{A}_1 - 2|\tilde{\phi}|^2 \tilde{A}_1 + 2 \left(\tilde{\phi}_1\partial_1\tilde{\phi}_2 - \tilde{\phi}_2\partial_1\tilde{\phi}_1\right) \,, \\
    \frac{\partial\tilde{A}_2}{\partial\tau} &= \partial_i^2 \tilde{A}_2 - 2|\tilde{\phi}|^2 \tilde{A}_2 + 2 \left(\tilde{\phi}_1\partial_2\tilde{\phi}_2 - \tilde{\phi}_2\partial_2\tilde{\phi}_1\right) \,,
\end{align}
where $i=1,2$. 
We set the initial configuration at $\tau=0$ as follows: 
\begin{align}
    \tilde{\phi}(\bar{x},0) &= F(\bar{x}) e^{i\theta_1+i\theta_2} \,, \\
    \tilde{A}_i(\bar{x},0) &= F(\bar{x}) \left[a_{i+}(\bar{x}) + a_{i-}(\bar{x})\right] \,,
\end{align}
where
\begin{align}
    \tan\theta_1\equiv\frac{\bar{x}_1-d/2}{\bar{x}_2}, \quad \tan\theta_2\equiv\frac{\bar{x}_1+d/2}{\bar{x}_2} \,
\end{align}
and
\begin{align}
    F(\bar{x}) &= \tanh\left(\sqrt{(\bar{x}_1-d/2)^2+\bar{x}_2^2}\right) \tanh\left(\sqrt{(\bar{x}_1+d/2)^2+\bar{x}_2^2}\right) \,,\\
    a_{1\pm}(\bar{x}) &= \left\{
    \begin{aligned}
        \frac{-\bar{x}_2}{(\bar{x}_1\mp d/2)^2 + \bar{x}_2^2} \qquad &((\bar{x}_1\mp d/2)^2 + \bar{x}_2^2 \neq 0) \\
        0 \qquad\qquad\quad &((\bar{x}_1\mp d/2)^2 + \bar{x}_2^2 = 0)
    \end{aligned}
    \right. \,, \\
    a_{2\pm}(\bar{x}) &= \left\{
    \begin{aligned}
        \frac{\bar{x}_1\mp d/2}{(\bar{x}_1\mp d/2)^2 + \bar{x}_2^2} \qquad ((\bar{x}_1\mp d/2)^2 + \bar{x}_2^2 \neq 0) \\
        0 \qquad\qquad\quad ((\bar{x}_1\mp d/2)^2 + \bar{x}_2^2 = 0)
    \end{aligned}
    \right. \,.
\end{align}
Note that these initial configurations have a winding number $n=2$. 
They also satisfy the boundary conditions such that $\tilde{\phi}(\bar{x},0)\rightarrow e^{i\theta_1+i\theta_2}$ and $|(\partial_1-i\tilde{A}_i)\tilde{\phi}(\bar{x},0)|\rightarrow 0$ at $\sqrt{\bar{x}_1^2+\bar{x}_2^2}\rightarrow\infty$. 
In addition, the gauge symmetry is restored at the cores of the strings. 
Therefore, these are suitable initial configurations of a two-string system. 
We evolve them from $\tau=0$ to $\tau=15$ in steps of $\Delta\tau=0.0005$ by the standard Euler method. 

Following the approach in Refs. \cite{Jacobs:1978ch,Fujikura:2023lil}, we impose two constraints during the flow time evolution. 
First, we fix $\tilde{\phi}(\pm d/2,0,\tau)=0$ at each step to maintain the separation of the strings. 
Second, we fix the complex phase of $\tilde{\phi}$ at each step of the evolution to conserve the total winding number of the system. 
Specifically, we first evolve $\tilde{\phi}(\bar{x},\tau)$ to $\tilde{\phi}(\bar{x},\tau+\Delta\tau)$ by the standard Euler method, and then replace $\tilde{\phi}(\bar{x},\tau+\Delta\tau)$ with $|\tilde{\phi}(\bar{x},\tau+\Delta\tau)|e^{i\theta_1+i\theta_2}$. 
As a result, the phases of $\tilde{\phi}$ at each point remain unchanged. 
Due to these constraints, we can obtain the field configuration of a two-string separated by any distance.

\let\doi\relax
\bibliographystyle{utphys28mod}
\bibliography{Ref}

\end{document}